\documentclass[journal]{rmaa}

\usepackage{paralist}
\usepackage{natbib}

\usepackage{psfrag,color}
\usepackage{soul} 

\usepackage[latin1]{inputenc}

\usepackage{threeparttable}
\usepackage{eqnarray,amsmath}

\def\mha{\mbox{$M_{\rm HI}$}}
\def\mhm{\mbox{$M_{\rm H_{2}}$}}
\def\mg{\mbox{$M_{\rm gas}$}}
\def\msun{\mbox{M$_{\odot}$}}
\def\RHI{\mbox{$R_{\rm HI}$}}
\def\RH2{\mbox{$R_{\rm H_{2}}$}}
\def\Rgas{\mbox{$R_{\rm gas}$}}
\def\MH2MHIMs{\mbox{$M_{\rm H_{2}}/M_{\rm H_{I}}$}}
\def\cotoh2{\mbox{$\rm CO\mbox{-to-}H_{2}$}}
\def\a_co{\mbox{$\alpha_{\rm CO}$}}

\def\HI{\mbox{$\rm H_{I}$}}
\def\H2{\mbox{$\rm H_{2}$}}

\def\ms{\mbox{$M_{*}$}}
\def\mha{\mbox{$M_{HI}$}}
\def\mhm{\mbox{$M_{H_{2}}$}}
\def\mg{\mbox{$M_{\rm gas}$}}
\def\msun{\mbox{M$_{\odot}$}}
\def\h2toco{\mbox{$\rm H_{2}-\mbox{to}-CO$}}
\def\RHI{\mbox{$R_{\rm HI}$}}
\def\RH2{\mbox{$R_{\rm H_{2}}$}}
\def\Rgas{\mbox{$R_{\rm gas}$}}
\def\HI{\mbox{$\rm HI$}}
\def\H2{\mbox{$\rm H_{2}$}}
\def\lMH2{\mbox{$\log_{10}(M_{\rm H_{2}}/M_{\odot})$}}
\def\Pjhi{\mbox{$P_j(\RHI|\ms)$}}
\def\Pjhii{\mbox{$P_j(\mhm|\ms)$}}

\def\gsmf{\mbox{GSMF}}

\def\gamf{\mbox{G\HI MF}}
\def\gmmf{\mbox{G\H2MF}}

\defcitealias{Yang+2012}{Y12}
\defcitealias{RAD13}{RAD13}
\defcitealias{RDA12}{RDA12}

\def\aap{\mbox{A\&A}}
\def\apjl{\mbox{ApJ}}
\def\apjs{\mbox{ApJS}}

\def\plotancho#1{\includegraphics[trim = 0mm 8mm 0mm 40mm, clip, width=17cm]{#1}} 
\def\plotanchomfs#1{\includegraphics[trim = 0mm 85mm 9mm 42mm, clip, width=17cm]{#1}}
\def\plotsix#1{\includegraphics[trim = 4mm 60mm 50mm 40mm, clip, width=17cm]{#1}}
\def\plotsixsep#1{\includegraphics[trim = 0mm 70mm 40mm 40mm, clip, width=17cm]{#1}}
\def\plotdelgadofig6#1{\includegraphics[trim = 1mm 67mm 120mm 40mm, clip, width=8.5cm]{#1}}
\def\plotfig8#1{\includegraphics[trim = 0mm 85mm 120mm 40mm, clip, width=9cm]{#1}}
\def\plotfigten#1{\includegraphics[trim = 6mm 20mm 88mm 35mm, clip, width=8.3cm]{#1}}
\def\plotfigD2#1{\includegraphics[trim = 0mm 20mm 80mm 35mm, clip, width=9cm]{#1}}

\def\plotPDFHILTG#1{\includegraphics[trim = 3mm 1.5mm 120mm 3mm, clip, width=8.3cm]{#1}}
\def\plotPDFH2LTG#1{\includegraphics[trim = 3mm 70.5mm 120mm 4.5mm, clip, width=8.3cm]{#1}}

\title{The \HI- and \H2-to-stellar mass correlations of late- and early-type galaxies and their consistency with the observational mass functions} 

\author{A. R. Calette\altaffilmark{1}, Vladimir Avila-Reese\altaffilmark{1}, Aldo Rodr\'iguez-Puebla\altaffilmark{1,2,3}, H\'ector
Hern\'andez-Toledo\altaffilmark{1}, Emmanouil Papastergis\altaffilmark{4,5}}

\altaffiltext{1}{Instituto de Astronom\'ia, Universidad Nacional Aut\'onoma de M\'exico}
\altaffiltext{2}{Department of Astronomy \&\ Astrophysics, University of California at Santa Cruz, USA}
\altaffiltext{3}{Center for Astronomy and Astrophysics, Shanghai Jiao Tong University, China}
\altaffiltext{4}{Kapteyn Astronomical Institute, University of Groningen, The Netherlands}
\altaffiltext{5}{Credit Risk Modeling Department, Co\"{o}perative Rabobank U.A., The Netherlands}
\shortauthor{Calette et al.}
\shorttitle{On the \HI- and \H2-to-stellar mass correlations of local galaxies} 

\fulladdresses{
\item Avila-Reese, Vladimir;  Calette, Angel R.; Hern\'andez-Toledo, H\'ector;  Rodr\'iguez-Puebla, Aldo: Instituto de Astronom\'ia, Universidad Nacional Aut\'onoma de M\'exico, A. P. 70-264, 04510, Ciudad de M\'exico, M\'exico (avila@astro.unam.mx, acalette@astro.unam.mx, hector@astro.unam.mx, apuebla@astro.unam.mx).
\item Papastergis, Emmanouil: Kapteyn Astronomical Institute, University of Groningen, Landleven 12, Groningen NL-9747AD, The Netherlands, and
 Credit Risk Modeling Department, Co\"{o}perative Rabobank U.A., Croeselaan 18, Utrecht NL-3521CB, The Netherlands (papastergis@astro.rug.nl).
}

\abstract{
We compile and carrefully homogenize local galaxy samples with available information on stellar, \HI\ and/or \H2\ masses, and morphology. After processing the information on upper limits in the case of non gas detections, we determine the \HI- and \H2-to-stellar mass relations and their $1\sigma$ scatter for both late- and early-type galaxies. The obtained relations are fitted to single or double power laws. Late-type galaxies are significantly gas richer than early-type ones, specially at high masses. The respective \H2-to-\HI\ mass ratios as a function of \ms\ are discussed. Further, we constrain the full mass-dependent distribution functions of the \HI- and \H2-to-stellar mass ratios. We find that they can be described  by a Schechter function for late types and a (broken) Schechter + uniform function for early types. By using the observed galaxy stellar mass function and the volume-complete late-to-early-type galaxy ratio as a function of \ms,  these empirical distribution functions are mapped into \HI\ and \H2\ mass functions. The obtained mass functions are consistent with those inferred from large surveys.
The empirical gas-to-stellar mass relations and their distributions for local late- and early-type galaxies presented here can be used to constrain models and simulations of galaxy evolution. 
}

\resumen{
Compilamos y homogeneizamos muestras locales de galaxias que contienen informaci\'on de la masa estelar, de \HI\ y/o \H2, y morfolog\'ia. Procesamos adecuadamente la informaci\'on relacionada con las no detecciones en gas y determinamos la relaciones de masa estelar a masa de \HI\ y \H2\ y sus dispersiones, tanto para galaxias tard\'ias como tempranas. Las relaciones se describen por leyes simples o doble de potencias; las respectivos cocientes de masa \H2\ a \HI\ son presentados. Contre\~nimos tambi\'en las distribuciones completas de los cocientes de masa de \HI\ y \H2\ a masa estelar, encontrando que se describen bien por una funci\'on de Schechter (galaxias tard\'ias) y una funci\'on Schechter (cortada) + uniforme (galaxias tempranas). Usando la funci\'on de masa estelar y el cociente de galaxias tempranas a tard\'ias en funci\'on de \ms, estas distribuciones son mapeadas en funciones de masa de \HI\ y \H2. Las funciones de masa obtenidas son consistentes con aquellas inferidas de catastros. Las relaciones emp\'iricas de masa de gas a estrellas y sus distribuciones para galaxias tard\'ias/tempranas presentadas aqu\'i pueden ser usadas para constre\~nir modelos y simulaciones de evoluci\'on de galaxias.
  
  }

\addkeyword{galaxies: general}
\addkeyword{galaxies: ISM}
\addkeyword{galaxies: mass functions}
\addkeyword{galaxies: statistics}

\begin{document}
\maketitle

\section{Introduction}

Galaxies are complex systems, formed mainly from the cold gas captured by the gravitational potential of dark matter halos 
and transformed into stars, but also reheated and eventually ejected from the galaxy by feedback processes \citep[see for a recent
review][]{Somerville+2015}. Therefore, the
content of gas, stars, and dark matter of galaxies provides key information to understand their evolution and present-day
status, as well as to constrain models and simulations of galaxy formation 
\citep[see e.g.,][]{Zhang+2009,Fu+2010,Lagos+2011,Duffy+2012,Lagos+2015}. 

Local galaxies fall into two main populations, according to the dominion of the disk or bulge 
component (late- and early-types, respectively; a strong segregation is also observed by 
color or star formation rate). The main properties and evolutionary paths of these components are 
different. Therefore, the present-day stellar, gaseous, and dark matter fractions are expected to be different among 
late-type/blue/star-forming and early-type/red/passive galaxies of similar masses.  
The above demands the gas-to-stellar mass relations to be determined separately for each population.
Morphology, color and star formation rate correlate among them, though there is a fraction of galaxies
that skips the correlations. In any case, when only two broad groups are used to classify galaxies, 
the segregation in the resulting correlations for each group is expected to be similar for any
of these criteria. Here we adopt the morphology as the criterion for classifying galaxies into two broad populations. 

With the advent of large homogeneous optical/infrared surveys, the statistical distributions of galaxies, for example
the galaxy stellar mass function (\gsmf), are very well determined now. In the last years, using these surveys and direct 
or statistical methods, the relationship between the stellar, \ms, and halo masses has been constrained 
\citep[e.g.,][]{Mandelbaum+2006, Conroy+2009, More+2011,Behroozi+2010,Moster+2010,Rodriguez-Puebla+2013,Behroozi+2013,Moster+2013,Zu+2015}. 
Recently, the stellar-to-halo mass relation has been even inferred for (central) galaxies separated into blue and red ones 
by \citet{Rodriguez-Puebla+2015}.  These authors have found that there is a segregation by color in this relation 
\citep[see also][]{Mandelbaum+2016}. The semi-empirical 
stellar-to-halo mass relation and its scatter provide key constraints to models and simulations of galaxy evolution.
These constraints would be stronger if the relations between the stellar and atomic/molecular gas contents of galaxies are included.
With this information, the galaxy baryonic mass function can be also constructed and  the baryonic-to-halo mass relation can 
be inferred, see e.g, \citet{Baldry+2008}.

While the stellar component is routinely obtained from large galaxy surveys in optical/infrared bands, the information about the
cold gas content is much more scarce due to the limits in sensitivity and sky coverage of current radio telescopes. In fact, the 
few blind \HI\ surveys, obtained with a fixed integration time per pointing, suffer of strong biases, and for \H2\ (CO) there
are not such surveys. 
For instance, the \HI\ Parkes All-Sky Survey \citep[HIPASS;][]{Barnes+2001,Meyer+2004} or the
Arecibo Legacy Fast ALFA survey \citep[ALFALFA;][]{Giovanelli+2005,Haynes+2011,Huang+2012a}, 
miss galaxies with low gas-to-stellar mass ratios, specially at low stellar masses. 
Therefore, the \HI-to-stellar mass ratios inferred from the crossmatch of these surveys with 
optical ones should be regarded as an upper limit envelope \citep[see e.g.,][]{Baldry+2008,Papastergis+2012,Maddox+2015}.  
In the future, facilities as the Square Kilometre Array \citep[SKA;][]{Carilli+2004,Blyth+2015} or precursor instruments as
the Australian SKA Pathfinder \citep[ASKAP;][]{Johnston+2008} and the outfitted Westerbork Synthesis Radio Telescope
(WSRT), will bring extragalactic gas studies more in line with optical 
surveys. Until then, the gas-to-stellar mass relations of galaxies can be constrained:  i) from limited studies of radio follow-up 
observations of large optically-selected galaxy samples or by cross-correlating some radio surveys with optical/infrared surveys
 \citep[e.g.,][]{Catinella+2012,Saintonge+2011, Boselli+2010, Papastergis+2012}; and
ii) from model-dependent inferences based, for instance, on the observed metallicities of galaxies or from calibrated correlations 
with photometrical properties \citep[e.g.,][]{Baldry+2008,Zhang+2009}. 

While this paper does not present new observations, it can be considered as an extension of previous 
efforts in attempting to determine the \HI-, \H2- and cold gas-to-stellar mass correlations of local galaxies over a 
wide range of stellar masses.  Moreover, here we separate galaxies into at least two broad populations, 
late- and early-type galaxies (hereafter LTGs and ETGs, respectively).
These empirical correlations are fundamental benchmarks for models and simulations of galaxy evolution. 
Our main goal here is to constrain these correlations by using and uniforming large galaxy samples of good quality 
radio observations with confirmed optical counterparts. 
Moreover, the well determined local GSMF combined with these correlations can be used 
to construct the galaxy \HI\ and \H2\ mass functions, \gamf\ and \gmmf, respectively.  As a test of consistency, we
compare these mass functions with those reported in the literature for \HI\ and CO (\H2).

Many of the samples compiled here suffer of incompleteness and selection effects or in many cases the radio observations 
provide only upper limits to the flux (non detections). To provide reliable determinations of the \HI- and \H2-to-stellar mass correlations, 
for both LTGs and ETGs, here we homogenize as much as possible the data, check them against selection effects that could 
affect the calibration of the correlations, and take into account the upper limits adequately.  
We are aware on the limitations of this approach. Note, however, that
in absence of large homogeneous galaxy surveys reporting gas scaling relations over a wide dynamical 
range and separated into late- and early-type galaxies, the above approach is well supported as well as their, fair, use.

The plan of the paper is as follows. In Section \ref{compilation} and Appendices \ref{App:HI} and  \ref{App:H2}, 
we present our compilation and homogenization of local galaxy samples from the literature with available information 
on stellar mass, morphological type, and \HI\ and/or \H2\ masses. In Section \ref{correlations_sec}, we test the different compiled samples
against possible biases in the gas contents due to selection effects. In Section \ref{gass_mass_relations}, 
we describe the strategy to infer the gas-to-stellar mass correlations taking into account
upper limits, and present the determination of these correlations for the LTG and ETG populations  (mean and standard deviations).
Further, in Section \ref{scatter-distribution} we constrain the full distributions of the gas-to-stellar mass ratios as a 
function of \ms. In Section \ref{mass-functions} 
we explore the consistency of the determined correlations with the observed \HI\ and \H2\ mass functions,
by using the GSMF as an interface.   In subsection \ref{H2-HIratio} we discuss the \H2-to-\HI\ mass ratios of LTGs and
ETGs inferred from our correlations; subsection \ref{environment} is devoted to a discussion on the 
role of environment, and  subsection \ref{comparison} presents comparisons with some previous attempts to determine
the gas scaling relations.  A summary of our results and the conclusions are presented in Section \ref{conclusions}.
Finally, Table \ref{T_acron} lists all the acronyms used in this paper, including the ones of the surveys/catalogs used here. 

\begin{table}[h]
	\centering
	\caption{List of acronyms used in this paper} 
	 \resizebox{8.0cm}{!} {
		\begin{threeparttable}
			\begin{tabular}{cc}
				\hline
				\hline
				BCD &  Blue compact dwarf \\
				ETG & Early-type galaxy \\
				\gamf\ & Galaxy \HI\ Mass Function\\ 
				\gmmf\ & Galaxy \H2\ Mass Function\\ 
				\gsmf\ & Galaxy Stellar Mass Function\\ 
				IMF & Initial Mass Function\\
				LTG & Late-type galaxy \\ 
				MW & Milky Way\\ 
				\RHI\ and \RH2 & \HI- and \H2-to stellar mass ratio\\
				SB & Surface brightness\\
				SFR & Star formation rate\\
				\hline
				ALFALFA & Arecibo Legacy Fast ALFA survey\\
				ALLSMOG & APEX Low-redshift Legacy Survey for MOlecular Gas\\
				AMIGA & Analysis of the interstellar Medium of Isolated GAlaxies \\ 
				ASKAP & Australian SKA Pathfinder\\
				ATLAS$^{\rm 3D}$ & (A volume-limited survey of local ETGs)\\ 
				COLD GASS & CO Legacy Database for GASS\\ 
				FCRAO & Five College Radio Astronomy Observatory\\
				GALEX & Galaxy Evolution EXplorer\\
				GAMA & Galaxy And Mass Assembly\\
				GASS & GALEX Arecibo SDSS Survey\\ 
				HERACLES & HERA CO-Line Extragalactic Survey\\
				HIPASS & \HI\ Parkes All-Sky Survey\\ 
				HRS & Herschel Reference Survey\\ 
				NFGS & Nearby Field Galaxy Catalog \\ 
				NRTA & Nancay Radio Telescope\\
				SDSS & Sloan Digital Sky Survey\\
				SINGS & Spitzer Infrared Nearby Galaxies Survey \\ 
				SKA & Square Kilometre Array\\
				THINGS & The \HI\ Nearby Galaxy Survey \\ 
				UNAM-KIAS & UNAM-KIAS survey of SDSS isolated galaxies\\
				UNGC & Updated Nearby Galaxy Catalog \\ 
				WRST & Westerbork Synthesis Radio Telescope\\ 
				\hline
			\end{tabular}
		\end{threeparttable}
		}
     	\label{T_acron}
\end{table}

\section{Compilation of Observational Data}
\label{compilation}

The main goal of this Section is to present our extensive compilation of observational studies (catalogs, surveys or small samples) 
that meet the following criteria:

\begin{itemize}
\item Include \HI\ and/or \H2\ masses from radio observations, and luminosities/stellar masses from optical/infrared observations. 
\item Provide the galaxy morphological type or a proxy of it.
\item Describe the selection criteria of the sample and provide details about the radio observations, flux limits, etc.
\item Include individual distances to the sources and corrections for peculiar motions/large-scale structures for the nearby galaxies. 
\item In the case of non-detections, provide estimates of the upper limits for \HI\ or \H2\ masses.
\end{itemize}

The observational samples that meet the above criteria are listed in Table \ref{data_obs}. In Appendices \ref{App:HI} 
and \ref{App:H2}, we present a summary of each one of them. 
We have found information on colors  ($g-r$ or $B-K$) for most of the samples. For $\ms> 10^9$ \msun, the galaxies 
in the color--mass diagram segregate into the so-called red sequence 
and blue cloud. Excluding those more inclined than 70 degrees, we find that $\sim 83\%$ of LTGs ($\sim 80\%$ of ETGs) have colors that can 
be classified as blue (red) by using a mass-dependent $(g-r)$ criterion for defining blue/red galaxies. 
At masses lower than $\ms\approx 10^9$ \msun,  the overwhelming majority of galaxies are of late types and classify as blue.

\begin{table*}[h]
	\centering
	\caption{Observational samples} 
	\resizebox{16cm}{!} {
		\begin{threeparttable}
			\begin{tabular}{ccccccccc}
				\hline
				\hline
				Sample & Selection & Environment &  \HI\  & Detections / Total  &  \H2\   & Detections / Total & IMF & Category\\ \hline 
				UNGC & ETG+LTG & local 11 Mpc &  Yes & 407 / 418 & No & -- & diet-Salpeter & Gold\\
				GASS/COLD GASS & ETG+LTG & no selection &  Yes & 511 / 749 & Yes &  229 / 360  & \citet{Chabrier2003} & Gold\\	
				HRS--field & ETG+LTG & no selection &  Yes & 199 / 224 & Yes & 101 / 156 & \citet{Chabrier2003} & Gold\\
				ATLAS$^{\rm 3D}$--field  & ETG & field &  Yes & 51 / 151 & Yes & 55 / 242 & \citet{Kroupa2001} & Gold\\
				\hline 
				NFGS & ETG+LTG & no selection &  Yes & 163 / 189 & Yes & 27 /  31   & \citet{Chabrier2003} & Silver\\
				\citet{Stark+2013} compilation$^*$ & LTG & no selection &  Yes & 62/62 & Yes & 14 / 19 & diet-Salpeter & Silver\\
				Leroy+08 THINGS/HERACLES & LTG  & nearby &  Yes & 23 / 23 & Yes & 18 / 20 & \citet{Kroupa2001} & Silver\\ 
				Dwarfs-Geha+06 & LTG  & nearby &  Yes & 88 / 88 & No & --  & \citet{Kroupa+1993} & Silver\\
				ALFALFA dwarf &  ETG+LTG  & no selection &  Yes & 57 / 57 & No & --  & \citet{Chabrier2003} & Silver\\
				ALLSMOG  & LTG & field & No & --  & Yes & 25 / 42 & \citet{Kroupa2001} & Silver\\
				\citet{Bauermeister+2013} compilation & LTG & field &  No & --  & Yes & 7 / 8 & \citet{Kroupa2001} & Silver\\
				\hline 
				ATLAS$^{\rm 3D}$--Virgo  & ETG & Virgo core &  Yes & 2 / 15 & Yes & 4 / 21 & \citet{Kroupa2001} & Bronze\\
				AMIGA& ETG+LTG  & isolated &  Yes & 229 / 233 & Yes & 158 / 241 & diet-Salpeter & Bronze\\
				HRS--Virgo  & ETG+LTG & Virgo core &  Yes & 55 / 82 & Yes & 36 / 62 & \citet{Chabrier2003} & Bronze\\
				UNAM-KIAS & ETG+LTG  & isolated &  Yes & 352 / 352 & No & --  & \citet{Kroupa2001} & Bronze\\
				Dwarfs-NSA & LTGs  & isolated &  Yes & 124 / 124 & No & --  & \citet{Chabrier2003} & Bronze\\
				\hline
			\end{tabular}
			\begin{tablenotes}
				\item $^*$ From this compilation, we considered only galaxies that were not in GASS, COLD GASS and ATLAS$^{\rm 3D}$ samples.
			\end{tablenotes}		\end{threeparttable}
		}
		\label{data_obs}
	\end{table*}

\subsection{Systematical Effects on the \HI- and \H2-to-stellar mass correlations}

To reduce potential systematical effects that can bias how we derive the \HI- and \H2-to-stellar mass correlations 
we homogenize all the compiled observations to a same basis. Following, we discuss some potential sources of bias/segregation
and the calibration that we apply to the observations.
It is important to stress that for inferring scaling correlations, as those of the gas fraction as a function of stellar mass,
what is important is to have a statistically representative and not biased population of galaxies {\it at each mass bin}. Thus,
it is not a need to have mass limited volume-complete samples (see also subsection \ref{strategy}). However, a volume-complete 
sample assures that possible biases on the measure in question due to selection functions in galaxy type, color, environment, surface brightness, 
etc., are not introduced. The main expected bias in the gas content at a given stellar mass is due to the galaxy type/color; this is why
we need to separate the samples at least into two broad populations, LTGs and ETGs. 

\subsubsection{Galaxy type}

The gas content of galaxies, at a given \ms, segregates significantly with galaxy morphological type 
\citep[e.g.,][]{Kannappan+2013,Boselli+2014b}. Thus, information on morphology 
is necessary in order to separate galaxies at least into two broad populations, LTGs and ETGs. 
Besides of its physical basis, this separation is important for not introducing biases in the obtained 
correlations due to selection effects related to the morphology in the different samples used here. 
For example, some samples are only for late-type or star-forming galaxies, others only for early-type galaxies, etc., 
so that by combining them without a separation by morphology would yield correlations that are not statistically 
representative. We consider ETGs those classified as ellipticals (E), lenticulars (S0), dwarf E, and 
dwarf spheroidals or with $T<1$, and LTGs those classified as Spirals (S), Irregulars (Irr), dwarf 
Irr, and blue compact dwarfs or with $T\ge 1$. The morphological classification criteria used in the different samples are diverse,
from individual visual evaluation to automatic classification methods as the one by \citet{Huertas-Company+2011}.
We are aware of the high level of uncertainty introduced by using different morphological classification methods.
However, in our case the morphological classification is used for separating galaxies just into two broad groups. 
Therefore, such an uncertainty is not expected to affect significantly any of our results. 
It is important to highlight that the terms LTG and ETG are useful only as qualitative descriptors. These descriptors 
{\it should not be applied to individual galaxies, but instead to two distinct populations of galaxies in a statistical sense.
}

\subsubsection{Environment}

The gas content of galaxies is expected to depend on environment 
\citep[e.g.,][]{Zwaan+2005,Geha+2012,Jones+2016,Brown+2017}. 
In this study we are not in position of studying in detail such a dependence, 
though our separation into LTG and ETG populations partially takes into account this dependence because 
these populations segregate by environment 
\citep[e.g.,][and more references therein]{Dressler1980,Kauffmann+2004,Blanton+2005c,Blanton+2009}. 
In any case, in our compilation we include three samples specially selected to contain very isolated galaxies
and one subsample of galaxies from the Virgo Cluster central regions. We will check whether their \HI\ and \H2\
mass fractions significantly deviate or not from the mean relations. 
   
\subsubsection{Systematical Uncertainties on the Stellar Masses}   
\label{errorMs}

There are many sources of systematic uncertainty in the inference of the stellar masses related to
the choices of: initial mass function (IMF), stellar population synthesis and dust attenuation models, 
star formation history parametrization, metallicity, filter setup, etc.  For inferences from broad-band spectral energy 
distribution fitting and using a large diversity of methods and assumptions,  \citet{Pforr+2012} estimate a maximal 
variation in stellar mass calculations of $\sim 0.6$ dex. The major contribution to these uncertainties comes from the IMF. 
The IMF can introduce a systematical variation up to $\approx 0.25$ dex \citep[see e.g.,][]{Conroy2013}. 
For local normal galaxies and from UV/optical/IR data 
(as it is the case of our compiled galaxies), \citet[][]{Moustakas+2013} find a mean systematic differences 
between different mass-to-luminosity estimators (fixed IMF) less than $0.2 $ dex. We have seen that in most of 
the samples compiled here, the stellar masses are calculated using roughly similar mass-to-luminosity estimators, but the IMF 
are not always the same.Therefore, we homogenize the reported stellar masses in the different compiled samples to the 
mass corresponding to a \citet{Chabrier2003} initial mass function (IMF), and neglect other sources of systematic differences. 

\subsubsection{Other effects}

We also homogenize the distances to the value of $H_0= 70$ kms$^{-1}$ Mpc$^{-1}$.  
In most of the samples compiled here (at least the most relevant ones for our study), {\it distances were
corrected for peculiar motions and large-scale structure effects}.
When the authors included helium and metals to their reported \HI\ and \H2\ masses, we take care
in subtracting these contributions. When we calculate the total cold gas mass, then helium and metals
are explicitly taken into account.

\subsubsection{Categories}
\label{categories}

The different \HI\ and \H2\ samples used in this paper are wide in diversity, in particular they were
obtained with different selection functions, radio telescopes, exposure times, etc. We have divided the different samples into
three categories according to the feasibility of each one for determining robust and statistically representative \HI- or \H2-to-stellar mass
correlations for the LTG and ETG populations. We will explore whether the less feasible categories should be included or not
for determining these correlations. The three categories are:

\begin{enumerate}
\item { \bf Golden:} It includes datasets based on volume-complete (above a given luminosity/mass) 
samples or on representative galaxies selected from volume-complete samples. The Golden datasets, by construction, 
are unbiased samples of the galaxy properties distribution.
\item { \bf Silver:} It includes datasets from galaxy samples that are not volume complete but that are attempted to be
statistically representative at least for their morphological groups, i.e., these samples do not present obvious or strong selection effects.
\item { \bf Bronze:} This category is for samples selected deliberately by environment,
and it will be used to explore the effects of environment on the LTG and ETG \HI- or \H2-to-stellar mass correlations.
\end{enumerate}

\subsection{The compiled \HI\ sample} 
\label{HI}

\begin{table}[h]
	\centering
	\caption{Number of galaxies with detections and upper limits by morphology} 
	\resizebox{8.0cm}{!} {
		\begin{threeparttable}
			\begin{tabular}{cccc}
				\hline
				\hline
				Morphology(\%) & Detections(\%) & Upper limits(\%)  & Total\\ \hline 
				\multicolumn{4}{c}{\HI\ data} \\ 
				LTG (78\%) & 1975 (94\%) & 121 (6\%) & 2096 \\
				ETG (22\%) & 292 (50\%) & 288 (50\%) & 580	 \\	\\		
				\hline
				\hline
				\multicolumn{4}{c}{\H2\ data} \\  
				LTG (63\%) & 533 (75\%) & 180 (25\%) & 713\\
				ETG (37\%) & 124 (29\%) & 298 (71\%) & 422\\				
				\hline
			\end{tabular}
		\end{threeparttable}
	}
	\label{Tmorphology}
\end{table}

\begin{table}[h]
	\centering
	\caption{Number of galaxies with detections and upper limits by category} 
	\resizebox{8.0cm}{!} {
		\begin{threeparttable}
			\begin{tabular}{cccc}
				\hline
				\hline
				Category (\%) & Detections (\%) & Upper limits (\%) & Total\\ \hline 
				\multicolumn{4}{c}{\HI\ data} \\ 
				Golden (58\%) & 1168 (76\%) & 374 (24\%) & 1542 \\
				Silver (16\%) & 391 (94\%) & 26 (6\%) & 417	\\			
				Bronze (26\%) & 708 (99\%) & 9 (1\%) & 717	\\	\\		
				\hline
				\hline
				\multicolumn{4}{c}{\H2\ data} \\  
				Golden (67\%) & 385 (51\%) & 373 (49\%) & 758 \\
				Silver (10\%) & 91 (76\%) & 29 (24\%) & 120 \\			
				Bronze (23\%) & 181 (70\%) & 76 (30\%) & 257	\\	\\		
				\hline
			\end{tabular}
		\end{threeparttable}
	}
	\label{Tcateg}
\end{table}

Appendix \ref{App:HI} presents a summary of the \HI\ samples compiled in this paper (see also Table \ref{data_obs}). 
Table \ref{Tmorphology} lists the total numbers and fractions of compiled galaxies with detection and non detection
for each galaxy population. Table \ref{Tcateg} lists the number of detected and non-detected galaxies for the golden, silver,
and bronze categories listed above (\S\S \ref{categories}). 

\begin{figure*}[ht!]
\begin{center}
\plotancho{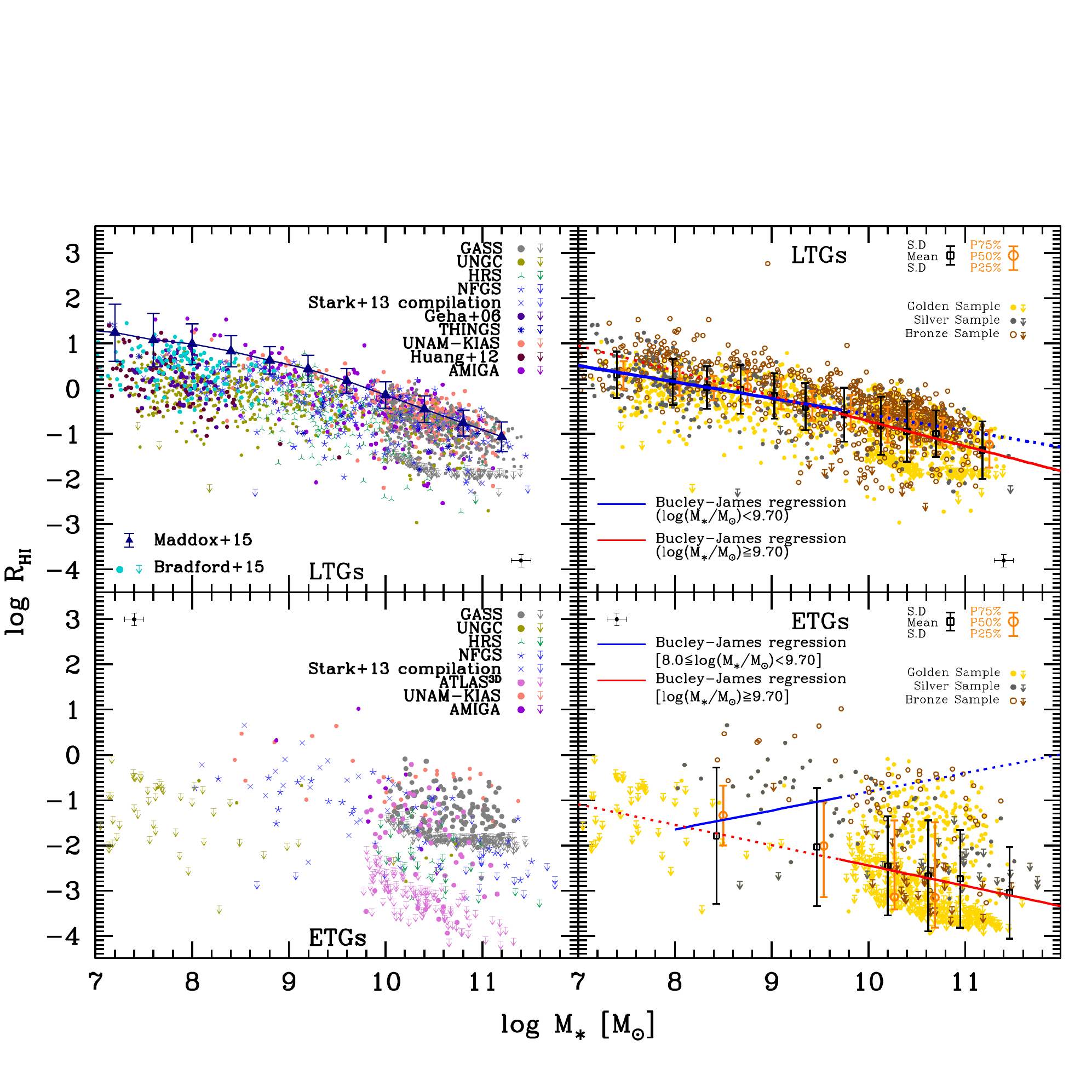}
\caption{Atomic gas-to-stellar mass ratio as a function of \ms. {\it Upper panels:} Compiled and homogenized data with information on 
\RHI\ and \ms\ for LTGs (the different sources are indicated inside the left panel; see Appendix \ref{App:HI} for the acronyms and
authors); downward arrows show the reported upper limits for non detections. 
The blue triangles with thin error bars are mean values and standard deviations 
from the v.40 ALFALFA and SDSS crossmatch according to \citet{Maddox+2015}; the ALFALFA galaxies are
biased to high values of \RHI\ (see text). Right panel is the same as left one, but with the data separated into three categories:
Golden, Silver, and Bronze (yellow, gray, and brown symbols, respectively). The red and blue lines are Buckley-James linear 
regressions (taking into account non-detections) for the high- and low-mass sides, respectively; the dotted lines
show extrapolations from these fits. Squares with error bars are the mean and standard deviation of the data in different
mass bins, taking into account non-detections by means of the Kaplan-Meier estimator. Open circles with error bars are
 the corresponding median and 25-75 percentiles. Estimates of the observational 
uncertainties are showed in the panel corners (see text). 
{\it  Lower panels:} Same as in upper panels but for ETGs. In the right panel, we have corrected by distance the galaxies 
with upper limits from GASS to make them consistent with the distances of the ATLAS$^{\rm 3D}$ sample (see text), and the upper limits
from the latter, where increased by a factor of two to homogenize them to the ALFALFA instrument and signal-to-noise criteria.  
For the bins where more than 50\% of the data are upper limits, the median and percentiles are not calculated.  }
\label{RHI} 
\end{center}
\end{figure*}

Figure \ref{RHI} shows the mass ratio $\RHI\equiv \mha/\ms$ vs. \ms\ for the compiled samples. Note that we have 
applied some corrections to the reported samples (see above) to homogenize all the data.  
The upper and bottom left panels of Figure \ref{RHI} show, respectively, the compilations for 
LTGs and ETGs. The different symbols indicate the source reference of the data and the 
downward arrows are the corresponding upper limits on the \HI-flux for non-detections. 
We also reproduce the mean and standard deviation in different mass bins as reported 
in \citet{Maddox+2015} for a cross-match of the ALFALFA and SDSS surveys. As mentioned in the Introduction, the ALFALFA survey
is biased to high \RHI\ values, specially towards the low mass side. Note that the small ALFALFA subsample of dwarf galaxies 
by \citet[][dark purple dots]{Huang+2012b} was selected namely as an attempt to take into account low-\HI\ mass galaxies in the low-mass end.

 \subsection{The compiled \H2\ sample} 
 \label{H2}
 
 Since the emission of cold \H2\ in the ISM is extremely weak, a tracer of the \H2\ abundance should be used. The best tracer
 from the observational point of view is the CO molecule due to its relatively high abundance and its low excitation energy.
 The \H2\ mass is related to the CO luminosity through a \cotoh2~conversion factor: $\mhm = \alpha_{\mbox{\tiny CO}}L_{\mbox{\tiny CO}}$.
 This factor has been determined in molecular clouds in the Milky Way (MW),
 $\alpha_{\mbox{\tiny CO,MW}}=3.2$ (\mbox{K km s$^{-1}$ pc$^{-1}$})$^{-1}$, with a systematic uncertainty of $30$\%. 
 It was common to assume that this conversion factor is the same for all galaxies. However, several pieces of evidence 
 show that \a_co\ is not constant, and it depends mainly on the gas-phase metallicity, increasing as the 
 galaxy metallicity decreases \citep[e.g.,][and more references therein]{Boselli+2002,Schruba+2012,Narayanan+2012,Bolatto+2013}.  As first-order, \a_co\ changes slowly for metallicities larger than $12+\log_{10}(\mbox{O}/\mbox{H})\sim 8.4$ 
 (approximately half the solar one) and increases considerably as the metallicity decreases. 
Here, we combine the dependence of \a_co\ on metallicity given by \citet{Wolfire+2010} and the observed mass--metallicity relation 
to obtain an approximate estimation of the dependence of \a_co\ on \ms\ for LTGs; see Appendix \ref{conv-factor} for details.
We are aware that the uncertainties involved in any metallicity-dependent correction remain substantial \citep[][]{Bolatto+2013}. 
Note, however, that our aim is to introduce and explore at a statistical level a reasonable mass-dependent correction to the \cotoh2\ factor, which 
must be better than ignoring it. In any case, we present results both for $\a_co=\alpha_{\mbox{\tiny CO,MW}}$ 
and our inferred mass-dependent \a_co\ factor. In fact, the mass-dependent factor is important only for 
LTGs with $\ms<\lesssim 3\times 10^{10}$ \msun; for higher masses and for all ETGs, 
$\a_co\approx \alpha_{\mbox{\tiny CO,MW}}$\footnote{This is well justifyied since massive LTGs are metallic
with typical values larger than $12+\log_{10}(\mbox{O}/\mbox{H})\sim 8.7$ while ETG have high metallicities at all masses.}

Appendix \ref{App:H2} presents a description of the CO (\H2) samples that we utilize in this paper.
Table \ref{Tmorphology} lists the number of galaxies with detections and upper limits of the compilation sample in terms of morphology. 
Table \ref{Tcateg} lists the number of detections and upper limits for the golden, silver, and bronze categories mentioned above (\S\S \ref{categories}). 

\begin{figure*}[ht!]
\plotancho{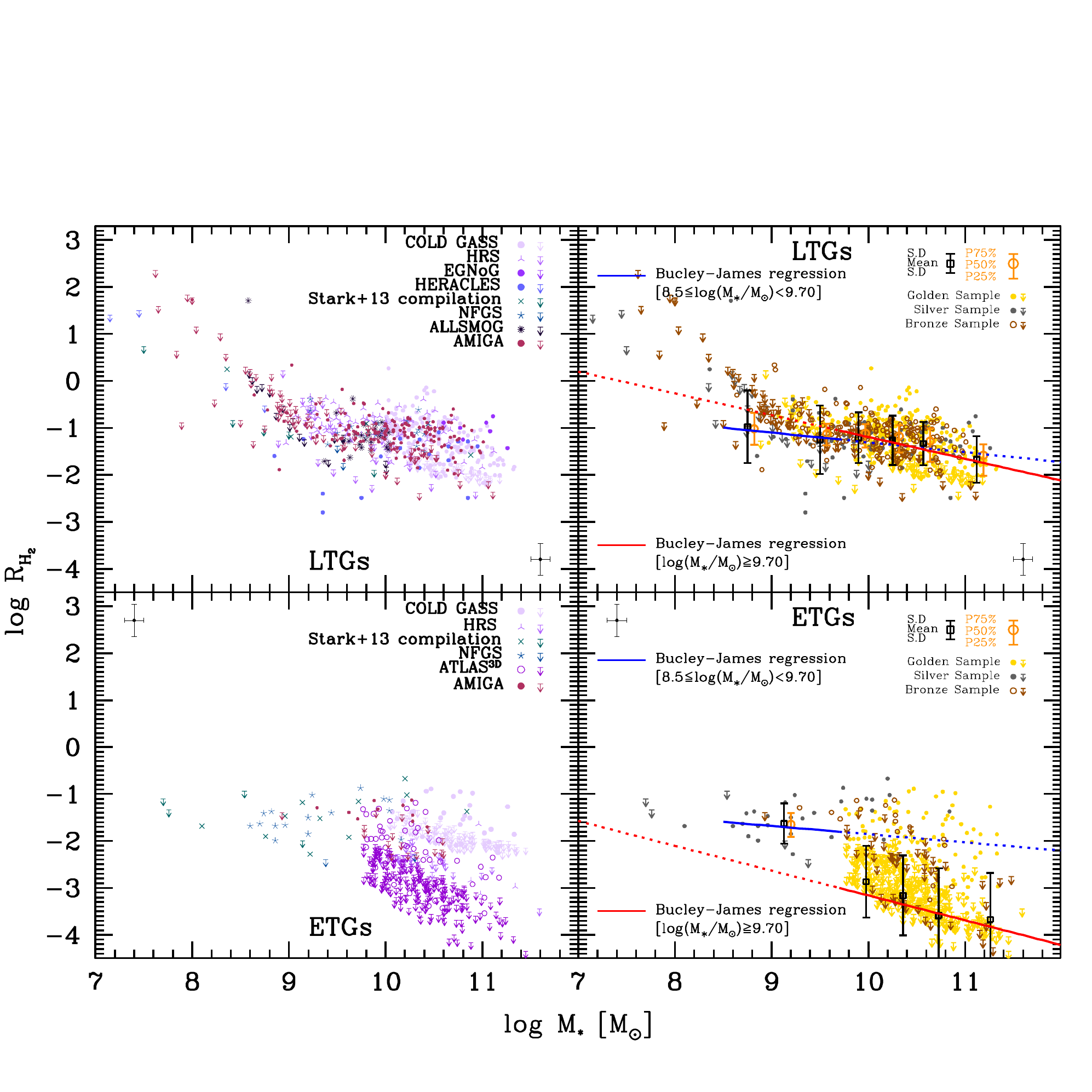}
\caption{Molecular gas-to-stellar mass ratio as a function of \ms. {\it  Upper panels:} Compiled and homogenized data with information 
on \RH2\ and \ms\ for LTGs (see inside the panels for the different sources; see Appendix \ref{App:H2} for the acronyms and
authors); downward arrows show the reported upper limits for 
non detections. Right panel is the same as left one, but with the data separated into three categories:
Golden, Silver, and Bronze (yellow, gray, and brown symbols, respectively). The red and blue lines are Buckley-James linear 
regressions (taking into account non-detections). The dotted lines show extrapolations from these fits. The green dashed line shows 
an estimate for the \RH2--\ms\ relation inferred from 
combining the empirical SFR--\mhm\ and SFR--\ms\ correlations for blue/star-forming galaxies (see text for details).
Squares with error bars are the mean and standard deviation of the data in different
mass bins, taking into account non-detections by means of the Kaplan-Meier estimator. Open circles with error bars 
are the corresponding median and 25-75 percentiles. Estimates of the observational/calculation 
uncertainties are showed in the panel corners (see text). {\it  Lower panels:} The same as in upper panels but for ETGs. 
In the right panel, we have corrected by distance the galaxies 
with upper limits from COLD GASS to make them consistent with the distances of the ATLAS$^{\rm 3D}$ sample (see text).  
For the bins where more than 50\% of the data are upper limits, the median and percentiles are not calculated.  }
\label{RH2} 
\end{figure*}

Figure \ref{RH2} shows the mass ratio $\RH2\equiv \mhm/\ms$ vs. \ms\ for the compiled samples. Similarly to the  \RHI-- \ms\ relation, 
we applied some corrections to observations in order to homogenize our compiled sample and to have this way 
a more consistent comparison between the different samples. The upper and bottom left panels of Figure \ref{RH2} show, 
respectively, the compiled datasets for LTGs and ETGs. 

\section{Tests against selection effects and preliminary results}
\label{correlations_sec}

In this Section we check the gas-to-stellar mass correlations from the different compiled samples
against possible selection effects. We also introduce, when possible,  an homogenization in the upper limits of ETGs. 
The reader interested only on the main results can skip to  Section \ref{gass_mass_relations}.

As seen in Figs. \ref{RHI} and \ref{RH2} there is a significant 
fraction of galaxies with no detections in radio, for which the authors report an upper limit flux (converted into an \HI\ or \H2\ mass). 
{\it The non detection of observed galaxies gives information that we cannot ignore,} otherwise a bias towards high gas 
fractions would be introduced in the gas-to-stellar mass relations to be inferred.
To take into account the upper limits in the compiled data, we resort to survival analysis methods for combining censored and 
uncensored data \citep[i.e., detections and upper limits for non detections; see e.g.,][]{Feigelson-Babu2012}.
We will use two methods: the Buckley-James linear regression \citep{BJ79} and the Kaplan-Meier product limit estimator 
\citep{Kaplan+1958}. Both are survival analysis methods commonly applied in Astronomy.\footnote{We use the ASURV 
(Astronomy SURVival analysis) package developed by 
T. Isobe, M. LaValley and E. Feigelson in 1992 \citep[see also][]{Feigelson+1985}, and implemented in the stsdas package 
(Space Telescope Science Science Data Analysis) in IRAF. In particular, we make use of the {\it buckleyjames} (Buckley-James linear regression) and 
{\it kmestimate} (Kaplan-Meier estimator) routines.}   
The former is useful for obtaining a linear regression from the censored and uncensored data. Alternatively, for data
that can not be described by a linear relation, we can bin them by mass, use the Kaplan-Meier estimator to  calculate 
the mean, standard deviation,\footnote{The IRAF package 
provides actually the standard error of the mean, $SEM= s/\sqrt{n}$, where $s=\sqrt{\frac{1}{n}\sum_{i=1}^{n}(x_{i}-\bar{x})^2}$ 
is the sample standard deviation, $n$ is the number of observations, and $\bar{x}$ is the sample mean. In fact,
$s$ is a biased estimator of the (true) population standard deviation $\sigma$. 
For small samples, the former underestimates the true population standard deviation. A commonly used rule of thumb to 
correct the bias when the distribution is assumed to be normal, is to introduce the term $n-1.5$ in the computation of $s$ instead
of $n$. In this case, $s\rightarrow \sigma$. Therefore, an approximation to the population standard
deviation is $\sigma= (n/\sqrt{n-1.5})\times SEM$. This is the expression we use
to calculate the reported standard deviations.} 
median, and 25-75 percentiles in each stellar mass bin, and fit these results to a function 
by using conventional methods, e.g., the Levenberg-Marquardt algorithm. 
For the latter case, the binning in $\log\ms$ is started with a width of $\approx 0.25$ dex but if the data is too scarce in the bin, then its width
is increased as to have not less than $25\%$ of galaxies than in the most populated bins. 
Note that, for detection fractions smaller than 50\%, the median and percentiles are very uncertain or impossible to be calculated 
with the Kaplan-Meier estimator \citep{Lee-Wang2003}, while the mean can be yet estimated for fractions as small 
as $\sim 20\%$, though with a large uncertainty. 
In the case of the Bukley-James linear regression, reliable results are guaranteed for detection fractions larger than $70-80\%$.

When the fraction of non detections is significant, the inferred correlations could be affected by
{\it selection effects in the upper limits} reported in the different samples. This is the case for ETGs, where a clear systematical
segregation between the upper limits of the GALEX Arecibo SDSS Survey (GASS) and ATLAS$^{\rm 3D}$ or Herschel Reference 
Survey (HRS) surveys is  observed in the $\log\RHI-\log\ms$ plane (see the gap in the left lower panel of Fig. \ref{RHI}), 
as well as between the CO Legacy Database for GASS (COLD GASS) and ATLAS$^{\rm 3D}$ or HRS surveys in the 
$\log\RH2-\log\ms$ plane (see the gap in the left lower panel of Fig. \ref{RH2}). The determination of the upper limits depends on 
distance and instrumental/observational constrains (telescope sensitivity, integration time, spatial coverage, signal-to-noise 
threshold, etc.). The \HI\ observations of GASS and ATLAS$^{\rm 3D}$ were carried out with different radio telescopes: 
the single-dish Arecibo Telescope and the Westerbork Synthesis Radio Telescope (WRST) interferometer array, respectively.  
\citet{Serra+2012} discussed about differences regarding detections between single- and multiple-beam observations.  
For some galaxies from ATLAS$^{\rm 3D}$ that they were able to observe also with the Arecibo telescope, they conclude that their upper limits
should be increased by a factor of $\sim 2$ in order to agree with the ALFALFA survey sensitivity and the signal-to-noise threshold they 
use for declaring non detections in their multiple-beam observations.  Thus, to homogenize the upper limits, we correct the 
ATLAS$^{\rm 3D}$ upper limits by this factor. In the case of \RH2, the CO observations in the ATLAS$^{\rm 3D}$ and COLD GASS 
samples were taken with the same radio telescope (IRAM). 

The GASS (COLD GASS) samples are selected to include galaxies at distances between $\approx 109$ and 222 Mpc, while 
the  ATLAS$^{\rm 3D}$ and HRS surveys include only nearby galaxies, with average distances of 25 and 19 Mpc, respectively. 
Since the definition of the upper limits {\it depends on distance}, for the same radio telescope and integration time, more 
distant galaxies have systematically higher upper limits than closer galaxies. This introduces a clear
selection effect.  In the case we have information for a sample of galaxies closer than other sample, and under
the assumption that both samples are roughly representative of the same local galaxy population, a distance-dependent correction
to the upper limits of the non-detected galaxies in the more distant sample should be introduced.  In Appendix \ref{upper-limit-corrections}, 
we describe our approach to apply such a correction to GASS (COLD GASS) ETG upper limits with respect to the ATLAS$^{\rm 3D}$ ETGs.  
We test our corrections by using a mock catalog. This correction by distance is an approximation based on the assumption
that the (COLD)GASS and ATLAS$^{\rm 3D}$ ETGs are statistically similar populations. In any case, we will present the 
correlations for ETGs for both cases, taking and  do not taking into account this correction.

Note that after our corrections by distance and instruments, the upper limits of the massive ETGs in the GASS/COLD GASS sample are now
consistent with those in the ATLAS$^{\rm 3D}$ (as well as HRS) samples, as seen in the right panels of Figs. \ref{RHI} and \ref{RH2} 
to be described below, and in Fig. \ref{MHI-histogram} in Appendix \ref{upper-limit-corrections}. 
In the case of LTGs, there is no evidence of much lower values of \RHI\ and \RH2\ than the upper limits given in GASS and 
COLD GASS for galaxies closer than those in these samples. 

In the right panels of Figs. \ref{RHI} and \ref{RH2}, all the compiled data shown in the left panels are again
plotted with dots and arrows for the detections and non detections, respectively. The yellow, dark gray, and brown
colors correspond to galaxies from the Golden, Silver, and Bronze categories, respectively (see \S\S \ref{categories}).
The above mentioned corrections to the upper limits of GASS/COLD GASS and ATLAS$^{\rm 3D}$ ETG samples 
were applied. Observe that the large gaps in the upper limits between the GASS/COLD GASS and ATLAS$^{\rm 3D}$ 
(or HRS) samples tend to disappear after the corrections we have applied. 

We further group the data in logarithmic mass bins and calculate in each mass bin the mean and standard deviation of 
$\log\RHI$ and $\log\RH2$ (black circles with error bars), taking into account the upper limits with the Kaplan-Meier estimator
as described above.  The orange squares with error bars are for the corresponding medians and 25-75 
percentiles, respectively. In some mass bins, the fraction of detections are smaller than 50\% for ETGs, therefore,  
the median and percentiles can not be estimated (see above). However, the mean and standard deviations can be yet 
calculated, though they are quite uncertain. 

As seen in the right panels of Figs. \ref{RHI} and \ref{RH2}, the logarithmic mean and median values tend to coincide and 
the 25-75 percentiles are roughly symmetric in most of the cases. Both facts suggest that the scatter around the mean 
relations (at least for the LTG population) tend to follow a nearly symmetrical distribution, for instance,
a normal distribution in the logarithmic values (for a more detailed analysis of the scatter distributions see 
section \ref{scatter-distribution}). 

In the following, we check whether each one of 
the compiled and homogenized samples deviate significantly or not from the mean trends. This could happen due to
selection effects in the given sample. For example, we expect systematical deviations in the gas contents
for the Bronze samples, because they are selected to contain galaxies in extreme environments. 
As a first approximation, we apply the Buckle-James linear regression to each one of the compiled individual samples, taking into
account this way upper limits. When the data in the given sample are too scarce and/or dominated by non detections, the linear
regression is not performed but the data are plotted.

\begin{figure*}[ht!]
\begin{center}
\plotsix{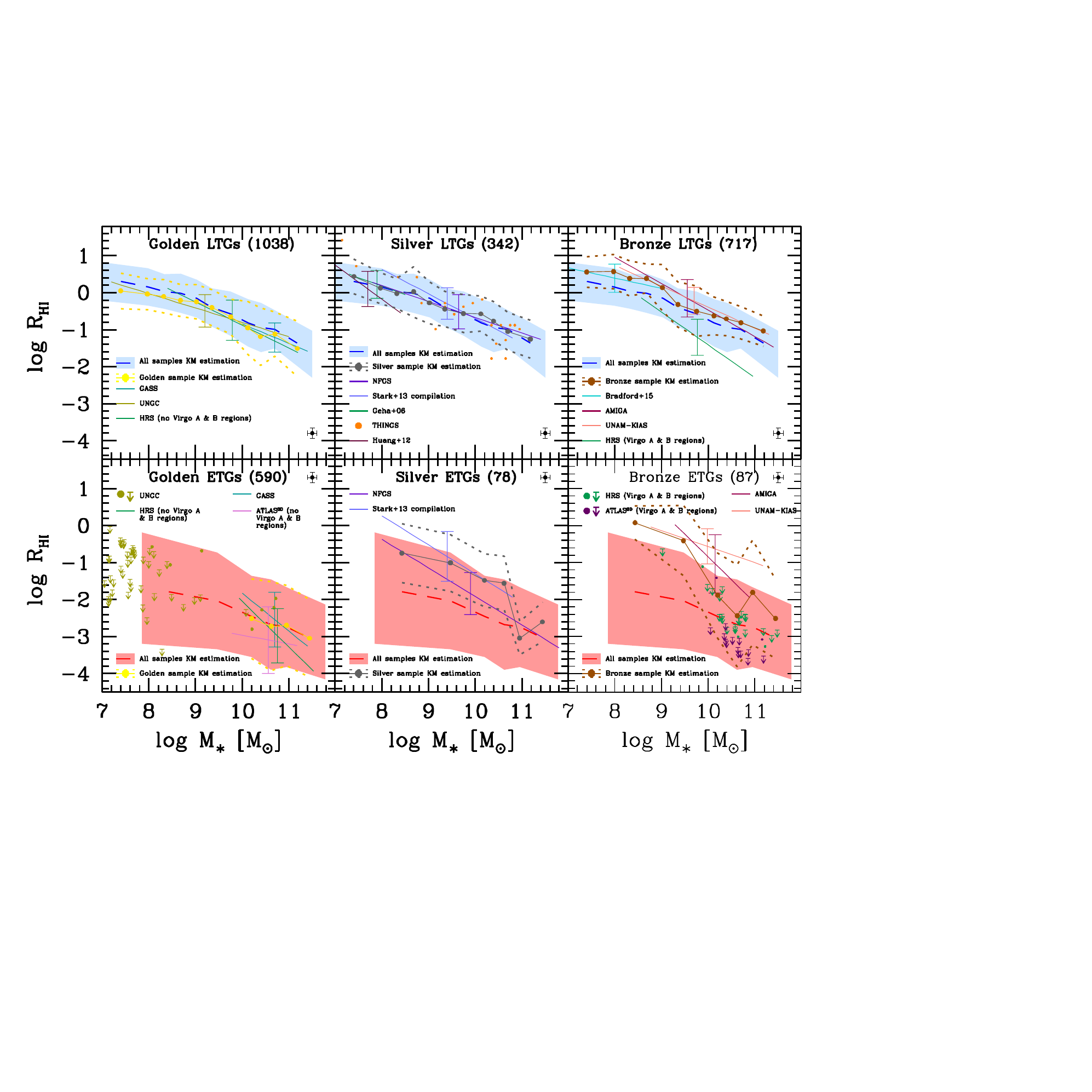}
\caption{Atomic gas-to-stellar mass ratio as a function of \ms\ for the Golden, Bronze, and  Silver LTGs (upper panels) 
and ETGs (lower panels). The mean and standard deviation in different mass bins, taking into account upper limits by means of the Kaplan-Meier 
estimator, are plotted for each case (filled circles connected by a dotted line and dotted lines around, respectively).  
For comparison, the mean and standard deviation (dashed lines and shaded area) 
from all the LTG (ETG) samples are reproduced in the corresponding upper (lower) panels. For each sample 
compiled and homogenized from the literature, the Buckley-James linear regression is applied, taking into account upper
limits. The lines show the result, covering the range of the given sample; the error bars show the corresponding standard
deviations obtained from the regression. When the data are too scarce and dominated by upper limits, the linear regression 
is not applied but the data are plotted. The number of LTG and ETG objects in each category are indicated in the respective panel.
 }
\label{checksRHI} 
\end{center}
\end{figure*}

\subsection{\RHI\ vs. \ms}
\label{HI-preliminary}

In Fig. \ref{checksRHI}, results for $\log\RHI$ vs. $\log\ms$ are shown for LTGs (upper
panels) and ETGs (lower panels). From left to right, the regressions for samples in the Golden, Silver, and Bronze categories are plotted. 
The error bars correspond to the $1\sigma$ scatter of the regression. Each line covers the mass range of the corresponding sample. 
 The blue/red dashed lines and shaded regions in each panel correspond to the mean and standard 
deviation values calculated with the Kaplan-Meier estimator in mass bins for {\it all} the compiled LTG and ETG samples and previously plotted 
in Figs. \ref{RHI} and \ref{RH2}, respectively. On the other hand, the yellow, gray, and brown dots connected with thin solid lines in each panel
are the mean values in each mass bin calculated {\it only} for the Golden, Silver, and Bronze samples, respectively. 
The standard deviation are plotted with dotted lines. In the following, we discuss the results shown in Fig. \ref{checksRHI}.

\textbf{Golden category:} For LTGs, the three samples grouped in this category 
agree well among them in the mass ranges where they overlap; even the $1\sigma$ scatter 
of each sample do not differ significantly among them.\footnote{Note also that the $1\sigma$ scatter provided by the Buckle-James 
linear regression is consistent with the standard deviations in the mass bins obtained with the Kaplan-Meier estimator.} 
Therefore, as expected, these samples provide unbiased information for determining the \RHI--\ms\ relation of LTGs from 
$\log$(\ms/\msun)$\approx 7.3$ to 11.4. For ETGs, the deviations of the Golden linear regressions among them 
and with respect to all galaxies are within their $1\sigma$ scatter, which are actually large. 
If no corrections to the upper limits of the  GASS and ATLAS$^{\rm 3D}$ are applied, then the regression to the former 
would be significantly above than the regression to the latter.  
Within the large scatter, the three Golden samples of ETGs seem not to be particularly biased, and they cover a mass 
range from $\log$(\ms/\msun)$\approx 8.5$ to 11.5. At smaller masses, the Updated Nearby Galaxy Catalog (UNGC) 
sample provides mostly only upper limits to \RHI. 

\textbf{Silver category:} The LTG and ETG samples in this category, as expected, show a more dispersed
distribution in their respective \RHI--\ms\ planes than those from the Golden category. However, the deviations of the Silver
linear regressions among them and with respect to all the galaxies are within the corresponding $1\sigma$ scatter. 
If any, there is a trend of the Silver samples to have mean \RHI\ values above the mean values of all galaxies 
in special for ETGs. Since the samples in this category are not from complete volumes, but they were specially constructed for studying
\HI\ gas contents, a selection effect towards objects with non-negligible or higher than the mean \HI\ contents can be expected. In any case,
the biases are small. Thus, we decide to include the Silver samples to infer the \RHI--\ms\ correlations below in order to increase 
slightly the statistics (the number of galaxies in this category is actually much lower than in the Golden category), specially
for ETGs of masses lower than $\log$(\ms/\msun)$\approx 9.7$ (see Table \ref{Tcateg}).  

\textbf{Bronze category and the effects of environment:} The very isolated LTGs (from the UNAM-KIAS and Analysis 
of the interstellar Medium of Isolated GAlaxies -AMIGA- samples) have \HI\ contents higher than the mean 
of all the galaxies, specially at lower masses: $\log\RHI$ is $0.1-0.2$ dex higher than the average at 
$\log$(\ms/\msun)$\ga 10$ and these differences increase up to $0.6-0.3$ dex for $8<\log(\ms/\msun)<9$,
though the number of galaxies at these masses is very small. The \HI\ content of the \citet{Bradford+2015} isolated dwarf galaxies
is also higher than the mean of all the galaxies but not by a factor larger than 0.4 dex.    
For isolated ETGs, the differences can attain an order of magnitude and are in the limit of the upper standard deviations around
the means of all the ETGs. Thus, while isolated LTGs have somewhat higher \RHI\ ratios on average
than galaxies in other environments, in the case of isolated ETGs, this difference is very large; isolated ETGs can be
almost as gas rich as LTGs.
In the Bronze group we have included also galaxies from the central regions of the Virgo Cluster as reported in 
HRS and ATLAS$^{\rm 3D}$ (only ETGs for the latter). According to Fig. \ref{checksRHI}, the LTGs in this high-density 
environment are clearly \HI\ deficient with respect to LTGs in less dense environments.
For ETGs, the \HI\ content is very low but only slightly lower on average than the \HI\ content of all ETGs. 
It should be noted that ETGs, in particular the massive ones, tend to be located in high-density environments.

We conclude that the \HI\ content of galaxies is affected by the effects of extreme environments. 
The most remarkable effect is for ETGs, which in the very isolated environment can be as rich in \HI\ as LTGs. 
Therefore, we decide do not include galaxies from the Bronze category to determine the \RHI--\ms\ correlation of ETGs. 
In fact, our compilation in the Golden and Silver categories includes galaxies from a range of environments (for instance, 
in the largest compiled catalog, UNGC, $58\%$ of the galaxies are members of groups and $42\%$ are field galaxies, 
see \citealp{Karachentsev+2014}) in such a way that the \RHI--\ms\ correlation determined below should represent an 
average of different environments. Excluding the Bronze category for the ETG population, we avoid biases due to 
effects of the most extreme environments. For LTGs, the inclusion of the Bronze category does not introduce significant biases
to the \RHI--\ms\ correlation of all galaxies but it helps to improve the statistics. The mean values of \RHI\ in mass bins
above $\sim 10^9$ \msun\ are actually close to the mean values of all the sample (compare the brown solid and blue dashed
lines); at lower masses the deviation increases, but the differences are well within the $1\sigma$ dispersion.

\begin{figure*}[ht!]
\begin{center}
\plotsix{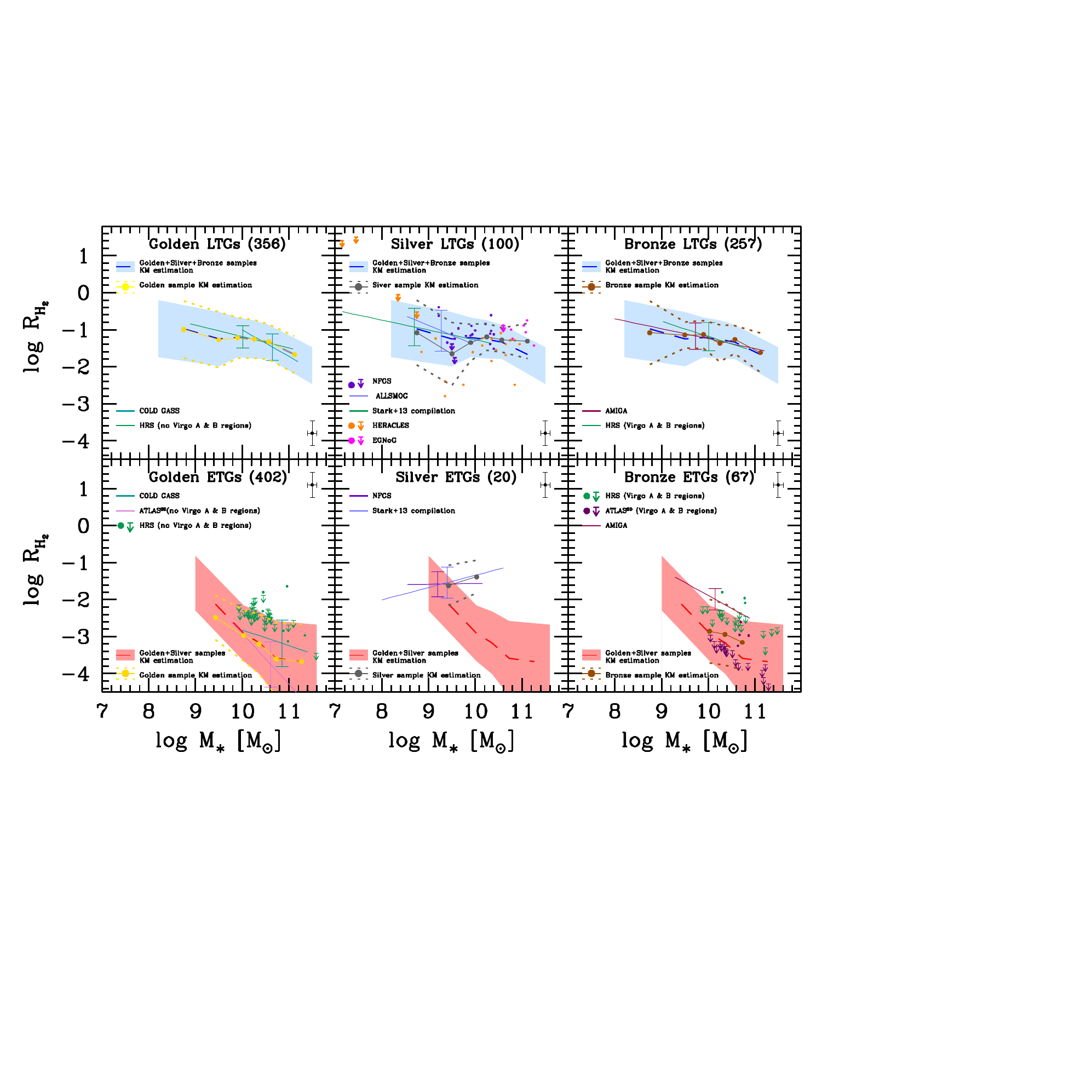}
\caption{Same as Fig. \ref{checksRHI} but for the molecular gas-to-stellar mass ratio.}
\label{checksRH2} 
\end{center}
\end{figure*}

\subsection{\RH2\ vs. \ms}

In Fig. \ref{checksRH2}, we present similar plots as in Fig. \ref{checksRHI} but for $\log\RH2$ vs. $\log\ms$. The symbol and
line codes are the same in both figures. In the following, we discuss the results shown in Fig. \ref{checksRH2}.

\textbf{Golden category:}   
For LTGs, the two samples grouped in this category agree well among them and with the overall sample, though for masses 
$<10^{10}$ \msun, where the Golden galaxies are only those from the HRS 
sample, the average \RH2\ values are slightly larger than those from the overall LTG sample (compare the solid yellow and dashed blue lines), 
but yet well within the $1\sigma$ scatter (shaded area). 
For ETGs, the deviations of the linear regressions of the Golden samples among them, and with respect to all ETGs, 
are within the respective $1\sigma$ scatters, which are actually large. If no corrections to the upper limits of the  GASS 
and ATLAS$^{\rm 3D}$ are applied, then the regression to the former would be significantly above than the regression to the latter.
Summarizing, the Golden samples of LTGs and ETGs do not show particular shifts in their respective \RH2-- \ms\ correlations. Therefore,
the combination of them are expected to provide reliable information for determining the respective \RH2-- \ms\ correlations; 
for LTGs, in the $\approx10^{8.5}-10^{11.5}$ \msun\ mass range, and for ETGs, only for $\ms\gtrsim 10^{10}$ \msun.

\textbf{Silver category:} The LTG samples present a dispersed distribution in the $\log$\RH2--$\log$\ms\ plane but well within the
$1\sigma$ scatter of the overall sample (shaded area). The mean values in mass bins from samples of the Silver category are in reasonable
agreement with the mean values from all the samples (compare the gray solid  and blue dashed lines). Therefore,
the Silver samples, though scattered and not complete in any sense, seem not to suffer a clear systematical shift in their \H2\ content.
We include then these samples to infer the \RH2-\ms\ correlation of LTGs.
For ETGs, the two Silver samples provide information for masses below $\ms\sim 10^{10}$ \msun, and both are consistent with each other.
Therefore, we include these samples to infer the ETG \RH2-\ms\ correlation down to $\ms\sim 10^{8.5}$ \msun. 

\textbf{Bronze category and the effects of environment:} The isolated (from the AMIGA sample) and Virgo central (from the HRS catalog)
LTGs have \H2\ contents similar to the mean in different mass bins of all the galaxies. If any, the 
Virgo LTGs have on average slightly higher values of \RH2\ than the isolated LTGs, specially at masses lower than $\ms\sim 10^{10}$ \msun. 
Given that LTGs in extreme environments do not segregate from the average \RH2\ values at different masses of all galaxies,
we include them for calculating the \RH2--\ms\ correlation of LTGs. 
For ETGs, the AMIGA isolated galaxies have on average significantly higher values of \RH2\ than the mean of other galaxies, while those ETGs from
the Virgo central regions (from HRS and ATLAS$^{\rm 3D}$; mostly upper limits), seem to be on average consistent with the mean of all the galaxies, though the 
scatter is large. Given the strong deviation of isolated ETGs from the mean trend, we prefer to exclude galaxies from the Bronze category 
for determining the ETG \RH2--\ms\ correlation. 
We conclude that the \H2\ content of LTGs is weakly dependent on the environment of galaxies, but in the case of ETGs, very isolated
galaxies have systematically higher \RH2\ values than galaxies in more dense environments.

\section{The gas-to-stellar mass correlations of the two main galaxy populations}
\label{gass_mass_relations}

\subsection{Strategy for constraining the correlations}
\label{strategy}
 
In spite of the diversity in the compiled samples and their different selection functions, 
the exploration presented in the previos Section shows that the \HI\ and \H2\ contents as a function of \ms\ from most of 
the samples compiled here do not segregate significantly among them.
The exception are  the Bronze samples for ETGs. Therefore, {\it the Bronze ETGs are excluded from our analysis.}
The strong segregation is actually by morphology (or color or star formation rate), 
and this is why we have separated since the beginning the compiled data into two broad galaxy groups, LTGs and ETGs.  

To determine gas-to-stellar mass ratios as a function of \ms\ we need (1) to take into account the 
upper limits of undetected galaxies in radio, and (2) to evaluate the correlation independently of the number of data points at each mass bin. 
If we have many data points at some mass bins and only a few ones in other mass bins (as it would happen if we use, for instance, 
a mass-limited volume complete sample, with much more data points at lower-masses than at large masses), then the overall correlation of 
\RHI\ or \RH2\ with \ms\ will be dominated by the former, giving probably incorrect values of \RHI\ or \RH2\ at other
masses. In view of these two requirements, our strategy to determine the $\log$\RHI--$\log$\ms\ and $\log$\RH2--$\log$\ms\  
correlations is as follows:
 
\begin{enumerate}
\item Calculate the logarithmic means and standard deviations (scatter) in stellar mass bins obtained from the compiled data taking into account 
the non detections (upper limits) by means of the Kaplan-Meier estimator.  
\item Get an estimate of the intrinsic standard deviations (scatter), taking into account estimates of the observational errors.
\item Propose a function to describe the relation given by the mean and intrinsic scatter as a function of mass (e.g., a single or double power law). 
\item Constrain the parameters of this function by performing a formal fit to the mean and scatter calculated 
at each mass bin; note that in this case {\it the fitting gives the same weight to each mass bin, in spite of the number
of galaxies in each bin.} 
\end{enumerate}

\begin{figure*}[ht!]
\begin{center}
\plotsix{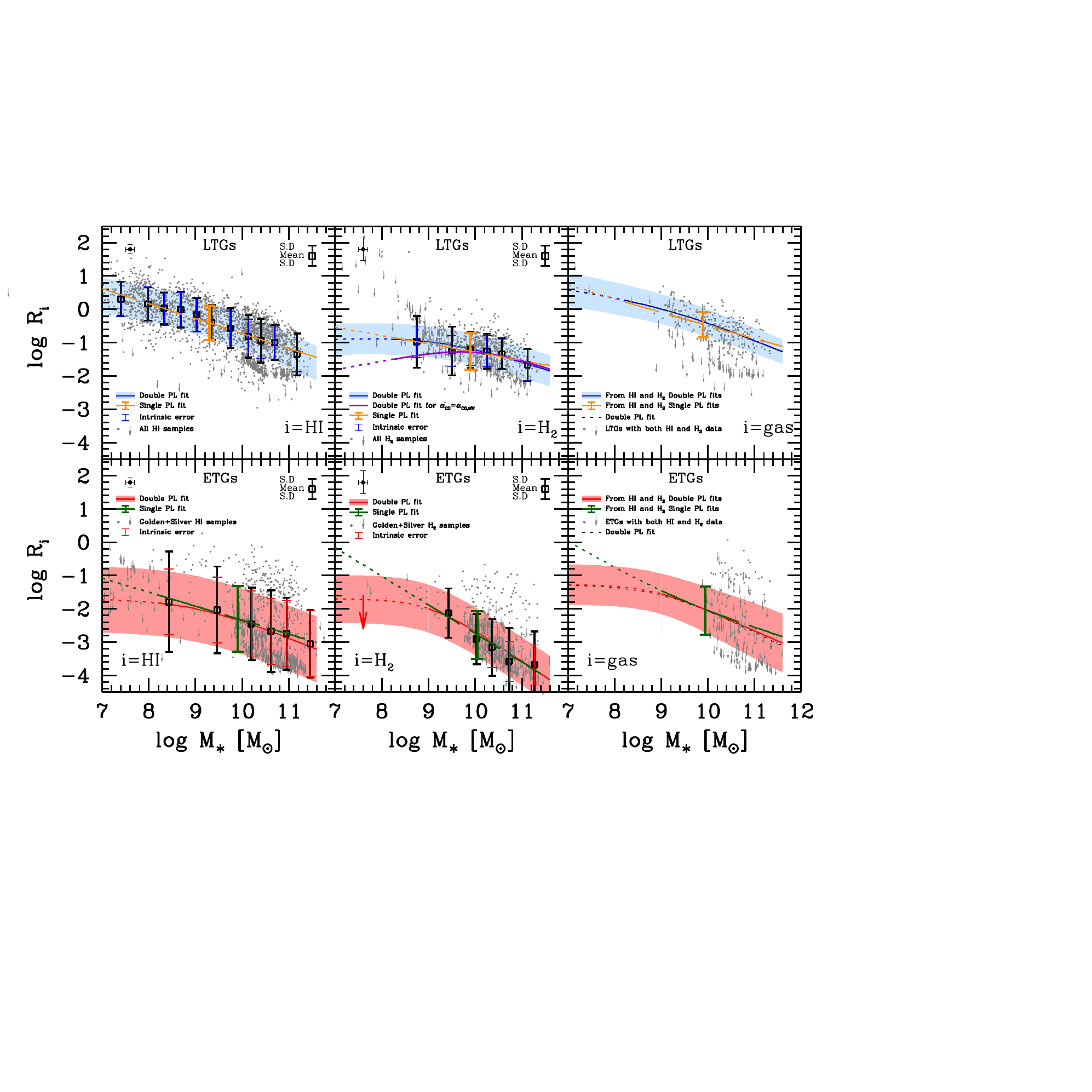}
\caption{ {\it Left panels:} The \RHI--\ms\ correlation for LTGs (upper panel) and ETGs (lower panel). 
Dots are detections and arrows are upper limits for non detections (for ETGs the Bronze sample were excluded). 
The squares and error bars are the mean and standard deviation in different mass bins calculated
by means of the Kaplan-Meier estimator for censored and uncensored data. The thin error bars 
correspond to our estimate of the {\it intrinsic} scatter after taking into account the observational errors (showed in the panel corners). 
The solid and long-dashed lines in each panel are respectively the best double- and single-power law fits. 
The shaded areas show the intrinsic scatter; to avoid
overcrowding, for the single power-law fit, the intrinsic scatter is plotted only at one point. The dotted lines are extrapolations of the correlations 
to low masses, where the data are scarce and dominated by upper limits.
 {\it Middle panels:} Same as in the left panels but for \RH2. For the ETG population, the double power-law fit was
 performed with the conservative constrain that below $\ms=10^9$ \msun, the low-mass slope is 0. {\it Right panels:} The \Rgas--\ms\ 
 correlations for LTGs and ETGs as calculated from combining the respective double- and single-power law
 \RHI--\ms\ and \RH2--\ms\ correlations and taking into account helium and metals 
 (see text). The shaded area and error bar are the ($1 \sigma$) intrinsic scatter obtained by error propagation of the 
 intrinsic scatter around the corresponding \RHI--\ms\ and \RH2--\ms\ relations. For completeness, 
 the data from our compilation that have determinations of both \HI\ and \H2\ masses are also plotted (the obtained 
correlations are not fits to these data). Dotted lines are extrapolations of the inferred relations to lower masses.  
The short dashed lines show the best fits using the double power-law function.
 }
\label{correlations} 
\end{center}
\end{figure*}

\subsection{The \HI-to-stellar mass correlations}
\label{RHI-Ms}

In the upper left panel of Fig. \ref{correlations}, along with the data from the Golden, Silver, and Bronze LTG samples,
the mean and standard deviation (squares  and black error bars) calculated in each mass bin with the Kaplan-Meier method are plotted.
In the lower left panel, the same is plotted but for the Golden and Silver ETG samples (recall that the Bronze samples are
excluded in this case). We see that the total standard deviations in $\log\RHI$, $\sigma_{\rm dat}$, do not evidence a 
systematical dependence on mass both for LTGs and ETGs. Then, we can use a constant value for each case. 
For LTGs, the standard deviations have values around 0.45--0.65 dex with an average 
of $\sigma_{\rm dat}\approx 0.53$ dex. For ETGs, the standard deviations are much larger and disparate among them 
than for LTGs (see subsection \ref{Mgas-Ms} below for a discussion on why this could be). We assume an average value of
$\sigma_{\rm dat}=1$ dex for ETGs. 

The {\it intrinsic} standard deviation (scatter) can be estimated as  
$\sigma_{\rm intr}^2 \approx \sigma_{\rm dat}^2 - \sigma_{\rm err}^2$
(this is valid for normal distributions), where $\sigma_{\rm err}$ is the mean statistical error in the $\log\RHI$ 
determination due to the observational uncertainties.  In Appendix \ref{observational-error} we present an estimate 
of this error, $ \sigma_{\rm err}\approx 0.14$ dex. Therefore, $\sigma_{\rm intr}\approx 0.52$
and $0.99$ dex for LTGs and ETGs, respectively. These estimates should be taken only as indicative values 
given the assumptions and rough approximations involved in their calculations.
For example, we will see in section \ref{scatter-distribution} that the distributions of $\log\RHI$  
(detections and non-detections) in different mass bins tend to deviate from a normal distribution, in particular for ETGs

\begin{table}[h]
	\centering
	\caption{Best fit parameters to the single power law (Eq. 1, \lowercase{$a=b$)}} 
	\resizebox{8.0cm}{!} {
		\begin{threeparttable}
			\begin{tabular}{ccccc}
				\hline
				\hline
				& $\log C'$   &  $a$  &  $\sigma_{\rm dat}$  & $\sigma_{\rm intr}$  \\ \hline
				\multicolumn{5}{c}{\RHI-\ms} \\ \hline 
				LTG & { 3.77} $\pm$ {0.22} &  {-0.45} $\pm$ {0.02} & 0.53 & 0.52 \\ 
				ETG & { 1.88} $\pm$ {0.33} &  {-0.42} $\pm$ {0.03} & 1.00 &  0.99 \\ 
				ETG$^{\rm ndc}$ & { 1.34} $\pm$ {0.46} &  {-0.37} $\pm$ {0.05} & 1.35 &  1.34 \\ \hline
				\multicolumn{5}{c}{\RH2-\ms} \\ \hline 
				LTG & { 1.21} $\pm$ {0.53} &  {-0.25} $\pm$ {0.05} & 0.58 & 0.47 \\ 
				ETG & { 5.86} $\pm$ {1.45} &  {-0.86} $\pm$ {0.14} & 0.80 &  0.72 \\ 
				ETG$^{\rm ndc}$ & { 5.27} $\pm$ {1.78} &  {-0.80} $\pm$ {0.17} & 0.95 &  0.88 \\ \hline
				\multicolumn{5}{c}{\Rgas-\ms} \\ \hline 
				LTG & { 4.76} $\pm$ {0.05} &  {-0.52} $\pm$ {0.03} & -- & 0.44 \\ 
				ETG & { 3.70} $\pm$ {0.07} &  {-0.58} $\pm$ {0.01} & -- &  0.68 \\ \hline
			\end{tabular}
			\begin{tablenotes}
				\item $\bullet$ The suffix ``ndc'' indicates when for the ETG correlations, no distance correction was
				applied to the upper limits in the (COLD) GASS samples.
				\item $\bullet$  $\sigma_{\rm dat}$  and $\sigma_{\rm intr}$ are in dex.
			\end{tablenotes}
		\end{threeparttable}
	}
	\label{linear-parameters}
\end{table}


\begin{table}[ht!]
	\centering
	\caption{Best fit parameters to the double power law (Eq. 1, \lowercase{$a\ne b$)}} 
	\resizebox{8.0cm}{!} {
		\begin{threeparttable}
			\begin{tabular}{ccccccc}
				\hline
				\hline
				& $C$   &  $a$  & $b$  &  $\log(M_*^{\rm tr}/\msun)$  & $\sigma_{\rm dat}$  & $\sigma_{\rm intr}$  \\ \hline
				\multicolumn{6}{c}{\RHI-\ms} \\ \hline 
				LTG & { 0.98} $\pm$ {0.06} &  { 0.21} $\pm$ {0.04} & 0.67  $\pm$ {0.03} & 9.24 $\pm$ {0.04} & 0.53  & 0.52 \\ 
				ETG & { 0.02} $\pm$ {0.01} & 0.00 $\pm$ {0.15} & 0.58 $\pm$ {0.03} & 9.00 $\pm$ {0.30} & 1.00 & 0.99 \\ 
				ETG$^{\rm ndc}$ & { 0.02} $\pm$ {0.01} & 0.00 $\pm$ {0.55} & 0.51  $\pm$ {0.05} & 9.00 $\pm$ {0.60} & 1.35 & 1.34 \\ \hline
				\multicolumn{6}{c}{\RH2-\ms} \\ \hline 
				LTG & { 0.19} $\pm$ {0.02} &  { -0.07} $\pm$ {0.18} & 0.47 $\pm$ {0.04} & 9.24 $\pm$ {0.12}  & 0.58 &  0.47 \\ 
				ETG & { 0.02} $\pm$ {0.01} & { 0.00} $\pm$ {0.00} & 0.94 $\pm$ { 0.15} & 9.01 $\pm$ {0.12} & 0.80 & 0.72 \\ 
				ETG$^{\rm ndc}$ & { 0.02} $\pm$ {0.03} & 0.00 $\pm$ {0.00}& 0.88 $\pm$ {0.18} & 9.01 $\pm$ {0.15} & 0.95 & 0.88\\ \hline
				\multicolumn{6}{c}{\Rgas-\ms} \\ \hline 
				LTG  & { 1.69} $\pm$ {0.02}& { 0.18} $\pm$ {0.01} & 0.61 $\pm$ {0.02} & 9.20 $\pm$ {0.04} & -- & 0.44 \\ 
				ETG & { 0.05} $\pm$ {0.02} & 0.01 $\pm$ {0.03} & 0.70 $\pm$ {0.01} & 9.02 $\pm$ {0.05} & -- & 0.68 \\ \hline
			\end{tabular}
			\begin{tablenotes}
				\item $\bullet$ The suffix ``ndc'' indicates when for the ETG correlations, no distance correction was
				applied to the upper limits in the (COLD) GASS samples.
				\item $\bullet$  $\sigma_{\rm dat}$  and $\sigma_{\rm intr}$ are in dex.
			\end{tablenotes}
		\end{threeparttable}
	}
	\label{parameters}
\end{table}


Next, we propose that the \HI-to-stellar mass relations can be described by the general function: 
\begin{equation}\label{function} 
y(\ms) = \frac{C}{\left(\frac{\ms}{M_*^{\rm tr}}\right)^{a} + \left(\frac{\ms}{M_*^{\rm tr}}\right)^{b}}
\end{equation} 
where $y=\RHI $, $C$ is the normalization factor, $a$ and $b$ are the low- and high-mass slopes of the function and 
$M_*^{\rm tr}$ is the transition mass. This function is continuous and differentiable. 
If $a=b$, then Eq. (\ref{function}) describes a single power law or a linear relation in logarithmic scales.
In this case, the equation remains as $y(\ms) = C'(\ms/\msun)^{-a}$. 
For $a\ne b$, the function corresponds to a double power law.

We fit the logarithm of function Eq. (\ref{function}) to the mean values of $\log\RHI$ as a function of mass (squares 
in the left panels of Fig. \ref{correlations}) with the corresponding (constant) intrinsic standard deviation as estimated above 
(thin blue/red error bars). 
For LTGs, the fit is carried out in the range $7.3\lesssim\log$(\ms/\msun)$\lesssim 11.2$, while 
for ETGs in the range $8.5\lesssim\log$(\ms/\msun)$\lesssim 11.5$. The Levenberg-Marquardt method is used for the fit 
\citep{Press+1996}. First, we perform the fits to the binned LTG and ETG data using a single power law, i.e., we fix $a=b$.  
The dashed orange and green lines with an error bar in the left  panels of Fig. \ref{correlations} show the results. The
fit parameters are given in Table \ref{linear-parameters}. We note that {\it these fits and those of the Buckley-James 
linear regression for all the data (not binned) in logarithm are very similar.} 

 Then, we fit to the binned data the logarithm of the double power-law function given in Eq. (\ref{function}). 
The corresponding best-fit parameters are presented in Table \ref{parameters}.  
We note that the fits are almost the same if the total mean standard deviation, $\sigma_{\rm dat}$, is used
instead of the intrinsic one. The reduced $\chi_{\rm red}^2$ are 0.01 and 0.03, respectively. 
The fits are actually performed to a low number of points (the number of mass bins) with large error bars; this is why the 
$\chi_{\rm red}^2$ are smaller than 1. Note, however, that the error bars are not related to measurement uncertainties but
 correspond to the population scatter of the data. Therefore, in this case $\chi_{\rm red}^2<1$ implies that while the best fit is good,
other fits could be also good within the scatter of the correlations. 
In the case of the single power-law fits, the $\chi_{\rm red}^2$ were 0.03 and 0.01, respectively for LTG and ETG.

The double power-law \RHI--\ms\ relations and the estimated intrinsic ($1\sigma$) scatter for the LTG (ETG) population are plotted in the 
left upper (lower) panel of Fig. \ref{correlations} with solid lines and shaded areas, respectively.
From the fits, we find for LTGs a transition mass $M_*^{\rm tr}=1.74\times 10^9$ \msun, with $\RHI\propto \ms^{-0.21}$ and $\ms^{-0.67}$ 
at masses much smaller and larger than this, respectively.
For ETGs, $M_*^{\rm tr}=1\times 10^9$ \msun, and $\RHI\propto \ms^{0.0}$ and $\ms^{-0.58}$, at masses 
much smaller and larger than this, respectively.  

Both the double and single power laws describe well the \HI-to-stellar mass correlations.
However, the former could be more adequate than the latter. In Fig. \ref{RHI} we plot the Buckley-James linear regressions 
to the \RHI\ vs. \ms\ data for the low- and high-mass sides (below and above $\log$(\ms/\msun)$\approx 9.7$; for ETGs 
the regression is applied only for masses above $10^{8}$ \msun);  
the dotted lines show the extrapolation of the fits. The slope at low masses for LTGs, $-0.36$, is shallower than the one at high masses, 
$-0.55$.  For ETGs, there is even evidence of a change in the slope sign at low masses. 
 A flattening of the overall (late + early type galaxies) correlation at low masses has been also suggested by 
\citet{Baldry+2008}, who have used the empirical mass--metallicity relation coupled with a metallicity-to-gas 
mass fraction relation (which can be derived from a simple chemical evolution model) to obtain a gas-to-stellar 
mass correlation in a large mass range. Another evidence that at low masses the \RHI--\ms\ relation flattens 
comes from the work by \citet{Maddox+2015} already mentioned above \citep[see also][]{Huang+2012a}. While the sample used 
by these authors does not allow to infer the \RHI--\ms\ correlation of galaxies due to its bias towards 
high \RHI\ values (see above), the upper envelope of this correlation can be actually
constrained; the high-\RHI\ envelope does not suffer of selection limit effects. 
As seen for the data from \citet{Maddox+2015} reproduced in the left upper panel of our Fig. \ref{RHI}, this envelope 
tends to flatten at $\ms\lesssim 2\times 10^9$ \msun,\footnote{ In \citet{Huang+2014}, the $SDSS-GALEX-\alpha.40$ common
sample was weighted by $V/V_{\rm max}$ to correct for incompleteness and mimic then the scaling relations derived
from a volume-limited sample. However,  only galaxies with $\mha\gtrsim 10^{8.2}$ \msun\ are included in their plot 
of \RHI\ vs. \ms\ (Fig. 1);  at lower masses, the correlation likely continues being biased to 
high values of \RHI.  Even that a weak flattening below $\ms\approx 10^9$ \msun\ is observed in their average curve.} 
which suggests (but it does not demonstrate) that the mean relation can suffer also such a flattening. 
Another pieces of evidence in favor of the flattening can be found in \citet{Huang+2012b}, and more recently 
in \citet[][]{Bradford+2015} for their sample of low-mass galaxies combined with larger-mass galaxies from  the ALFALFA survey.

\subsection{The \H2-to-stellar mass correlations} 
\label{RH2-Ms}

In the upper middle panel of Fig. \ref{correlations}, along with the data from the Golden, Silver, and Bronze LTG samples,
the mean and standard deviation (error bars) calculated in each mass bin with the Kaplan-Meier method are plotted.
In the lower panel, the same is plotted but for the Golden and Silver ETG samples (recall that the Bronze samples are
excluded in this case). The poor observational information at stellar masses smaller than 
$\approx 5\times 10^8$ \msun\ does not allow us to constrain the correlations at these masses, both for LTG and ETGs. 
  Regarding the total standard deviations, for both LTGs and ETGs, they vary from mass bin to mass bin but without a clear
 trend. Then we can use a constant value for both cases. For LTGs, the total standard deviations have values around 0.5--0.8 dex with an 
average of $\sigma_{\rm dat}\approx 0.58$ dex. For ETGs, the average value is roughly 0.8 dex. 
As in the case of \HI\ (previous subsection), we further estimate indicative values for the {\it intrinsic} population standard deviations (scatter). 
For this, we present in Appendix \ref{observational-error} an estimate of the the mean observational error in the $\log\RH2$ 
determination, $ \sigma_{\rm err}\approx 0.34$ dex. Therefore, the estimated mean intrinsic scatters in $\log\RH2$ are 
$\sigma_{\rm intr}\approx 0.47$ and $0.72$ dex for LTGs and ETGs, respectively. Given the assumptions and approximations
involved in these estimates, they should be taken with caution.  For example, we will see in section \ref{scatter-distribution} 
that the distributions of $\log\RH2$  (detections and non-detections) in different mass bins tend to deviate from a normal distribution, 
in particular for the ETGs.

We fit the logarithm of function Eq. (\ref{function}; $y=\RH2$) to the mean values of $\log\RH2$ as a function of mass (squares 
in the left panels of Fig. \ref{correlations}) with their corresponding scatter as estimated above (thin blue/red error bars), 
assumed to be the individual standard deviations for the fit.
Again, the Levenberg-Marquardt method is used to perform the fit. 
The fits extend only down to $\ms\approx 5\times 10^8$ \msun. 
First, the fits are performed for a singe power law, i.e., we fix $a=b$. 
The dashed orange and green lines in the middle panels of Fig. \ref{correlations} show the results. The parameters 
of the fit and their standard deviations are given in Table \ref{linear-parameters}. {\it The fits are very similar to those
obtained using  the Buckley-James linear regression to the all (not binned) logarithmic data.}

Then, we fit the binned LTG and ETG data to the double power-law function Eq. (\ref{function}). 
In the case of the ETG population, we impose an extra condition to the fit: that the slope of the relation
at masses below $\sim 10^9$ \msun\ is flat. The few data at these masses clearly show that \RH2\
does not increase as \ms\ is smaller; it is likely that even decreases, so that our assumption of 
a flat slope is conservative. The corresponding best-fit parameters are presented in Table \ref{parameters}.  
As in the case of the $\RHI-\ms$ correlations, the reduced $\chi_{\rm red}^2$ are smaller than 1 (0.04 and 0.10, respectively), 
which implies that while the best fits are good, other fits could describe reasonably well the scattered data.
In the case of the single power-law fits, $\chi_{\rm red}^2$ were 0.04 and 0.07, respectively for LTG and ETG.
The double power-law \RH2--\ms\ relations and their ($1\sigma$) {\it intrinsic} scatter for the LTG (ETG) population are plotted in the 
middle upper (lower) panel of Fig. \ref{correlations} with solid lines and shaded areas, respectively. We note that the fits are 
almost the same if the total mean standard deviation, $\sigma_{\rm dat}$, is used instead of the intrinsic one.

From these fits, we find for LTGs, $M_*^{\rm tr}=1.74\times 10^9$ \msun, with
$\RH2\propto \ms^{-0.07}$ and $\ms^{-0.47}$ at much smaller and larger masses than this, respectively.
For ETGs, $M_*^{\rm tr}=1.02\times 10^9$ \msun, with $\RH2\propto \ms^{0.00}$ and $\ms^{-0.94}$ at much smaller 
and larger masses than this, respectively. 
In the middle upper panel of Fig. \ref{correlations}, we plot also the best double power-law fit to the \RH2--\ms\ correlation 
of LTGs in the case the \a_co\ factor is assumed constant and equal to the MW value (purple dashed line).

Both the single and double power-law functions describe equally well the \RH2 --\ms\ correlations for the LTG and ETG population,
but there is some evidence of a change of slope at low masses. In Fig. \ref{RH2}, the Buckley-James linear regressions 
to the \RH2\ vs. \ms\ data below and above $\log$(\ms/\msun)$\approx 9.7$ are plotted (in the former case the regressions are applied  
for masses only above $10^8$ \msun);  the dotted lines show the extrapolation of the fits. The slopes in the small mass range at low 
masses for LTGs/ETGs are shallower than those at high masses.  Besides, in the case of ETGs, if the single power-law fit shown in 
Fig. \ref{correlations} is extrapolated to low masses, ETGs of $\ms\approx 10^7$ \msun\ would be dominated in mass by \H2\ gas. 
Red/passive dwarf spheroidals are not expected to contain significant fractions of molecular gas.
Recently, \citet{Accurso+2017} have also reported a flattening in the \H2-to-stellar mass correlation at stellar masses 
below $\sim 10^{10}$ \msun.

\subsection{The cold gas-to-stellar mass correlations}
\label{Mgas-Ms}

Combining the \RHI--\ms~and \RH2--\ms~ relations presented above, we can obtain now the 
\Rgas--\ms~ relation, for both the LTG and ETG populations. Here, $\Rgas = \mg/\ms = 1.4(\RHI + \RH2)$, 
where \mg\ is the galaxy cold gas mass, including helium and metals (the factor 1.4 accounts for these components). 
The intrinsic scatter around the gas-to-stellar mass relation can be estimated by propagating the intrinsic scatter around the 
\HI- and \H2-to-stellar mass relations. Under the assumption of null covariance, the logarithmic standard 
deviation around the composed $\log$\Rgas--$\log$\ms\ relation is given by 
\begin{equation}
\label{eq:Sigma_logRgas}
\sigma_{\rm intr,R_{\rm gas}} =\frac{1}{R_{\rm H_{I}}+R_{\rm H_{2}}}\left(R_{\rm H_{I}}^{2}\sigma_{\rm intr,R_{\rm H_{I}}}^{2}+R_{\rm H_{2}}^{2}\sigma_{\rm intr,R_{\rm H_{2}}}^{2}\right)^{\frac{1}{2}}
\end{equation}

The obtained cold gas-to-stellar mass correlations for the LTG  and ETG populations 
are plotted in the right panels of Fig. \ref{correlations}. The solid lines and shaded bands (intrinsic scatter given 
by the error propagation) were obtained from the double power-law correlations, while the solid green lines and the 
error bars were obtained from the single power-law correlations.   
For completeness, we plot in Fig. \ref{correlations} also those galaxies from our compilation that have determinations for 
{\it both} the \HI\ and \H2\ masses. Note that a large fraction of our compilation have not determinations for both
quantities at the same time.  
We fit the results obtained for the singe (double) power-law fits, taking into account the intrinsic scatter, to 
the logarithm of the single (double) power-law function given in Eq. (\ref{function}) with $y=\Rgas$ and 
report in Table \ref{linear-parameters} (Table \ref{parameters}) the obtained parameters for both the LTGs and ETGs. 
The fits for the double power-law fit are shown with dotted lines in Fig. \ref{correlations}. 
The standard deviations $\sigma_{\log R_{\rm gas}}$ change slightly with mass; we report an average value for them
in Tables \ref{linear-parameters} and \ref{parameters}. Both for LTGs and ETGs, the mass at which the 
\Rgas--\ms\ correlations change of slope is $M_*^{\rm tr}\approx 1.7\times 10^9$ \msun, {\it the mass that roughly separates
dwarf from normal galaxies.}

According to Fig. \ref{correlations}, the LTG and ETG \Rgas--\ms~correlations are significantly different among them. 
The gas content in the former is at all masses larger than in the latter, the difference being maximal at the
largest masses. 
For the LTG population, $\mg\approx \ms$ on average at $\log(\ms/\msun)\sim 9$, and at lower masses, these galaxies
are dominated by cold gas; at stellar masses around $2\times 10^7$ \msun, \mg\ is on average three times larger than \ms. 
For ETGs, there is a hint that at $\sim 10^9$ \msun$, \Rgas$  changes from increasing as \ms\ is smaller to decrease. 

\section{The distributions of the scatter around the gas-to-stellar mass relations}
\label{scatter-distribution}

To determine the correlations presented above, we have made use only of the mean and standard deviation of the data
in different mass bins. It is also of interest to learn about the scatter distributions around the main relations. Even more, in 
the next Section we will require the full distributions of \RHI(\ms) and \RH2(\ms) in order to generate a mock galaxy catalog through which the 
 \HI\ and \H2\ mass functions will be calculated. The Kaplan-Meier estimator provides information for constructing
 the probability density function (PDF) at a given stellar mass including the uncensored data. By using these PDFs we explore the 
 distribution of the \RHI\ and \RH2\ data (detections + upper limits). Given the 
heterogeneous nature of our compiled data, these ``scatter'' distributions  should be taken just as a rough approximation. 
On the other hand, when the uncensored data dominate  (this happens in most of the mass bins for the ETG samples), 
the Kaplan-Meier estimator can not predict very well the distribution of the uncensored data. 
\begin{figure} [ht!]
\plotPDFHILTG{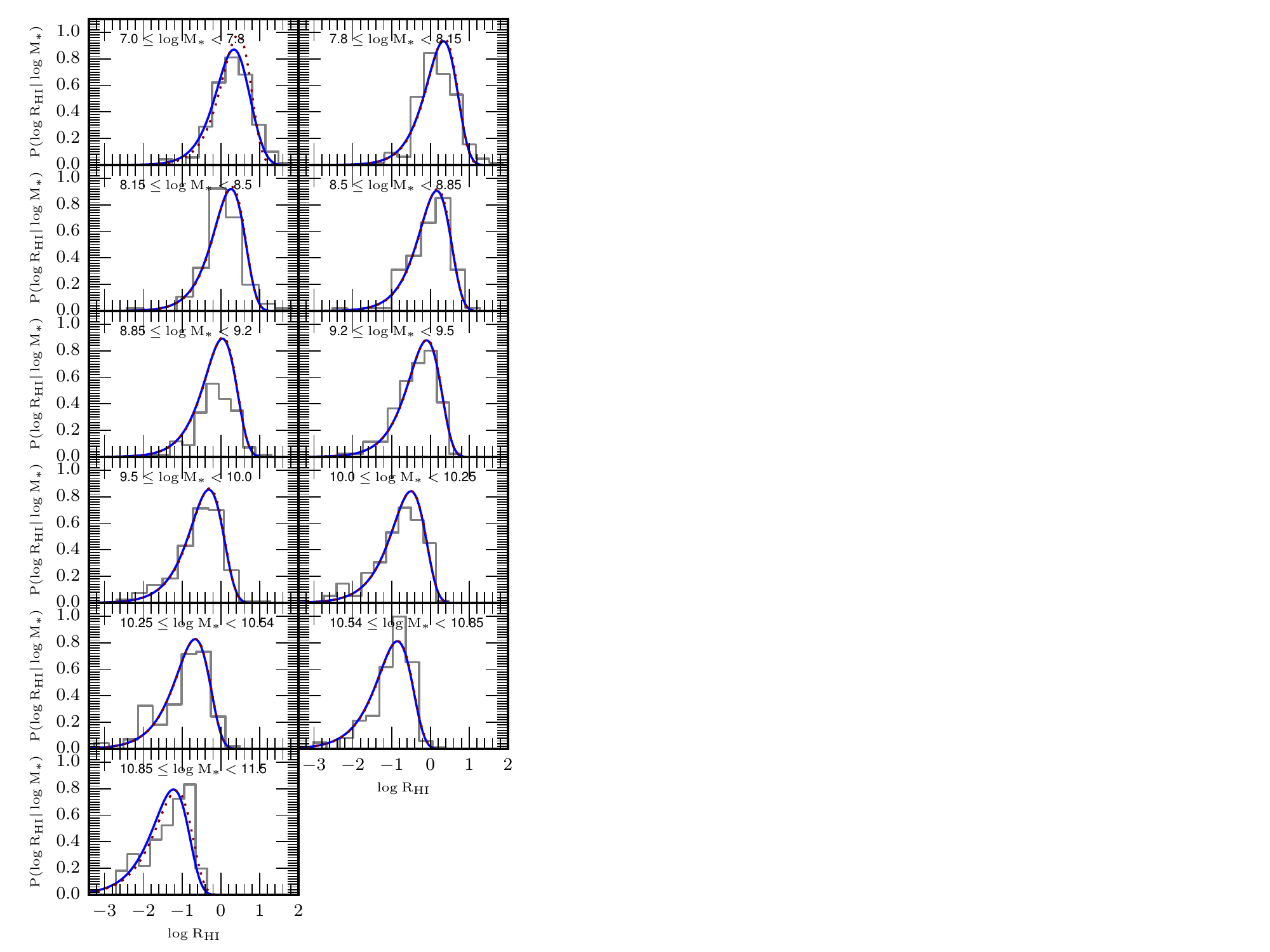}
\caption{Distributions (PDFs) of the LTG \HI-to-stellar mass ratios in different stellar mass bins (indicated inside the panels). The gray histograms show results from the Kaplan-Meier estimator applied to the data (detections + upper limits), and the solid blue line corresponds to the best fitted number density-weighted distribution within the given mass bin (eq. \ref{ave-PDF-LTGs}); the constrained parameters of the mass-dependent PDF (Eq. \ref{full-fit}) are given in Table \ref{full-distributions}. The red dotted line shows the constrained function Eq. (\ref{full-fit}) evaluated at the mass corresponding to the logarithmic center of each mass bin.}
\label{pdfHI-ltg} 
\end{figure}

\begin{figure} [ht!]
\plotPDFH2LTG{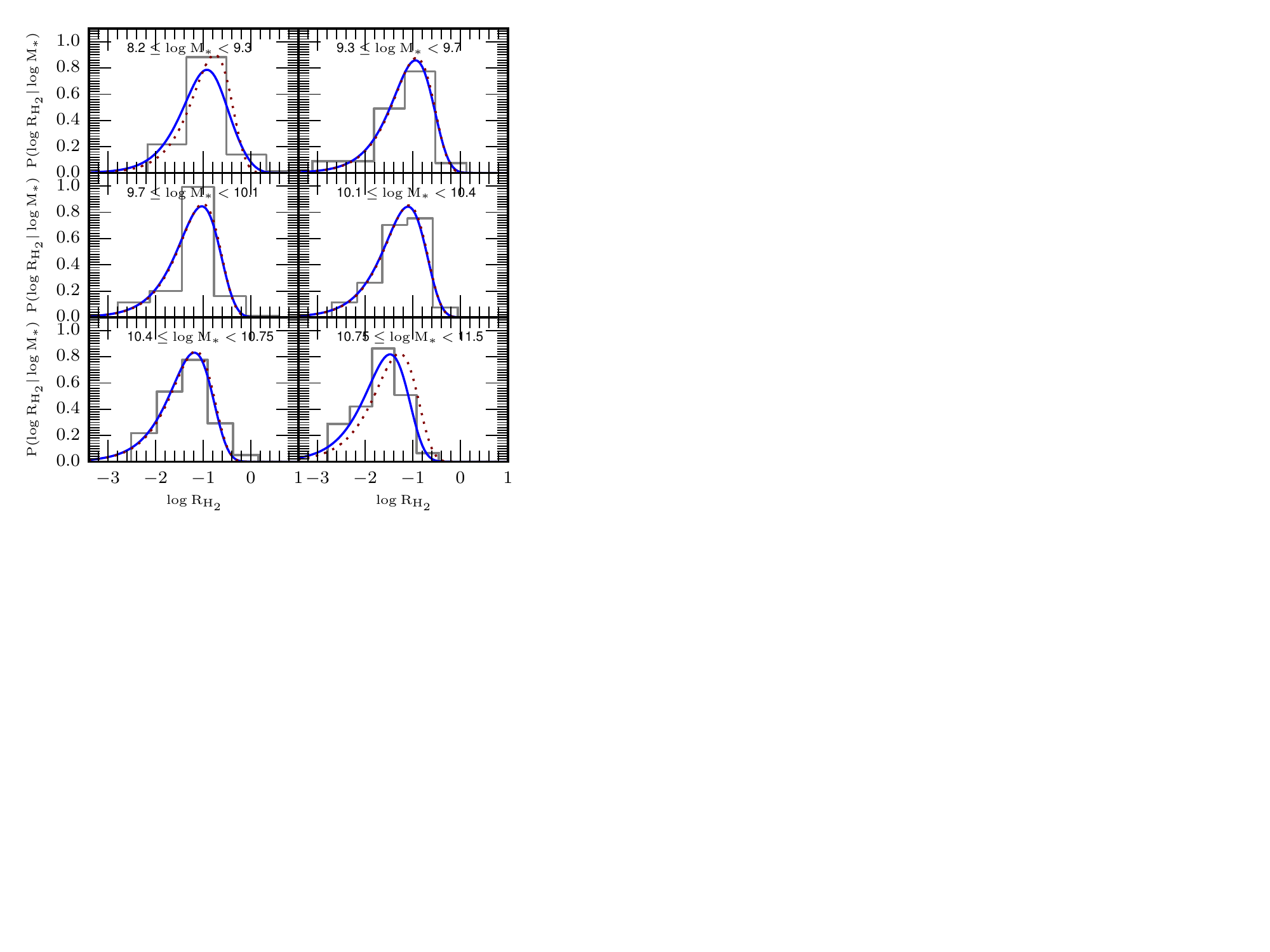}
\caption{Same as Figure \ref{pdfHI-ltg} but for the \H2-to-stellar mass ratios.} 
\label{pdfH2-ltg} 
\end{figure}

\begin{table*}[ht!]
	\centering
	\caption{Best fit parameters to the full distributions} 
	\resizebox{18.0cm}{!} {
		\begin{threeparttable}
			\begin{tabular}{ccccccccccc}
				\hline
				\hline
				& $c$   &  $d$  & $x_0$  &  $\log(m_*^{\rm tr}/\msun)$  & $e$ & $f$ & $g$ & $h$ & $i$ & $j$ \\ \hline
				\multicolumn{11}{c}{$P(\RHI|\ms)$ distributions} \\ \hline 
				LTG & {1.11$\pm$0.35}  &  { -0.11$\pm$0.04}  & {2.45$\pm$0.76}  & { 8.77$\pm$0.45} & {0.002$\pm$0.10} &  0.61$\pm$0.07 & -- & -- & -- & --  \\ 
				ETG & {-0.42$\pm$0.80}    &   {-0.02$\pm$0.08}    &  {2.15$\pm$0.55} & { 8.30$\pm$0.38} & -0.43$\pm$1.10 & 0.52$\pm$0.09 & -0.22$\pm$0.37 & 0.07$\pm$0.04 & -1.62$\pm$1.08 & -0.13$\pm$0.11 \\  \hline
				\multicolumn{11}{c}{$P(\RH2|\ms)$ distributions} \\ \hline 
				LTG & {0.70$\pm$1.28}  &  {-0.07$\pm$0.13}  & {0.15$\pm$0.03}  & {10.37$\pm$0.31} & { 0.19$\pm$0.17} &  { 0.19$\pm$0.16} & -- & -- & -- & --  \\ 
				ETG & {-0.52$\pm$1.19}    &   {-0.01$\pm$0.11}     &  {0.71$\pm$0.27} & { 7.90$\pm$1.09} & 0.42$\pm$0.50 & 0.21$\pm$0.28 & 0.24$\pm$0.97 & 0.04$\pm$0.09 & 5.74$\pm$3.17 & -0.86$\pm$0.29 \\  \hline
			\end{tabular}
			\begin{tablenotes}
				\item For LTGs the distributions are given by Eq. (\ref{full-fit}), while for ETGs, by
				Eq. (\ref{full-fit2}). 
			\end{tablenotes}
		\end{threeparttable}
	}
	\label{full-distributions}
\end{table*}

{\bf Late-type galaxies.- }
Figures \ref{pdfHI-ltg} and \ref{pdfH2-ltg} present the \RHI\ and \RH2\ PDFs in different \ms\ bins 
for LTGs. Based on the bivariate \HI\ and stellar mass function analysis of \citet{Lemonias+2013}, who used the 
GASS sample for (all-type) massive galaxies, we propose that the PDFs of \RHI\ and \RH2\ for LTGs 
can be described by a Schechter (Sch) function (Eq. \ref{full-fit} below; $x$ denotes either \RHI\ or \RH2).  
By fitting this function to the \RHI\ data in each stellar mass bin we find that the power-law index $\alpha$
weakly depends on \ms\ with most of the values being around $-0.15$  \citep[see also][]{Lemonias+2013}, 
while the break parameter $x^*$ varies with \ms. A similar behavior was found for \RH2\ with most of the
values of $\alpha$ around $-0.10$.  We then perform for each case (\RHI\ and \RH2) a 
continuous fit across the range of stellar-mass bins rather than fits within independent bins. 
The general function proposed to describe the \RHI\ and \RH2\ PDFs of LTGs, at a fixed \ms\ and within the range $\log x\pm d\log x/2$, is:
\begin{equation}\label{full-fit}
P_{\rm Sch}(x|\ms)= \frac{\phi^*}{\log e} \left(\frac{x}{x^*}\right)^{\alpha+1} \exp\left(-\frac{x}{x^*}\right),
\end{equation}
and with the normalization condition, $\phi^* =1/\Gamma(1+\alpha)$, 
where $\Gamma$ is the complete gamma function, which guarantees that the integration over the full space in $x$ is 1. 
The parameters $\alpha$ and $x^*$ depend on \ms. We propose the following functions for these dependences:
\begin{equation}
\alpha(\ms) = c + d\log\ms,
\end{equation}
and
\begin{equation}
x^*(\ms)=\frac{x_0}{\left(\frac{\ms}{m_{\rm tr}}\right)^{e} + \left(\frac{\ms}{m_{\rm tr}}\right)^{f}}. 
\end{equation}

The parameters $c, d, x_0, m_{\rm tr}, e,$ and $f$ are constrained from a {\it continuous fit across all the mass bins} using a Markov Chain Monte Carlo method following \citet{Rodriguez-Puebla+2013}. Since the stellar mass bins from the data have a width, for a more precise determination, we convolve the PDF with the GSMF within a given bin. Therefore, the PDF of $x$ averaged within the bin $\Delta\ms=$[$\rm M_{\ast 1}$,$\rm M_{\ast 2}$] is:
\begin{equation}\label{ave-PDF-LTGs}
\left\langle P_{\rm Sch}(x|\Delta\ms)\right\rangle=\frac{\int_{\rm M_{\ast 1}}^{\rm M_{\ast 2}}P_{\rm Sch}(x|\ms)\Phi_{late}(\ms)d\ms}{\int_{\rm M_{\ast 1}}^{\rm M_{\ast 2}}\Phi_{late}(\ms)d\ms},
\end{equation}
where $\Phi_{late}(\ms)$ is the GSMF for LTGs (see Section \ref{mass-functions}). The constrained parameters are reported in Table \ref{full-distributions}. The obtaiened mass-dependent
PDFs are plotted in each one of the panels of Figures \ref{pdfHI-ltg} and \ref{pdfH2-ltg}. The solid blue line corresponds to the number density-weighted distribution within the given mass bin (eq. \ref{ave-PDF-LTGs}), while the red dotted line is for the function Eq. (\ref{full-fit}) evaluated at the mass corresponding to the logarithmic center of each bin.  As seen, the Kaplan-Meier PDFs obtained from the data (gray histograms) are well described by the proposed Schechter function averaged within the different mass bins (blue lines), both for \RHI\ and \RH2.

\begin{figure} [ht!]
\plotPDFH2LTG{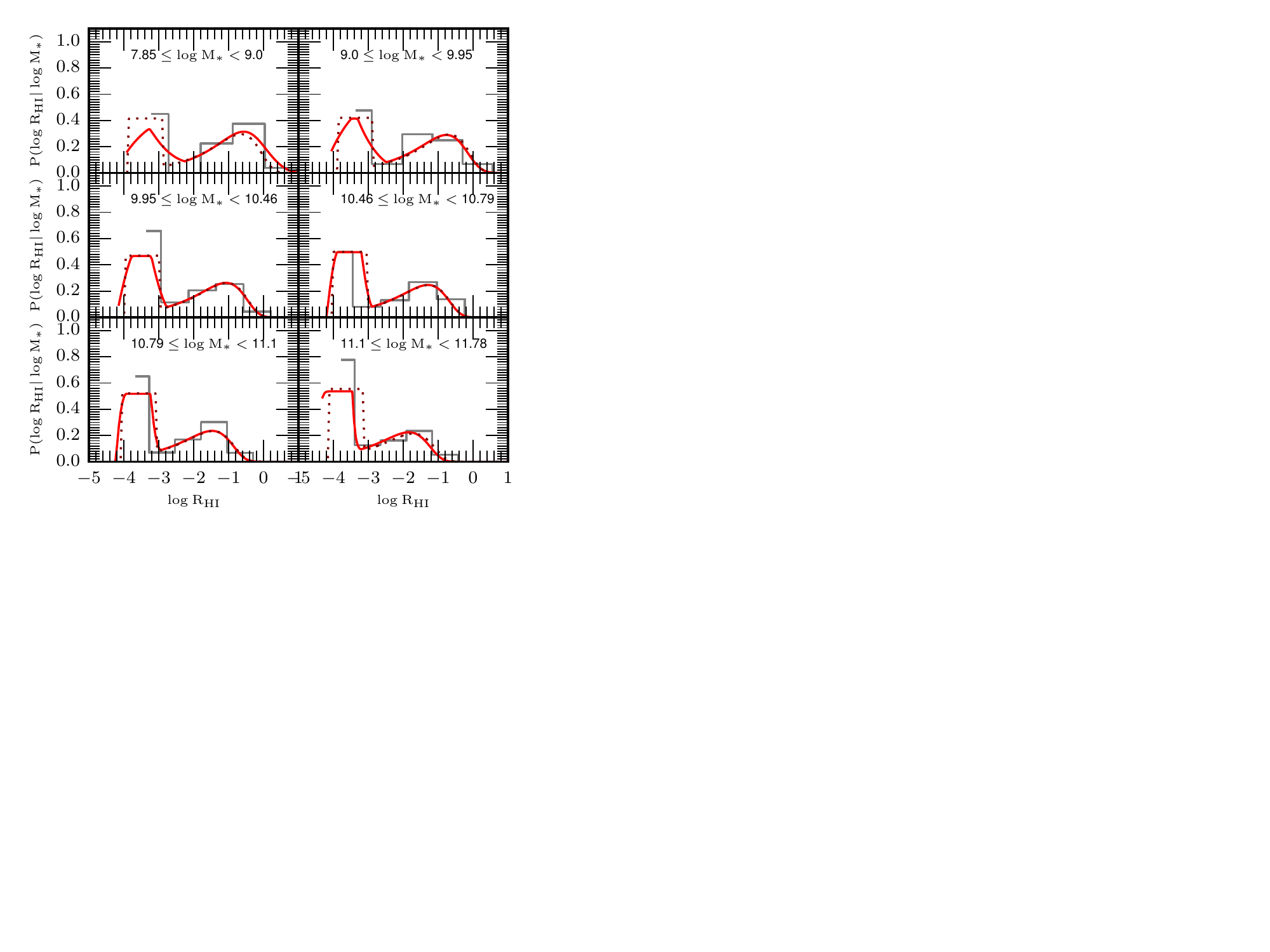}
\caption{Distributions (PDFs) of the ETG \HI-to-stellar mass ratios in different stellar mass bins (indicated inside the panels).  The gray histograms show results from the Kaplan-Meier estimator applied to the data (detections + upper limits), and the solid ted line corresponds to the best fitted number density-weighted distribution within the given mass bin (eq. \ref{ave-PDF-ETGs}); the constrained parameters of the mass-dependent PDF (Eq. \ref{full-fit2}) are given in Table \ref{full-distributions}. The red dotted line shows the constrained function Eq. (\ref{full-fit2}) evaluated at the mass corresponding to the logarithmic center of each mass bin.}
\label{pdfHI-etg} 
\end{figure}

\begin{figure} [ht!]
\plotPDFH2LTG{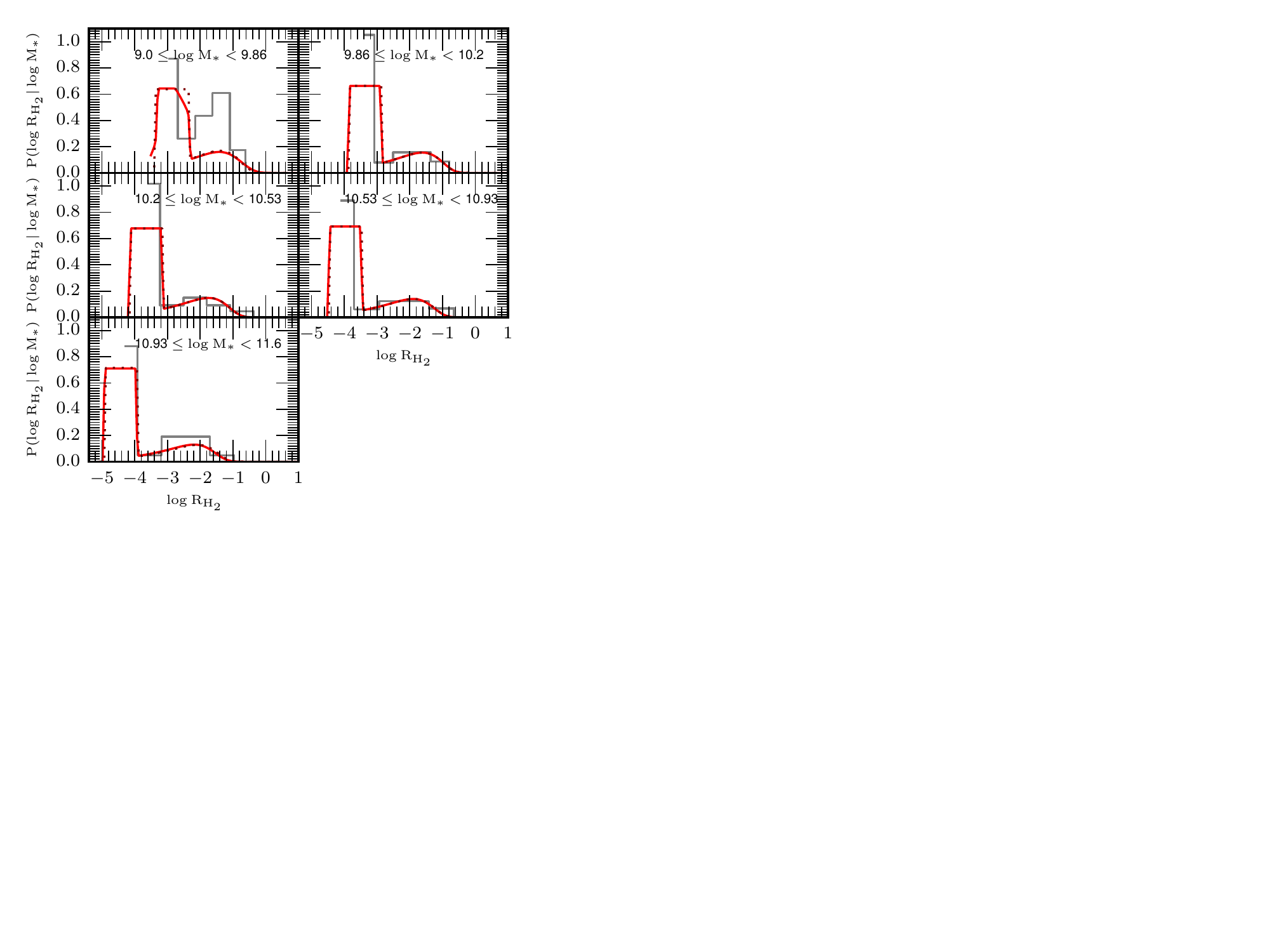}
\caption{Same as Figure \ref{pdfHI-etg} but for the \H2-to-stellar mass ratios.} 
\label{pdfH2-etg} 
\end{figure}

{\bf Early-type galaxies.-}
We present the \RHI\ and \RH2\ PDFs for ETGs in Figures \ref{pdfHI-etg} and \ref{pdfH2-etg}, respectively.
The distributions are very extended, implying a large scatter in the \RH2--\ms\ correlations as discussed in 
subsections \ref{RHI-Ms} and \ref{RH2-Ms}.\footnote{Given this large scatter, previous works, for small samples of massive galaxies, 
have suggested that red or early-type galaxies do not follow a defined correlation between \mha\ and \ms\ 
\citep[or luminosity; e.g.,][]{Welch+2010,Serra+2012}  and between \mhm\ and \ms\ \citep[e.g.,][]{Saintonge+2011,Lisenfeld+2011,Young+2011}.}
The distributions seem to be bimodal, with a significant fraction of ETGs having gas fractions around a low limit ($\sim 10^{-4}$) and the
remaining galaxies with higher gas fractions, following an asymmetrical distribution. The low limit is given by the Kaplan-Meier 
estimator and it is associated with the reported upper limits of non-detections. 
We should have in mind that when non-detections dominate, the Kaplan-Meier estimator can not provide a reliable PDF at the low end
of the distribution.  From a physical point of view, we know that ETGs are in general quiescent galaxies that likely exhausted
their cold gas reservoirs and did not accrete more gas. However, yet small amounts of gas can be available from the winds of old/intermediate-age
stars.  For instance, Sun-like stars can lose $\sim 10^{-4}-10^{-5}$ of their masses in 1 Gyr; more massive stars, lose higher fractions. 
A fraction of the ejected material is expected to cool efficiently and ends as \HI\ and/or \H2\ gas. 
On the other hand, those ETGs that have larger fractions of cold gas, could get it by radiative cooling from their hot halos or by accretion 
from the cosmic web, and/or by accretion from recent mergers \citep[see for a discussion ][and more references therein]{Lagos+2014}. 
The amount of gas acquired depends on the halo mass, the environment, the gas mass
of the colliding galaxy, etc. The range of possibilities is large, hence, the scatter around the 
ETG $\RHI-\ms$  and \RH2--\ms\ relations are expected to be large as semi-analytic models show \citep{Lagos+2014}.

To describe the PDFs seen in Figures \ref{pdfHI-etg} and \ref{pdfH2-etg}, we propose a (broken) Schechter
function plus a uniform distribution. The value of \RHI \ or \RH2\ where the Schechter function breaks and the uniform
distribution starts,  $x_2$, seems to depend on \ms\ (see Figs. \ref{pdfHI-etg} and \ref{pdfH2-etg}). The lowest values where
the distributions end, $x_{1}$, are not well determined by the Kaplan-Meier estimator, as mentioned above.  
To avoid unnecessary sophistication, we just fix $x_1$ as one tenth of $x_2$. This implies physical lowest values for \RHI\ and \RH2\ of $10^{-4\div -5}$, which are plaussible according to our discussion above.  
The value of the Schechter parameter $\alpha$ shows a weak dependence on \ms\ for both \HI\ and \H2. 
On the other hand, the fraction of galaxies between $x_1$ and $x_2$, $F$, seems to depend on \ms. For the uniform distribution, this fraction is given by $F = P(<x_2|\ms) - P(<x_1|\ms)=\int_{x_1}^{x_2}Cd\log x$, where $C=F/(\log x_2- \log x_1)$; given our assumption of $\log x_2- \log x_1 = 1$ dex, then $C=F(\ms)$.
We parametrize all these dependences on \ms\ and perform a continuous fit across the range of stellar-mass bins, 
both for the \RHI\ and \RH2\ data. The general function proposed to describe 
the PDFs of ETGs as a function of \ms\ within the range 
$\log x \pm d\log x/2$ is the sum of a 
Schechter function, $P_{\rm Sch}(x|\ms)$, and a uniform function
in $x$ but dependent on $\ms$, $C=F(\ms)$:
\begin{eqnarray}\label{full-fit2}
F(\ms) = g + h\log\ms, \ \ \ \ x_1\le x < x_2(\ms), \\ \nonumber
x_2(\ms)= i + j\log\ms, \\ \nonumber
P_{\rm Sch}(x|\ms), \ \ \ \ \ \ \ \ \ \ \ \ x\ge x_2(\ms),
\end{eqnarray}
where the parameters $x^*$ and $\alpha$ in $P_{\rm Sch}(x|\ms)$ are described by Eq. (\ref{full-fit}) with the normalization 
condition $\phi^* =(1-F)/\Gamma(1+\alpha)$, and $\log x_1= \log x_2 - 1$. 
The parameters $x_0$, $m_{\rm tr}$, $e$, and  $f$ of the broken Schechter function and the parameters 
$g$, $h$, $i$, and $j$ of the uniform distribution are constrained as described for LTGs above, from a continuous fit accross all the mass bins using the number density-weighted PDFs at each stellar mass bin:
$$\left\langle P_{\rm Sch}(x|\Delta \ms)+C\right\rangle=\hspace*{1.5in}$$
\vspace*{-0.3in}
\begin{eqnarray}\label{ave-PDF-ETGs}
 \frac{\int_{\rm M_{\ast 1}}^{\rm M_{\ast 2}}(P_{\rm Sch}(x|\ms)+C)\cdot\Phi_{early}(\ms)d\ms}{\int_{\rm M_{\ast 1}}^{\rm M_{\ast 2}}\Phi_{early}(\ms)d\ms},
\end{eqnarray}
where $\Phi_{early}(\ms)$ is the GSMF for ETGs (see Section \ref{mass-functions}). The constrained parameters are reported in Table \ref{full-distributions}, both for \RHI\ and \RH2. The obtained mass-dependent distribution function is plotted in each one of the panels of Figures \ref{pdfHI-etg} and \ref{pdfH2-etg}
The solid red line corresponds to the number density-weighted distribution within the given mass bin (eq. \ref{ave-PDF-ETGs}), while the red dotted line is for the proposed broken Schechter + uniform function evaluated at the mass corresponding to the logarithmic center of each bin.  As seen, the Kaplan-Meier PDFs obtained from the data (gray histograms) are reasonably well described by the proposed function (eq. \ref{full-fit2}) averaged within the different mass bins (red lines), both for \RHI\ and \RH2.

\begin{figure}[ht!]
\includegraphics[trim = 3mm 37mm 28mm 0mm, clip, width=\textwidth]{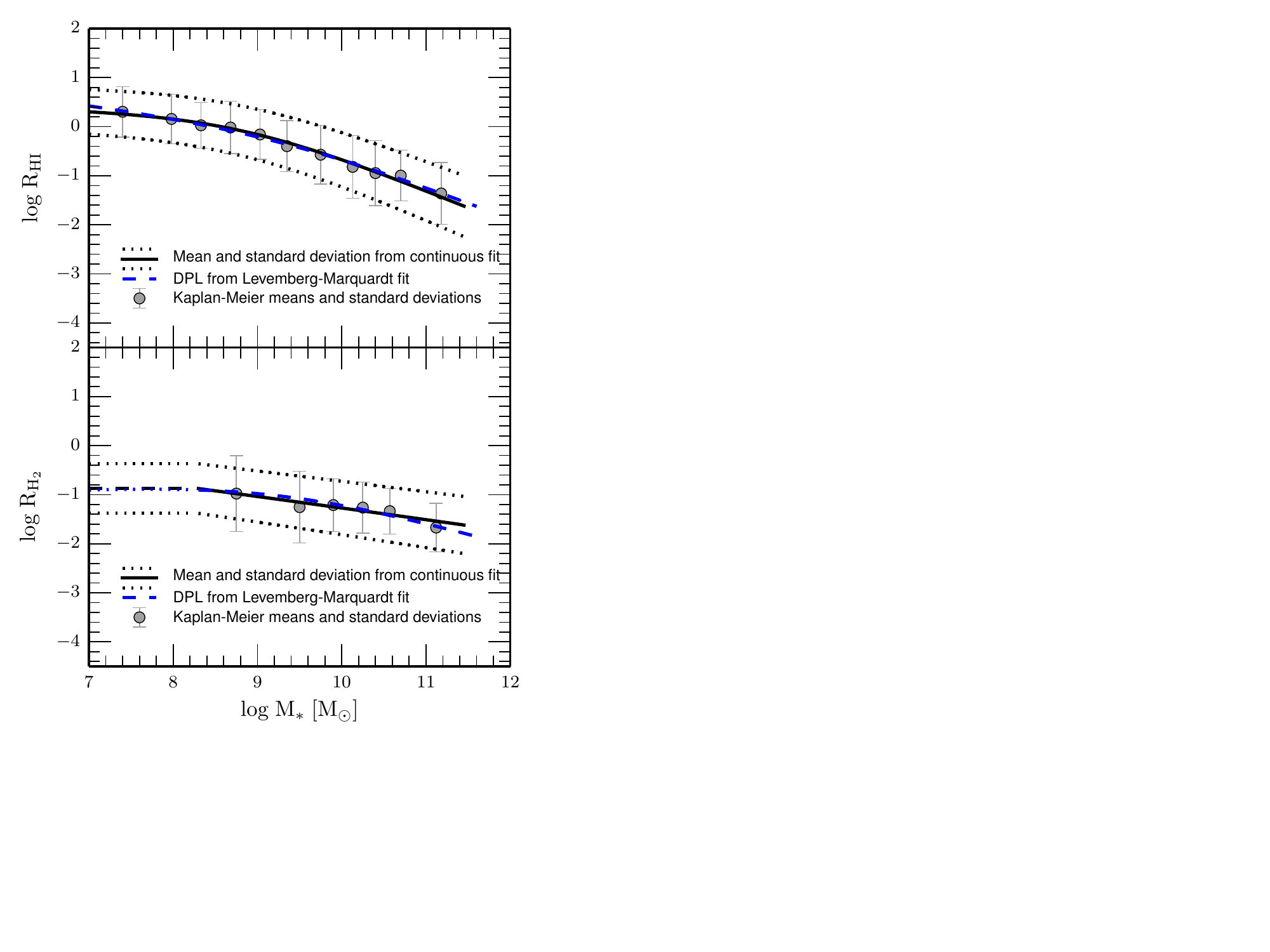}
\caption{Mean and standard deviation as a function of stellar mass (solid and dotted black lines) from the 
distributions of \RHI\ (upper panel) and \RH2\ (lower panel) for LTGs as given by Eq. (\ref{full-fit}) (see Table \ref{full-distributions}
for the constrained parameters). When the data are insufficient at low masses, the distributions are assumed the same
as in the last mass bin (dashed black lines). The gray dots with error bars are the mean and standard deviation obtained with the
Kaplan-Meier estimator applied to the data (detections + upper limits) in different mass bins, as shown in Figure \ref{correlations}.
The double-power law fits to these data as reported in Section 4 are reproduced with the blue dashed lines (the blue
dotted lines are extrapolations of these fits). 
 } 
\label{ltg-fulldistribution} 
\end{figure}

\begin{figure} [ht!]
\includegraphics[trim = 3mm 37mm 28mm 0mm, clip, width=\textwidth]{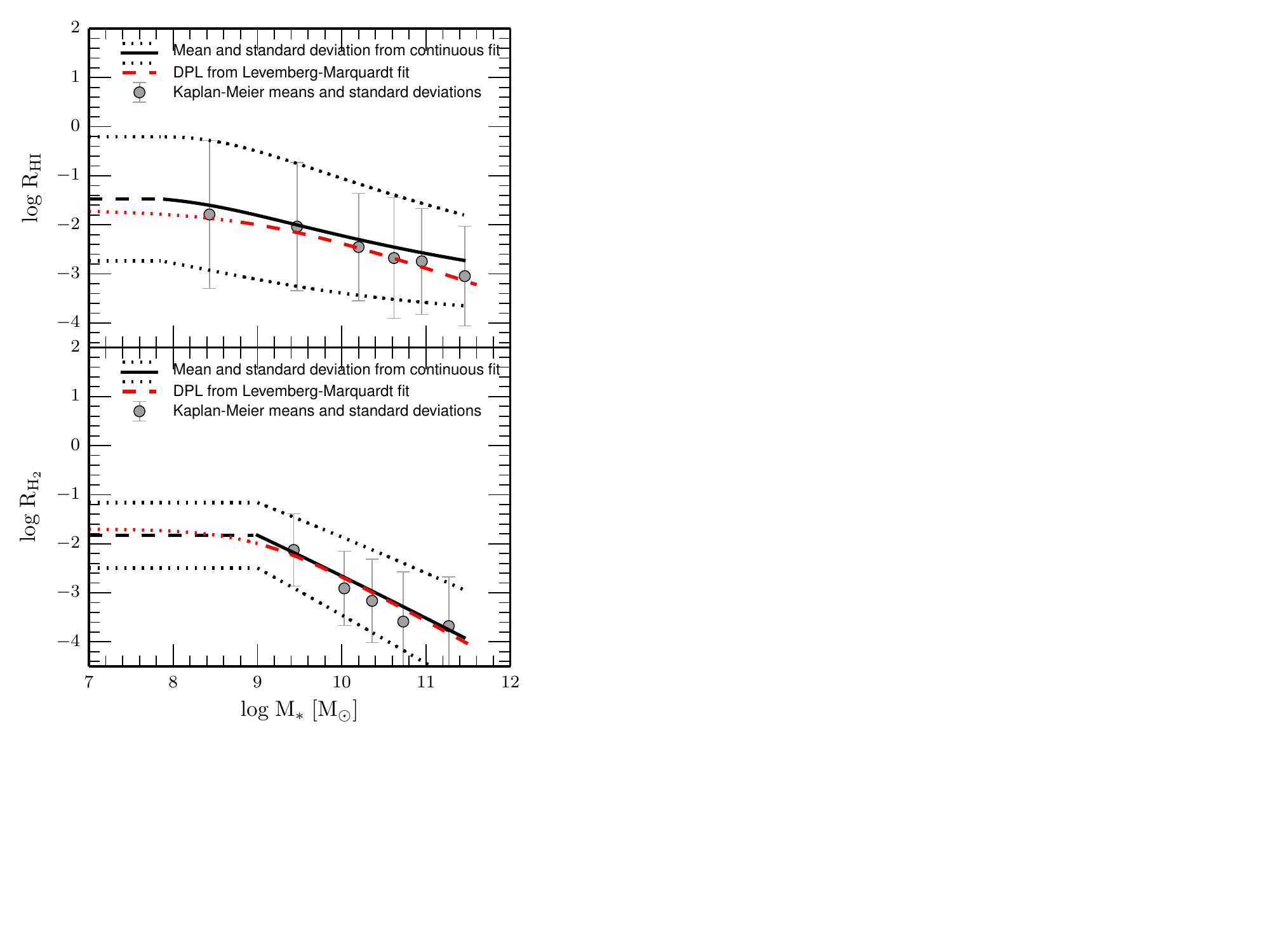}
\caption{Same as in Figure \ref{ltg-fulldistribution}  but for ETGs.} 
\label{etg-fulldistribution} 
\end{figure}

Finally, in Figures \ref{ltg-fulldistribution} and \ref{etg-fulldistribution} we reproduce from Figure \ref{correlations} the means and 
standard deviations obtained with the Kaplan-Meier estimator in different \ms\ bins (gray dots and error bars) for LTG and
ETGs, respectively, and compare them with the means and standard deviations of the general mass-dependent
distributions functions given in Equations  (\ref{full-fit}) and (\ref{full-fit2}) and constrained with the data 
(black solid line and the two dotted lines surrounding it). 
The agreement is rather good in the log-log \RHI--\ms\ and \RH2--\ms\ diagrams both for LTGs and ETGs. Black dashed lines are extrapolations of the mean and standard deviation inferences from the distributions mentioned above, assuming they are the same
as in the last mass bin with available gas observations.
We also plot in these Figures the respective mean double power-law relations determined in subsections \ref{RHI-Ms} and \ref{RH2-Ms} 
(dashed blue or red lines, for LTGs and ETGs respectively; dotted blue or red lines are extrapolations.).  

In conclusion, the \RHI\ and \RH2\ distributions as a function of \ms\ described by 
Equations (\ref{full-fit}) and (\ref{full-fit2}) (with the parameters given in Table \ref{full-distributions}) for LTGs and ETGs, 
respectively, are fully consistent with the corresponding
\RHI--\ms\ and \RH2--\ms\ correlations determined in subsections \ref{RHI-Ms} and \ref{RH2-Ms}. Therefore, 
{\it Equations  (\ref{full-fit}) and (\ref{full-fit2}) provide a consistent description of the \HI- and \H2-to-stellar mass relations and their
scatter distributions, for LTGs and ETGs, respectively.
}  

\section{Consistency of the gas-to-stellar mass correlations with the observed galaxy gas mass functions}
\label{mass-functions} 

The \HI- and \H2-to-stellar mass relations can be used to map the observed GSMF 
into the \HI\ and \H2\  mass functions (\gamf\ and \gmmf, respectively). This way, we can check whether the correlations we have
inferred from observations in subsectiona \ref{RHI-Ms} and \ref{RH2-Ms} are consistent or not with the \gamf\ and \gmmf\ obtained from 
\HI\ and CO (\H2) surveys, respectively.  In order to carry out this check of consistency, we need a GSMF, on one hand,
defined in a large enough volume as to include massive galaxies and to minimize cosmic variance, and on the other hand, 
complete down to very low masses. 
As a first approximation to obtain this GSMF, we follow here a procedure similar as in \citet[][see their Appendix A]{Kravtsov+2014}.
 We use the combination of two GSMFs: \citet{Bernardi+2013} for the large SDSS volume (complete from $\ms\sim 10^9$ \msun), 
and \citet[][]{Baldry+2012} for a local small volume but nearly complete down to $\ms\sim 10^7$ \msun\ (GAMA).
In Appendix \ref{ourGSMF} we describe how we apply some corrections and homogenize both samples to obtain an uniform 
GSMF from $\ms\sim 10^7$ to $\sim 10^{12}$ \msun.  

\begin{figure} [ht!]
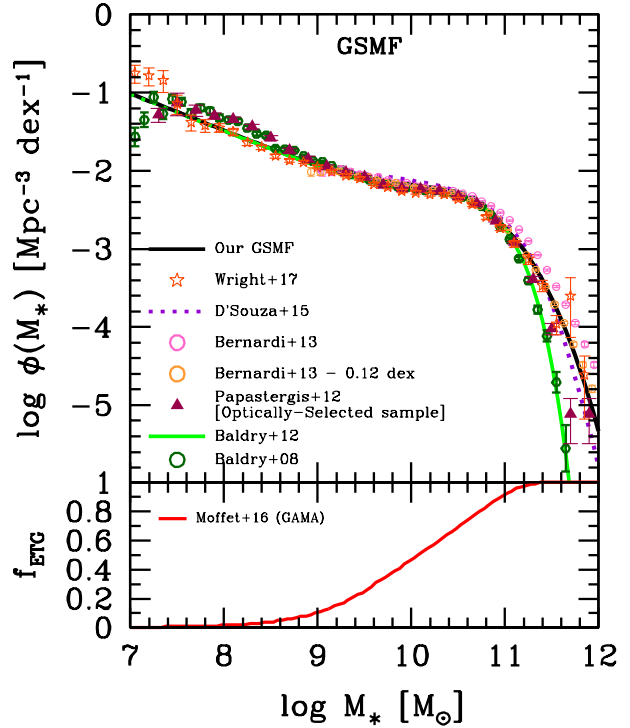

\plotdelgadofig6{figure-6.eps}
\caption{Our GSMF obtained from the combination of three observational GSMFs following \citet{Kravtsov+2014} (thick solid line): one
from the large SDSS DR7 volume but complete only down to $\sim 10^9$ \msun\ \citep[][pink open circles with error bars; the orange open 
circles with error bars are after correcting \ms\ by 0.12 dex, see text]{Bernardi+2013}, and two complete down to lower masses but 
in a very local volume (\citet{Wright+2017}, \citealp{Papastergis+2012} and \citealp{Baldry+2012}). We also plot for comparison, the GSMFs reported in 
\citet[][]{Baldry+2008} and \citet[][]{DSouza+2015}.  The lower panel shows the fraction of ETGs as a function of mass inferred by \citet{Moffett+2016}, using GAMA galaxies and their visual morphological classification.
} 
\label{GSMF} 
\end{figure}

Figure \ref{GSMF} presents our combined GSMF (solid line) and some GSMFs reported in the literature:
the two used by us (see above), and those from \citet{Wright+2017}, \citet[][]{Papastergis+2012}, and \citet{Baldry+2008} 
in small but deep volumes, and \citet{DSouza+2015} in a large volume. 
We plot both the original data from \citet{Bernardi+2013} (pink symbols) and after dismissing \ms\ by 0.12 dex (blue symbols)
to homogenize the stellar masses to the BC03 population synthesis model (see Appendix \ref{ourGSMF}). There is very good
agreement between our combined GSMF and the recent GSMF reported in \citet{Wright+2017} for the GAMA data.

Since the GSMF will be used as an interface for constructing the \HI\ and \H2\ mass functions, it is implicit the assumption 
that each galaxy with a given stellar mass has its respective \HI\ and \H2\ content. Hence, the gas mass functions presented 
below exclude the possibility of galaxies with gas content but not stars, and are equivalent to gas mass functions constructed 
from optically-selected samples \citep[as in e.g.,][]{Baldry+2008,Papastergis+2012}. In any case, it seems that the probability of
finding only-gas galaxies is very low \citep{Haynes+2011}.

We generate a volume complete mock galaxy catalog that samples the empirical GSMF presented above, and that takes 
into account the empirical volume-complete fraction of ETGs, $f_{\rm early}$, as a function of stellar mass (the complement 
is the fraction of LTGs, $f_{\rm late}=1-f_{\rm early}$). The catalog is constructed as follows:

\textbf{1.} A minimum galaxy stellar mass $M_{\rm *,min}$ is set ($=10^{7}$ \msun). 
From this minimum we generate a population of $5\times10^6$ galaxies that samples the
GSMF presented above. 

\begin{figure*}[ht!]
\plotanchomfs{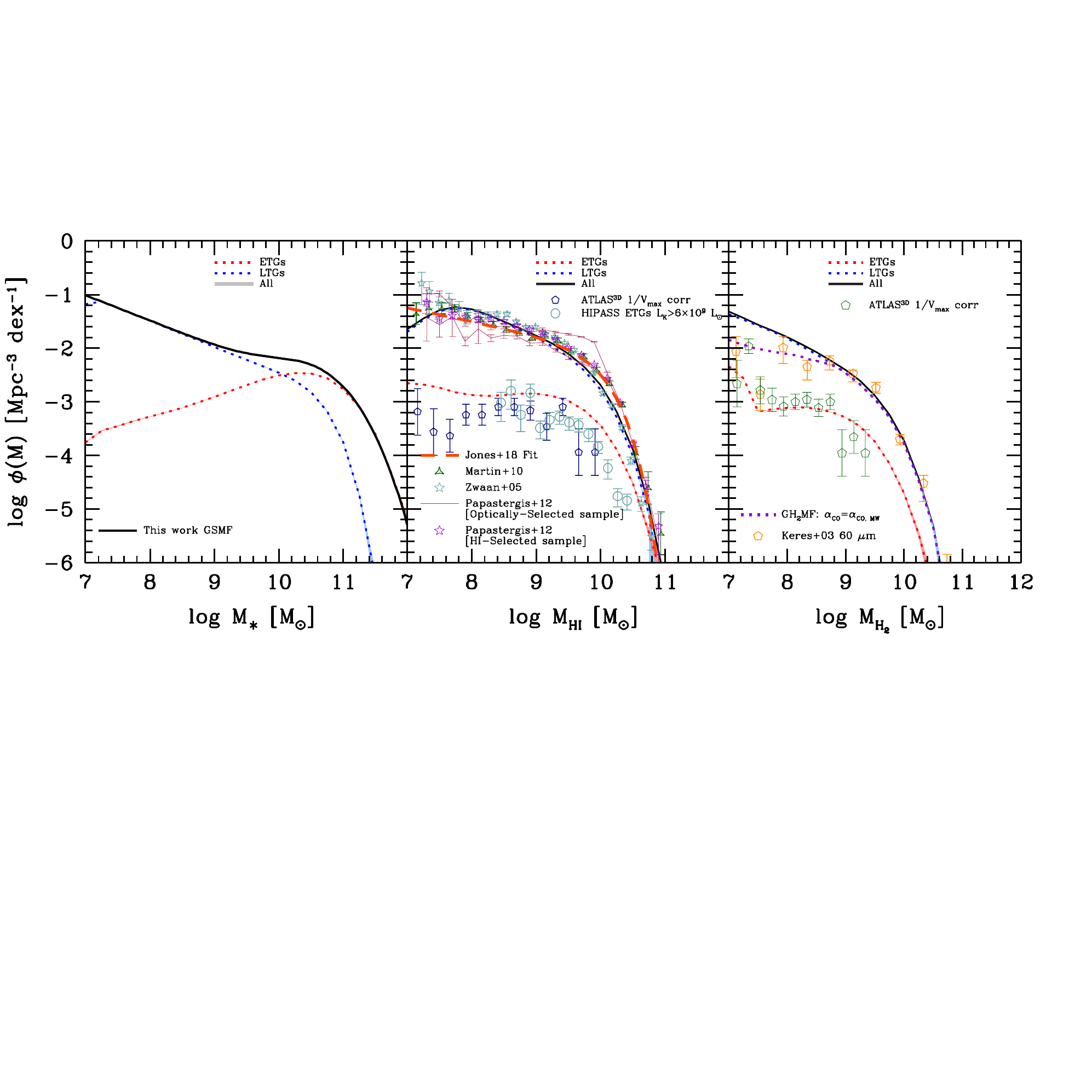}
\caption{ {\it Panel (a):} Total GSMF from the mock catalog that reproduces the empirical GSMF of Fig. \ref{GSMF} (solid line). 
The gray shadow represents the Poisson errors (except for large masses, these errors
are thinner than the line thickness).  The GSMF from the mock catalog samples very well the empirical GSMF used as input.
The blue/red dotted lines and shadows correspond to the LTG/ETG mass function components, using the empirical 
ETG fraction as a function of \ms\ shown in Fig. \ref{GSMF}. {\it Panel (b):}  Same as in panel (a) but for
atomic gas, using the mean \RHI--\ms\ relation and its scatter distribution as given in section \ref{scatter-distribution}. 
Several observational \gamf's from blind \HI\ samples, and the ETG \gamf\ from ATLAS$^{\rm 3D}$ and HIPASS surveys 
are reproduced (see labels inside the panel).
{\it Panel (c):} Same as in panel (a) but for molecular gas. The \gmmf\ calculated from the \citet{Keres+2003} L$_{\rm CO}$ function  
is reproduced.  The dotted purple line is the total \gmmf\ from the mock catalog when using a \RH2--\ms\  correlation obtained from 
our compilation but assuming that \a_co=$\alpha_{\rm CO,MW}$=const., as done in \citet{Keres+2003}.  
 }
\label{MFs} 
\end{figure*}

\textbf{2.} Each mock galaxy is assigned either as LTG or ETG. For this, we use the results reported in \citet{Moffett+2016}, who visually classified galaxies from the GAMA survey. They consider ETGs those classified as Ellipticals and S0-Sa galaxies. 
The $f_{\rm early}$ fraction as a function of \ms\ is calculated as $\Phi_{\rm early}(\ms)/\Phi_{\rm all}(\ms)$, 
with $\Phi_{\rm early}(\ms)=\Phi_{\rm Ell}(\ms)+\Phi_{\rm S0-Sa}(\ms)$, using the fits to the respective GSMFs reported in \citet{Moffett+2016}.\footnote{Note that  Sa galaxies are not included in our definition of ETGs, so that $f_{\rm early}$ is probably overestimated at masses
where Sa galaxies are abundant, making that $f_{\rm early}=0.5$ at masses lower than the break mass, $M^*$ (see figure \ref{MFs}).}

\textbf{3.} For each galaxy, \RHI\ is assigned randomly from the conditional probability distribution \Pjhi\
that a galaxy of mass \ms\ and type $j=$LTG or ETG lies in the $\RHI\pm d\RHI/2$
bin. Then, \mha=\RHI$\times$\ms. The probability distributions for LTGs and ETGs are given by 
the mass-dependent PDFs presented in Equations (\ref{full-fit}) and (\ref{full-fit2}), respectively (their
parameters are given in Table \ref{full-distributions}).

\textbf{4.} The same procedure as in the previous item is applied to assign \mhm=\RH2$\times$\ms, by using for the
\Pjhii\ probability distributions the corresponding mass-dependent PDFs for LTGs and ETGs presented in Equations 
(\ref{full-fit}) and (\ref{full-fit2}), respectively (their parameters are given in Table \ref{full-distributions}). 

Our mock galaxy catalog is a volume-complete sample of $5\times10^6$ galaxies above $\ms=10^7$ \msun, 
corresponding to a co-moving volume of $5.08\times 10^7$ Mpc$^{3}$. 
Since the \HI\ and  \H2\ mass functions are constructed from the GSMF, its mass limit
$M_{\rm *,min}$ will propagate in different ways to these mass functions. The co-moving volume 
in our mock galaxy catalog is big enough as to avoid significant effects from Poisson noise. This noise 
affects specially the counts of massive galaxies, which are the less abundant objects. 

\subsection{The mock galaxy mass functions}

\subsubsection{Stellar mass function} 

The mock GSMF is plotted in panel (a) of Fig. \ref{MFs} along with the Poisson errors given by the thickness of the gray line; 
except for the highest masses, the Poisson errors are actually thinner than the line. 
The mock GSMF is an excellent realization of the empirical GSMF (compare it with Fig. \ref{GSMF}). We also plot the
corresponding contributions to the mock GSMF from the LTG and ETG populations (blue and red dashed lines). As expected,
LTGs dominate at low stellar masses and ETGs dominate at high stellar masses. The contribution of both populations is equal 
($f_{\rm early} = f_{\rm late}=0.5$) at $\ms^{\rm cross}=10^{10.20}$ \msun\  (recall that the fraction $f_{\rm early}$ used here comes from
\citet{Moffett+2016}, who included Sa galaxies as ETGs; if consider Sa galaxies as LTGs, then $\ms^{\rm cross}$ would likely be higher).
In order to predict accurate gas and baryonic mass functions, the present analysis will be further refined in Rodriguez-Puebla et al. (in prep.), 
where several sources of systematic uncertainty in the GSMF measurement and in the definition of the LTG/ETG fractions will be taken into account.
Our aim here is only to test whether the empirical correlations derived in Section \ref{gass_mass_relations} are roughly consistent or not with the
total \HI\ and \H2\ empirical mass functions.

\subsubsection{\HI\ mass function} 

In panel (b) of  Fig. \ref{MFs}, we plot the predicted \gamf\ from our mock galaxy catalog using 
the mean (LTG+ETG) \RHI--\ms\ relations and their scatter distributions as given in section \ref{scatter-distribution} 
(black line, the gray shadow shows the Poisson errors). For comparison, we plot also the \HI\ mass functions 
estimated from the blind \HI\ surveys ALFALFA (\citealp{Martin+2010}; \citealp {Papastergis+2012}, 
for both their \HI- and optically-selected samples; and
the latest results from \citealp{Jones+2018}) and HIPASS \citep[][]{Zwaan+2005}.  
At masses larger than $\mha\sim 3\times 10^{10}$ \msun, our \gamf\ is in vey good agreement with those from 
the ALFALFA survey but significantly above than the HIPASS one. \citet{Martin+2010} 
argue that the larger volume of ALFALFA survey compared to the HIPASS one, makes ALFALFA more likely 
to sample the mass function at the highest masses, where objects are very rare. The volume of our mock 
catalog is even larger than the ALFALFA one. 
At intermediate masses, $9\lesssim \log$(\mha/\msun)$\lesssim 10.5$,
our \gamf\ is in reasonable agreement with the observed mass functions but it has in general
a slightly less curved shape than these functions. 
At low masses, $\log$(\mha/\msun)$\lesssim 8$, the observed \gamf's flatten more than our predicted mass function. It could
be that the blind surveys  start to be incomplete due to sensitivity limits in the radio observations.
Note that \citet{Papastergis+2012} imposed additional optical requirements to their \HI\ blind sample (see their Section 2.1), 
which make flatter the low-mass slope. Regarding the optically-selected sample of 
\citet{Papastergis+2012}, since it is constructed from a GSMF that starts to be incomplete below $\log$(\ms/\msun)$\sim 8$
(see Fig. \ref{GSMF}), one expects incompleteness in the \gamf\ starting at a larger mass in \HI.
Since our \gamf\ is mapped from a volume-complete GSMF from $M_{\rm *,min}\approx 10^7$ \msun,  
``incompleteness'' in \mha\ is expected to start from the \HI\ masses corresponding to $M_{\rm *,min}\times P(\RHI |M_{\rm *,min})$,
where the latter is the scatter around the \RHI--\ms\ relation. This shows that our \gamf\  
can be considered complete from $\log$(\mha/\msun)$\approx 8$.  The slope of the \gamf\ around this mass is $-1.52$,
steeper than the slope at the low-mass end of the corresponding GSMF ($\alpha=-1.47$). 

In Fig. \ref{MFs} are also plotted the LTG and ETG components of the \gamf\ as obtained from our mock catalog.  The \gamf\
is totally dominated by the contribution of LTGs. Our ETG \gamf\ is  compared with the ones obtained from observations by using the
ATLAS$^{\rm 3D}$ and HIPASS surveys as reported  in \citet[][]{Lagos+2014}. 

\subsubsection{\H2\ mass function} 

In panel (c) of  Fig. \ref{MFs}, we plot the predicted \gmmf\ from our mock galaxy catalog using 
 the mean (LTG+ETG) \RH2--\ms\ relations and their scatter distributions as given in section \ref{scatter-distribution} 
(black line, the gray shadow shows the Poisson errors). 
We compute the \H2\ mass function from the CO luminosity function derived by \citet{Keres+2003}, who
used the small and incomplete FCRAO CO survey \citep{Young+1995} and combined it with the volume-complete 
FIR survey. We adopt the MW \H2-to-$\rm CO$ conversion factor and correct their $h$ parameter to 0.7. 
Unfortunately, this derivation is highly uncertain since is based on a empirical correlation between the
60$\mu$m and CO luminosities, and the selection effects in both used surveys introduce several biases. 
The obtained \gmmf\ is plotted in Fig. \ref{MFs}. 
Our \gmmf\ decreases faster than the one by \citet{Keres+2003} at high masses, roughly agrees with 
it at intermediate masses, and for masses below $\log$(\mhm/\msun)$\sim 8.5$, our mass function is steeper.
The reason for this latter difference seems to be the mass-dependent \cotoh2~conversion factor
introduced by us (see Appendix \ref{conv-factor}). This factor increases as \ms\ is smaller while in the case of \citet{Keres+2003} 
it is constant. We recalculate the \gmmf\ by using in the conversion from $L_{\rm CO}$ to \mhm\ a constant  \cotoh2\ 
factor equal to the MW value, and plot it with the purple dotted line; the mass function at the low-mass side is now in
good agreement with that of \citet{Keres+2003}. 

In Fig. \ref{MFs} are also plotted the LTG and ETG components of the \gmmf\ as obtained from our mock catalog.  The \gmmf\
is totally dominated by the contribution of LTGs. Our ETG \gmmf\ is compared with the one 
obtained from observations by using the ATLAS$^{\rm 3D}$ survey as reported  in \citet[][]{Lagos+2014}.

\section{Discussion}
\label{discussion}

\subsection{The \H2-to-\HI\ mass ratio} 
\label{H2-HIratio}
\begin{figure*} [ht!]
\begin{center}
\plotsixsep{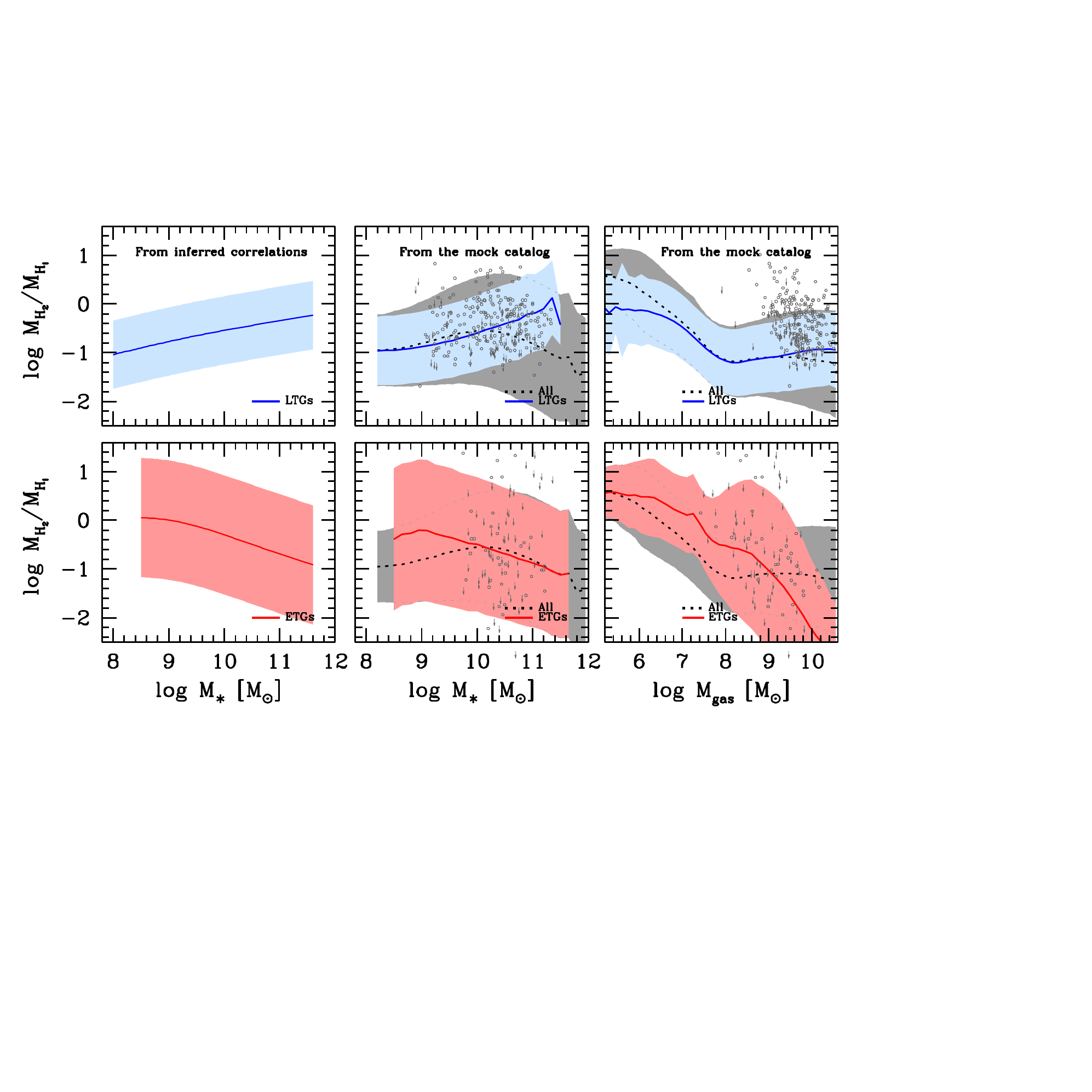}
\caption{{\it Left panels:} Molecular-to-atomic mass ratio, \mhm/\mha, for LTGs (upper panel) and ETGs (lower panel) inferred from our double power-law
fits to the \RHI-\ms\ and \RH2-\ms\ correlations. The shaded areas are the 1$\sigma$ scatter obtained by error propagation of the scatter 
around the \RHI-\ms\ relations. 
{\it Middle panels:} Same as in left panels but from our mock catalog generated to sample the empirical GSMF, volume-complete ETG/LTG fractions as
a function of mass, and \RHI--\ms\ and \RH2--\ms\ correlations. The dotted line surrounded by the gray area are the total \mhm/\mha\ ratio and 1$\sigma$ 
scatter as a function of stellar mass. {\it Right panels:}  Molecular-to-atomic mass ratio as a function of the cold gas mass, \mg, from the mock
catalog for LTGs (upper panel) and ETGs (lower panel). We plot available detected and undetected cold gas  observational data as gray unfilled circles and downward arrows respectively.}
\label{HIH2ratio} 
\end{center}
\end{figure*}

The global \H2-to-\HI\ mass ratio of a galaxy characterizes its global efficiency of converting atomic into molecular hydrogen.
This efficiency is tightly related to the efficiency of large-scale SF in the galaxy \citep[see e.g.,][]{Leroy+2008}.
From the empirical correlations inferred in Section \ref{gass_mass_relations}, we can calculate \mhm/\mha\ as a function of \ms\ for both 
the LTG and ETG populations. We do this by using our double power-law fits to the data.  Left panel of Fig. \ref{HIH2ratio} presents the 
obtained \mhm/\mha--\ms\ relations and their $1\sigma$ scatter calculated by propagating the dispersions in the assumption
of null covariance. In this sense, the plotted scatter are upper limits, since there is evidence of some (weak) correlation between
the \HI\ and \H2\ content of galaxies, in particular among those deficient in \HI\ and \H2\ \citep{Boselli+2014b}.
We can plot the same correlations from the mock catalog presented in Section \ref{mass-functions}, which samples the observed GSMF,
 the LTG and ETG fractions as a function of \ms, and the empirical correlations inferred by us. The middle panels of Fig. \ref{HIH2ratio}
 present what we measure from the mock catalog for LTG (blue), ETG (red), and all galaxies (gray). The lines are the logarithmic
 means in small mass bins and the shaded regions are the corresponding standard deviations. At low masses, LTGs dominate, so 
 the correlation of all galaxies is practically the one of LTGs. At high masses, ETGs become more important. 

According to Fig. \ref{HIH2ratio}, the molecular-to-atomic mass ratio of LTGs increases with \ms, albeit with a large
scatter.  On average, \mhm/\mha\ increases from $\approx 0.1$ to $\approx 0.8$ for masses ranking from 
$\ms= 10^8$ \msun\ to $3\times 10^{11}$ \msun. Given that the surface density of LTGs correlates significantly with \ms, 
one can expect this dependence of \mhm/\mha\ on \ms\ at least from two arguments: 
1) Disk instabilities, which drive the formation of molecular clouds (e.g., the Toomre criterion 
\citealp{Toomre}),  are more probable to occur as the disk surface density is higher. 2) The \H2-to-\HI\ mass ratio in galaxies 
has been shown to be directly related to the hydrostatic gas pressure \citep{Blitz+2006,Krumholz+2009}, and this
pressure depends on the (gas and stellar) surface density \citep{Elmegreen1989}. 
In fact, the physics of \H2\ condensation from \HI\ is very complex and it is expected to be driven by
local parameters of the ISM \citep[see e.g.,][]{Blitz+2006,Krumholz+2009,Obreschkow+2009}. Therefore, 
the dependence of  the \H2-to-\HI\ mass ratio on \ms\ should be understood as consequence of the correlations of
these parameters (their mean values along the galaxy) with \ms, introducing this actually a large scatter in the 
dependence of \mhm/\mha\ on \ms. 
Indeed, several authors have shown that \mhm/\mha\ correlates better with the mean gas-phase metallicity or mean stellar 
surface density than with \ms\ \citep[e.g.,][]{Saintonge+2011,Boselli+2014}.

For ETGs, the trend of the \H2-to-\HI\ mass ratio is inverse to the one of LTGs and with a very large scatter. The
ETGs more massive than $\sim 10^{11}$ \msun\ have mean ratios around $0.15$ and a 1-$\sigma$ scatter of $\sim \pm 1$ dex; 
for intermediate masses, this ratio increases on average, and for ETGs with masses $\ms\sim 10^9$ \msun, which are actually 
very rare, their mean \H2-to-\HI\ mass ratios are $\sim 1$ with the same scatter of $\sim \pm 1$ dex. Even though the gas fraction
in ETGs is much smaller than in LTGs at all masses (see Fig. \ref{correlations}), the former are also typically more compact than
the latter, resulting probably on average in similar or higher gas pressures, and consequently a similar or even higher \mhm/\mha\ ratios, 
specially at masses lower than $\ms\approx 10^{10}$ \msun. In fact, given the large scatter in \mhm/\mha\ for ETGs, this ratio 
depends likely on many other internal and external (mergers, environment, etc.) factors that do not correlate significantly with \ms. 

Regarding \mhm/\mha\ vs. \mg, for LTGs, which for $\ms> 10^7$ \msun\ have mostly gas masses $>10^8$ \msun,
there is not any significant dependence, while for ETGs, which are almost inexistent with $\mg\gtrsim 10^9$ \msun, \mhm/\mha\ is larger on average
for lower values of \mg. This can be seen in the right panel of Fig. \ref{HIH2ratio}, where the mock catalog has been used. 
Basically, for a given \mg, in the mass range $\mg\sim 10^7-10^9$ \msun), ETGs have typically larger \H2-to-\HI\ mass ratios than LTGs.
In combination, the \H2-to-\HI\ ratio
appears to be larger for lower values of \mg. Such a dependence has been reported by \citet{Obreschkow+2009} for their 
compiled sample of galaxies, and predicted by these authors from a physical model. 

The dependences of the \H2-to-\HI\ mass ratio on \ms, \mg, and morphological type discussed above are in qualitative
agreement with several previous observational works, which actually are part of our compilation 
\citep{Leroy+2008,Obreschkow+2009,Saintonge+2011,Boselli+2014,Bothwell+2014}. However, our results extend to a larger
mass range and separate explicitly the two main populations of galaxies.

\subsection{The role of environment} 
\label{environment}

There are several pieces of evidence that the atomic gas fraction of galaxies is lower in higher-density environments 
\citep[e.g.,][]{Haynes+1984, Gavazzi+2005,Cortese+2011,Catinella+2013,Boselli+2014b}. 
The fact that the ETG population has lower \HI\ gas fractions than the LTG one 
(Section \ref{gass_mass_relations}), being the former commonly found in higher-density environments, agrees with the 
mentioned trends with environment.  Thus, due to the morphology-density relation, our determinations 
of the \RHI--\ms\ (as well as \RH2--\ms\ and \Rgas--\ms) correlations for the LTG and ETG populations, account 
partially for the dependence of these correlations on environment. Moreover, for the very
isolated LTGs and for the subsample of LTGs
in the Virgo cluster central regions, we confirm higher and lower \HI-to-stellar
mass ratios than the average of the overall LTG sample, respectively (see subsection \ref{HI-preliminary}). However, this
systematical difference with the environment is within the $1\sigma$ scatter of the \RHI--\ms\
correlation of LTGs (see Fig. \ref{checksRHI}). Instead, in the case of ETGs, the isolated galaxies have much larger \RHI\ values than
the means of all ETGs, above the $1\sigma$ scatter;  isolated ETGs are almost as \HI\ gas rich as the mean of LTGs. 

For molecular gas fraction, the observational results are controversial in the literature. Recent studies seem to incline the 
controversy to the fact that galaxies in clusters are actually \H2--deficient with respect to similar galaxies in the field, however,
the deficiencies are smaller than in the case of \HI\ \citep[][and more references therein]{Boselli+2014b}.
Here, for isolated and Virgo-center LTGs, we do not see any systematical segregation of \RH2\ from the rest of our compiled 
LTGs (Fig. \ref{checksRH2}), but in the case of ETGs, the isolated galaxies have on average larger values of \RH2. 

In summary, the results from our compilation point out that the \HI\ content of LTGs has a (weak) dependence on environment,
mainly due to the fact that at high densities LTGs are \HI\ deficient. Instead, the \H2\ content of LTGs seems not to change on average
with the environment.  In the case of ETGs, those very isolated are significantly more  gas rich (both in \HI\ and \H2) than the 
average among ETGs at a given mass. 

An important aspect related to the environment is whether a galaxy is central or satellite. The local environmental effects once a galaxy 
becomes a satellite inside a halo (ram pressure and viscous stripping, starvation, harassment, tidal 
interactions) work in the direction of lowering the gas content of the galaxy, 
likely more as more massive is the halo  \citep[][]{Boselli+2006,Brown+2017}.
Part of the scatter in the gas-to-stellar mass correlations are probably due to the external processes produced by 
these local-environment mechanisms. A result in this direction has been recently shown for the \RHI--\ms\ correlation by \citet{Brown+2017}.
These authors have found that the \HI\ content of satellite galaxies in more massive halos have, on average, lower \HI-to-stellar mass 
ratios at fixed stellar mass and specific SFR. According to their analysis, the systematic environmental suppression of 
\HI\ content at both fixed stellar mass and fixed specific SFR in satellite galaxies begins in halo masses typical of the group 
regime ($>10^{13}$ \msun), and fast-acting mechanisms such as ram-pressure stripping are suggested to explain their results.
In a future work, we will attempt to characterize the central/satellite nature of our compiled galaxies, as well as to calculate a proxy to their 
halo masses, in order to study this question.

 \subsection{Comparisons with previous works} 
 \label{comparison}
 
\begin{figure}[ht!]
\plotfigten{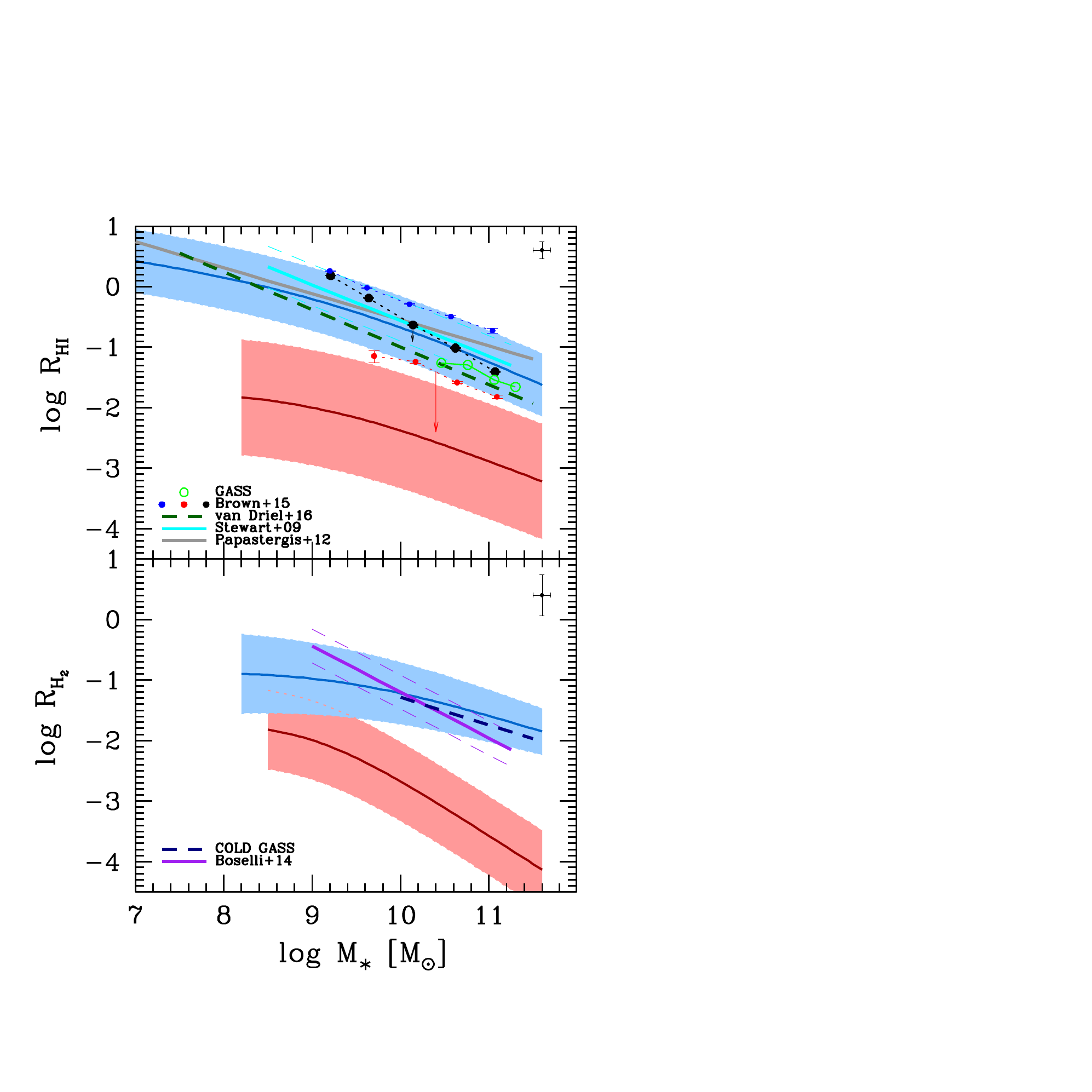}
\caption{{\it Upper panel:} Our empirical \HI-to-stellar mass correlations for LTGs and ETGs (blue and red shaded areas, respectively)
compared with some previous determinations (see labels inside the panel and details of each determination in the text). These
previous determinations are for compilations typically biased to late-type, blue galaxies, and/or do not take into account non detections. 
The blue and red arrows correspond  to estimates of the difference between the logarithm of the mean (the stacking technique provides
the equivalent of the mean value) and the logarithmic mean (our determinations are for this case) for standard deviations
of 0.52 and 0.99 dex, respectively (see text for more details). 
{\it Lower panel:} Our empirical molecular \H2-to-stellar mass correlations for LTGs and ETGs (blue and red shaded areas, respectively)
compared with very rough previous determinations not separated into LTGs and ETGs (see labels inside the panel and details of each 
determination in the text). }
\label{comparisons} 
\end{figure}

In Fig. \ref{comparisons} we compare our results with those of previous works. When necessary, the data are corrected to 
a \citet{Chabrier2003} IMF.  Most of the previous determinations of the \HI- and \H2-to-stellar mass correlations are not explicitly separated
into the two main galaxy populations as done here, and in several cases non detections are assumed to have the values 
of the upper limits or are not taken into account at all. 

In the upper panel, our empirical \RHI--\ms\ correlations for LTGs and ETGs are plotted along with
the linear relations given by \citet{Stewart+2009} (cyan line, the dashed lines show the $1\sigma$ scatter) and 
\citet{Papastergis+2012} (gray line). The former authors used mainly the observational data presented in \citet{McGaugh2005} for  
disk-dominated galaxies, and the latter authors used samples from \citet{Swaters&Balcells2002}, \citet{Garnett2002}, 
\citet{Noordermeer+2005}, and \citet{Zhang+2009}, which refer mostly to late-type galaxies.
Their fits are slightly above the mean of our LTG \RHI--\ms\ correlation. This is likely because they ignore non-detections. 
We also plot the logarithmic average values in mass bins reported by \citet{Catinella+2013} for GASS (green open circles). 
Since ETGs progressively dominate in number as the mass increase, our total (density-weighted) \RHI--\ms\ correlation would fall 
below the one by \citet{Catinella+2013}, specially at the highest masses. Note that for the data plotted from \citet{Catinella+2013},
the \HI\ masses of non-detections were set  equal to their upper limits. Therefore, the plotted averages are biased 
to high values of \RHI, specially for ETGs which are dominated by non-detections.  On the other hand, recall that we
have corrected by distance the upper limits of GASS to make them compatible with those of the closer ATLAS$^{3D}$ survey. 

More recently, \citet{Brown+2015} have used the \HI\ spectral stacking technique for a volume-limited, stellar mass
selected sample from the intersection of SDSS DR7, ALFALFA, and $GALEX$ surveys. With this technique 
the stacked signal of co-added raw spectra of detected and non-detected galaxies (about 80\% of the ALFALFA selected sample)
is converted into a (lineal) average \HI\ mass. The authors have excluded from their analysis \HI-deficient galaxies --typically found within clusters-- 
because of their significant offset to lower gas content. The black dots connected  by a dotted line show the logarithm of the average \RHI\ 
values reported at different stellar mass bins in \citet{Brown+2015}.  Since \HI-deficient galaxies --which typically are ETGs-- 
were excluded, then the \citet{Brown+2015} correlation should be compared with our correlation for LTGs. Note that with the
stacking technique is not possible to obtain the population scatter in \RHI\ because the reported mean values come from stacked 
spectra instead from averaging individual values of detections and non detections.
 However, the stacking can be applied to subsets of galaxies, for example, selected by color. \citet{Brown+2015} have divided
their sample into three groups by their NUV$-r$ colors: [1,3), [3,5), and [5,8]. The average \RHI\ values at different masses
corresponding to the bluest and reddest  groups are reproduced in Fig. \ref{comparisons} with the blue and red symbols, respectively. 
Note that the logarithmic mean is lower than the logarithm of the mean. For a lognormal distribution, 
$\langle \log x \rangle= \log\langle x \rangle -0.5\times \sigma^2_{\log x} \ln 10$ \citep[see e.g.,][]{Rodriguez-Puebla+2017}. Then, for the typical scatter of 0.44 dex corresponding
to LTGs, the logarithm of the stacked values of \RHI\ should be lowered by $\approx 0.2$ dex to compare formally with our reported values 
of logarithmic means; this is shown with a black arrow in Fig. \ref{comparisons}.  If the reddest galaxies in the \citet{Brown+2015} stacked 
sample are associated with ETGs (which is true only partially), then for them the correction to a logarithmic mean is of 
$\approx 1$ dex, shown with a red arrow.

Finally, recently \citet{vanDriel+2016} reported the results from
\HI\ observations at the Nancay Radio Telescope (NRT) of 2839 galaxies selected evenly from SDSS. The authors
present a Buckley-James linear regression to their data (long-dashed green line in Fig. \ref{comparisons}), 
taking into account this way upper limits for non-detections (though their upper limits are quite high given the 
low sensitivity of NRT). Their fit is for all the sample, that is, they do not separate 
into LTG/ETG or blue/red groups. In a subsequent paper \citep{Butcher+2016}, the authors obtained $\sim 4$ times 
more sensitive follow-up \HI\ observations at Arecibo for a fraction of the galaxies that were either not detected or marginally 
detected; 80\% of them were detected with \HI\ masses $\sim 0.5$ dex lower than the upper limits in \citet{vanDriel+2016}, 
and the rest, mostly luminous red galaxies, were not detected. If this trend is representative of the rest of the NRT undetected 
galaxies, \citet{Butcher+2016} expect the fit plotted in Fig. \ref{comparisons} to be offset toward lower \RHI\ values by 
about 0.17 dex and even more at the highest masses. This fit is in between a density-weighted fit to our two correlations when
taking into account that at high masses the fraction of ETG/red galaxies increases and at low masses LTG/blue galaxies dominate at all. 

The lower panel of Fig. \ref{comparisons} is similar to the upper panel but for the \RH2--\ms\ correlations.
In the case of the molecular gas content, in the literature there are only a few attempts to determine the relation between \mhm\ and \ms. 
In fact, those works that report approximate correlations are included in our compilation: \citet{Saintonge+2011} for COLD GASS,
and \citet{Boselli+2014} for HRS.  The former authors report a linear regression to their binned data assuming \H2\ masses
for non-detection set equal to their upper limits. The latter authors present a bisector fit using only detected,
late-type gas-rich galaxies. Therefore, in both cases the reported relations are clearly biased to LTGs and to the side 
of high \RH2\ values.

The differences we find between our correlations and those plotted in Fig.  \ref{comparisons}, as discussed above, can be understood 
on the basis of the different limitations that present each one of the previous works.  
Having in mind these limitations in each concrete case, we can conclude that the correlations presented here are 
in rough agreement with previous ones but with respect to them (i) extend the correlations to a larger mass range, (ii) separate 
explicitly galaxies into their two main populations, and (iii) take into account adequately the non detections.

\section{Summary and Conclusions}
\label{conclusions}

The fraction of stars and atomic and molecular gas in local galaxies is the result of complex 
astrophysical processes across their evolution. Thus, the observational determination 
of how these fractions vary as a function of mass provides key information
on galaxy evolution at different scales. Before the new generation of radio telescopes, 
which will bring extragalactic gas studies more in line with optical surveys, the
main way to get this kind of information is from studies based on radio follow-up observations 
of (small) optically-selected galaxy samples. In this work, we have compiled and homogenized 
from the literature samples with information on \ms\ and \mha\ and/or \mhm\ for galaxies that can
be identified belonging to two main operational (in a statistical sense) groups: the LTG and
ETG populations. For estimating \mhm\ from CO observations, we
have introduced a mass-dependent \cotoh2\ conversion factor in agreement with studies that 
show that this factor is not constant and depends on metallicity (hence, statistically on mass). 
Results using a constant \cotoh2\ factor were also presented. 
Figures \ref{RHI} and \ref{RH2} summarize our compilation in the \RHI\ vs. \ms\ and \RH2\ vs. \ms\ logarithmic diagrams.

Previous to infer the correlations, we have tested how much each one of the compiled samples deviate from the 
rest and classified them into three categories: (1) samples complete in limited volumes (or selected from them) 
without selection effects that could affect the calibration of the correlations (Golden), (2) samples that are not complete 
but are representative of the average galaxy population, without obvious selection effects (Silver), and (3) samples 
selected by environment (Bronze).  We showed that most of the samples, after
our homogenization, are suitable to infer the \RHI--\ms\ and \RH2--\ms\ correlations, except those from the
Bronze category in the case of ETGs. These galaxies in extreme environments show significant deviations from the mean trends, and 
then are not taken into account in our determinations. From the combination of all the chosen samples, we have calculated the 
mean, standard deviation, and percentiles of the logarithms of the $\RHI$ and $\RH2$ mass ratios in several 
stellar mass bins, taking into account non-detected galaxies and their reported upper limits, which are a non-negligible
fraction of the data, specially for the ETG population. The accounting of non-detected galaxies and their homogenization among different
samples are relevant for determining the gas-to-stellar mass correlations of ETGs.

The mean logarithmic values in mass bins, $\langle \log\RHI \rangle$ and $\langle \log\RH2 \rangle$, with
the corresponding (intrinsic) standard deviation calculated by means of the Kaplan-Meier estimator 
were fitted to the logarithm of single and double power-law functions (Eq. \ref{function}). 
The parameters of the best fits to these functions, both for LTGs and ETGs, are reported in Tables \ref{linear-parameters} and
\ref{parameters}, respectively. We highlight the following results from our analysis:

$\bullet$ The \RHI--\ms\ and \RH2--\ms\ correlations for the LTG and ETG populations, 
can be described roughly equally well by a single or double power law at masses larger 
than $\log$(\ms/\msun)$\gtrsim 9$. For smaller masses, we see some
hints of a flattening in these correlations.  LTGs have significantly higher 
\HI\ and \H2\ gas fractions than ETGs, the differences  increasing at the high- and low-stellar mass ends.  
For the ETG population, the scatter of the \RHI--\ms\ and \RH2--\ms\ correlations are much larger than for the LTG one.

$\bullet$ Combining the \RHI--\ms\ and \RH2--\ms\ correlations and propagating errors, we 
calculated the cold gas (\mg=1.4(\mha + \mhm))-to-stellar mass correlations of the LTG and ETG populations.
For the former, \Rgas\ is around 4 on average at $\ms=10^7$ \msun\ and  
$\approx 1$ at $\ms=1.60\times 10^9$ \msun. At larger masses, \Rgas\ continues decreasing sifnificantly. 
For the ETG population, \Rgas\ on average is smaller than 1 even for 
the smallest galaxies.  Galaxies as massive as $\ms=10^{11}$ \msun\ have on average 
\Rgas\ ratios smaller than $2.5\times 10^{-3}$. The intrinsic standard deviation of the \Rgas--\ms\ correlation
of the LTG population is $\approx 0.44$ dex while for the ETG one is larger, around 0.68 dex. 

$\bullet$ The \H2-to-\HI\ mass ratio implied by our correlations is such that for LTGs, increases on average
with \ms, from $\approx 0.1$ to $0.8$ for masses ranking from $\ms= 10^8$ \msun\ to $3\times 10^{11}$ \msun. 
For ETGs, the trend is the opposite but with large scatter (standard deviation of $\sim \pm1$ dex). While ETGs have much less
gas content than LTGs, the \H2-to-\HI\ mass ratio at intermediate and low masses is higher on average in the former than in the later,
and lower at large masses. 

$\bullet$ In an attempt to describe the full distributions of \RHI\ and \RH2\ as a function of \ms\ for both the LTG and ETG populations,
the respective PDFs from the censored+uncensored data in different mass bins provided by the Kaplan-Meier estimator were used. 
For LTGs, we have found that a Schechter function with their parameters depending on \ms\ offers a good description of the
\RHI\ and \RH2\ distributions as a function of \ms\ (Eq. \ref{full-fit}).
For ETGs, these distributions look bimodal, with a (broken) Schechter function and a uniform distribution at the low-end side providing an approximate description of them (Eq. \ref{full-fit2}). These mass-dependent PDFs offer a full description of the \RHI--\ms\ and \RH2--\ms\ 
relations and their scatter distributions for both LTGs and ETGs. Their first and second moments agree very well with our previously determined 
double power-law correlations (Figures \ref{ltg-fulldistribution} and \ref{etg-fulldistribution}). 

$\bullet$ The mass-dependent distribution functions of \RHI\ and \RH2\ were used to map the GSMF into the corresponding 
\HI\ and \H2\ mass functions, both for LTGs and ETGs.  
We use an empirical GSMF from the combination of GSMFs from a low-$z$ survey and from the overall DR7 sample, following
\citet{Kravtsov+2014}. The fractions of LTGs/ETGs as a function of \ms\ are calculated from the fitted mass functions of ETGs obtained by \citet{Moffett+2016} using the GAMA survey. The predicted total \HI\ and \H2\ mass functions agree with those obtained 
from empirical determinations in the mass ranges where these determinations are reliable.

Our (marginal) finding of a flattening in the \HI- and \H2-to-stellar mass correlations at low masses has 
been suggested in some previous works (see Section \ref{gass_mass_relations} for references). 
For our double power-law fits (Eq. \ref{function}), we find that the transition mass $M_*^{tr}$ is around $1-2\times 10^9$ \msun\  
for both the \RHI--\ms\ and \RH2--\ms\ correlations and for both the LTG and ETG populations. Interestingly enough, this is the mass 
that roughly separates normal and dwarf galaxies. 

We are aware that our determination of the gas-to-stellar mass relations come from an heterogeneous mix of samples.
However, we have shown that there are not significant differences in the \RHI\ and \RH2\ values as a function of \ms\
from volume-limited complete and incomplete samples. Significant differences are observed only for samples
selected by environment in the case of ETGs.
On the other hand, our correlations for ETGs (and LTGs in the case of molecular gas), are very limited
at low masses. They are actually just extrapolations for stellar masses below several $10^8$ \msun, but 
we have checked them to be consistent with the very few available determinations (mostly non detections)
below these masses. 

In spite of the mentioned shortcomings, it is encouraging that the correlations (in fact, the full mass-dependent
distributions), when mapped to the 
\HI\ and \H2\ mass functions using the observed GSMF as an interface, are consistent with the 
mass functions determined from observational radio surveys, at least in the mass ranges where these 
surveys do not suffer of strong selection, volume, and cosmic variance effects. Such a self-consistency 
between the gas-to-stellar correlations and mass functions supports the reliability of our results,
which help to pave the way for the next generation of radio telescopes.

The empirical gas-to-stellar mass correlations and their approximate scatter distributions presented in this paper 
for the two main populations of galaxies, are useful for understanding global aspects of galaxy evolution as a function of mass.
We encourage to use these correlations (or the full mass-dependent PDFs) for comparisons with 
predictions of models and simulations of galaxy formation and evolution. 

Finally, we provide upon request to A. R. Calette a Python-based code that allows to generate plots and electronic tables for both LTGs and ETGs of 1) the \RHI-\ms\ and \RH2-\ms\ double power-law relations and their $1\sigma$ intrinsic scatters as presented in Fig. 5 and Table 6; and 2) the mass-dependent full \RHI\ and \RH2\ PDFs as constrained in Section 5, including the first and second moments (mean and standard deviation) of these PDFs.

\acknowledgments
We thank Dr. David Stark for kindly making to us available his compilation of data in electronic form, and
Dr. Claudia Lagos for providing us the ETG data plotted in Figure \ref{MFs}.
We thank the anonymous referee for useful comments and suggestions, which improved the quality of the manuscript.
The authors acknowledge CONACyT grant (Ciencia B\'asica) 285721 for partial funding. 
ARC acknowledges a PhD Fellowship provided by CONACyT.  ARP has been suported 
by a UC-MEXUS Fellowship.  

\appendix

\section{The compiled galaxy samples with \HI\ information}
\label{App:HI}

\subsection{Golden category}
\label{HI-golden}

\textbf{Updated Nearby Galaxy Catalog} \citep[UNGC;][]{Karachentsev+2013,Karachentsev+2014}: It is the most 
representative and homogeneous sample of galaxies (869, most of them of low masses) in the Local Volume, located within 11 Mpc 
or with corrected radial velocities $V_{LG}<600$ Km s$^{-1}$. The authors mention that the sample is complete to $M_B\sim -11$ mag, 
spanning all morphologies.  
However, we take a more conservative limit, having in mind that at low luminosities the
fraction of hardly-to-detect low surface brightness (LSB) galaxies strongly increases. \citet{Karachentsev+2013} report the 
mean $B-$band surface brightness (SB) within the Holmberg isophote, $\bar{\mu}_{B,26}$ for the UNGC galaxies. The SB decreases
on average as lower is the luminosity. For LTGs, the distribution of SBs appears to be incomplete 
from $M_B\approx -13.5$ mag, in such a way that most of the galaxies could be lost at lower luminosities. This is in agreement
with the completeness limit suggested by \citet{Klypin+2015} for UNGC, based on the turnover that suffers the luminosity function 
constructed by them at this luminosity.  In view of these arguments, we consider complete 
the UNGC sample for LTGs, only from $M_B\approx -13.5$ mag ($\ms\approx 10^{7.2 - 7.4}$ \msun); the few
LTGs below this limit are of high SB and are expected then to contain less gas than the average. Since ETGs are of higher SBs 
than LTGs, the SB distribution for the small fraction of them seems not to be affected even at the lowest observed luminosities, 
$M_B\sim -11$ mag. There are 561 galaxies with available \HI\ data \citep[for details regarding the data sources on \HI\ fluxes, 
see Table 3 from][]{Karachentsev+2013}; 90 of them do not obey our completeness limit. 
We estimate stellar masses from the reported 
$K$-band luminosities and $B-K$ colors as in \citet{Avila-Reese+2008}, who calculated the mass-to-light ratios for HSB and LSB galaxies following
\citet{Bell+2003} and \citet{Verheijen+1997}, respectively. The obtained masses (assuming a diet Salpeter IMF) were corrected to the Chabrier IMF.
To separate HSB and LSB galaxies we use the reported $\bar{\mu}_{B,26}$, and transform it to a central surface brightness, $\mu_{0,B}$ assuming an exponential disk. Thus, the criterion $\mu_{B,0}>22.5$ mag/arcsec$^2$ for selecting LSB galaxies corresponds to $\bar{\mu}_{B,26}>24.6$ mag/arcsec$^2$.
\citet{Karachentsev+2013} apply corrections for peculiar motions in the determination of the distances of all the galaxies.

\textbf{GALEX Arecibo SDSS Survey} \citep[GASS;][]{Catinella+2013}: It is an optically-selected sub-sample of 760 galaxies 
more massive than $10^{10}$ \msun\ taken from a parent SDSS DR6 sample volume limited in the redshift range $0.025<z<0.05$
and cross-matched with the ALFALFA and GALEX surveys. The \HI\ information comes from follow-up observations carried 
out with the Arecibo 305 m telescope and 
detections taken from the ALFALFA survey or the Cornell \HI\ digital archive. The \RHI\ limit of the sample is well controlled:  
0.015 for $\log(\ms/\msun)>10.5$ and up to 0.05 for smaller masses. There are 473 detections and 287 non detections; for the
latter, upper limits are provided. 
For the morphological type, we use the \citet{Huertas-Company+2011} automatic classification
applied to the SDSS DR7. These authors, first of all, provide for each galaxy the probability of being early type, $PE$, i.e., E or S0. 
We have tested this probability in a catalog of galaxies with careful visual morphological classification 
\citep[UNAM-KIAS, see below;][]{Hernandez-Toledo+2010} and found that galaxies of types $T\le1$ are mostly those with 
$PE>0.65$, and those with $PE\le0.65$ correspond mostly to $T>1$.\footnote{\citet{Huertas-Company+2011} define as 
ETGs those with $T\le1$, but the $T$ index in their case is from the \citet{Fukugita+2007} notation, which assigns $T=1$ to 
lenticulars instead of $T=0$ as in the usual de Vacouleours notation.}
Thus, we consider here as ETGs those with $PE>0.65$, and the complement are LTGs. We find a good correlation between the
ETGs and LTGs this way defined with those defined using the concentration parameter $c=R_{90}/R_{50}$ to characterize 
the galaxy type, with the value of $c=2.85$ for separating the LTG population from the ETG one (for the latter, it is asked additionally 
to obey the color criterion $NUV-r>5$, \citealp{Deng2013}). 
The stellar masses in \citet{Catinella+2013} were calculated from the spectral energy distribution (SED) of the SDSS galaxies 
\citep{Salim+2007} and assuming a \citet{Chabrier2003} IMF.

\textbf{Herschel Reference Survey -- field galaxies} \citep[HRS; ][]{Boselli+2010,Boselli+2014,Boselli+2014a,Boselli+2014b}: It is a $K-$band
volume limited ($15\hspace{0.05in}\leq D/\hbox{Mpc} \leq 25\hspace{0.05in}$) sample of 323 galaxies complete to $K_s=-12$ and $-8.7$
mag for LTGs and ETGs, respectively. 
The authors collected and homogenized  from the literature \HI\ data for 315 galaxies, and CO data for most of them. The morphological type was
taken from NED or, if not available, from their own classification. Stellar masses are derived from $i$-band luminosities and $g-i$ colors 
\citep[from][]{Cortese+2012} by using stellar mass-to-light ratios as given in \citet{Zibetti+2009}, and assuming a Chabrier IMF. 
The distances were corrected for the peculiar motions and presence of clusters.
The sample includes objects in environments of different density, from the core of the Virgo cluster, to loose groups and fairly isolated systems.
To match the Golden category, we exclude the numerous galaxies from the Virgo Cluster center (regions A and B), which bias
the sample to high densities. 

\textbf{ATLAS$^{3D}$ \HI\ sample -- field ETGs} \citep{Serra+2012}: 
ATLAS$^{3D}$ is a sample of 166 local ETGs observed in detail with integral field
unities \citep[IFUs;][]{Cappellari+2011}. The distance range of the sample is in between 10 and 47 Mpc; the sample includes 39 galaxies  (24\% out
of the galaxies) from the Virgo Cluster. For the Golden category, we exclude those ETGs in the Virgo core. 
The sample is not complete, but after excluding the large number of Virgo core galaxies, it is expected to
be representative of the local population of ETGs since the galaxies were selected from a complete volume-limited parent sample. 
The masses range from $\approx 10^{9.8}$ to $10^{11.3}$ \msun; more massive galaxies are not found typically in small volumes.
We estimate stellar masses using the $\log(\ms)=\log(0.5)+\log(L_{K})$, where $L_{K}$ is the K-band luminosity inferred from the K-band absolute magnitude.
The \HI\ observations were carried out  in the Westerbork Synthesis Radio Telescope \citep{Serra+2012}. They use ALFALFA spectra to determine \mha\ upper limits using one resolution element and find that \mha\ limit is a factor $\sim$2 above the \HI\ mass limit obtained with their data. The 
\RHI\ limit detection increases with mass on average by more than 1.5 orders of magnitude,
attaining values a slow as $\sim 10^{-4}$ for the most massive systems. Because of the ATLAS$^{3D}$ galaxies are nearby, the upper
limits are much lower than in the case of the GASS galaxies in the same mass range.

\subsection{Silver category}
\label{HI-silver}

\textbf{Nearby Field Galaxy Survey} \citep[NFGS;][see more references therein]{Jansen+2000a,Jansen+2000b,Wei+2010a,Kannappan+2013}:
It is a broadly representative sample of 198 local galaxies spanning stellar masses $\ms\sim 10^8-10^{12}$ \msun\ and all the morphological types. 
Morphological classification was obtained from \citet{Jansen+2000b}.
The sample is not complete in volume; galaxies span distances from 2 to 306 Mpc. 
Distances were derived from the Virgo centric flow corrected velocities with respect to the centroid of the Local Group.
Stellar masses were estimated using a variant of the code described in \citet{Kannappan&Gawiser2007} and improved in
\citet{Kannappan+2009}, which fits the SED and integrated spectrum of a galaxy with 
a suite of stellar populations models. Both the diet Salpeter and the \citet{Chabrier2003} IMFs were used.
The single-dish \HI\ fluxes for most of the galaxies were taken from the HyperLeda database \citep{Paturel+2003} or were 
obtained by the authors with the Green Bank Telescope (GBT) Spectrometer. The sample provides strong upper limits
up to $\RHI\sim 0.1$; all galaxies with larger ratios are detected (139, and the rest have only upper limits).

\textbf{\citet{Stark+2013} compilation:} These authors compiled and homogenized from the literature 323 galaxies with 
available \HI,  CO, and multi-band imaging data. Most of the compiled galaxies are from the GASS,  NFGS and ATLAS$^{3D}$ surveys 
described above. We use here only those galaxies that are not in these surveys (67 galaxies).  
The authors use morphological type to separate galaxies into two groups, coincident with our morphology criterion for ETGs and LTGs. 
In their compilation are included some blue compact dwarfs (BCDs). We exclude those BCDs classified as early types. 
The stellar masses were calculated following \citet{Kannappan+2013}. The optical and NIR information required for this calculation
were taken from SDSS DR8 (for those galaxies outside the SDSS footprint, the $BVRI$ photometry from the SINGS sample is used)
and 2MASS, respectively.

\textbf{\citet{Leroy+2008} THINGS sample:} It is a sample of selected 23 nearby, star-forming galaxies, which 
we associate with LTGs; 11 are dwarf, H$_{\mbox{\mdseries\tiny I}}$-dominated galaxies and 12 are large well-defined spiral galaxies. 
The \HI\ information of the galaxies comes from ``The HI Nearby Galaxy Survey'' ~\citep[THINGS,][]{Walter+2008} and it was obtained
with the NRAO Very Large Array (VLA). The stellar masses
are calculated from 3.6 $\mu$m information taken from the Spitzer Infrared Nearby Galaxies Survey \citep[SINGS][]{Kennicutt+2003}.
To convert the 3.6 $\mu$m intensity to surface stellar mass density, they use a $K$-to-$3.6$ $\mu$m calibration and 
adopt a fixed $K-$band mass-to-light ratio, $\Upsilon_{\ast}^{\mbox{\tiny K}}=0.5M_{\odot}/L_{\odot}$, assuming a \citet{Kroupa2001} IMF;
\ms\ is calculated from integrating the surface stellar mass density.

\textbf{Dwarf LTGs} \citep{Geha+2006}: It is a sample of 101 dwarf galaxies, 88 out of them with \HI\ measurements and being of late types. 
Galaxies with absolute magnitudes $M_{r}-5\log_{10}(h_{70})>-16$ were selected from the low-luminosity spectroscopy 
catalog of \citet{Blanton+2005a}, based on the SDDS. 
Distances are estimated based on a model of the local velocity field \citep{Willick+1997}.
Possible selection effects related to the \citet{Blanton+2005a} catalog are that it does not span the full range of environments 
(there are not clusters), and LSB dwarfs are missed.
Stellar masses are based on the optical SDSS $i$-band magnitude and $g-r$ colors using the mass-to-light ratios of \citet{Bell+2003}. 
The \mha\ masses were obtained by \citet{Geha+2006} from the H$_{\mbox{\tiny I}}$ integrated fluxes measured with the Arecibo 305 m 
telescope and the GBT. 

\textbf{ALFALFA dwarf sample} \citep[][]{Huang+2012b}: It consists of 176 low \HI\ mass dwarf galaxies from the ALFALFA survey. The galaxies were selected to have $\mha<10^{7.7}$ \msun\ and \HI\ line widths $<80$ km s$^{-1}$ ({\it s-com} sample).  This
sample is not complete in a volume-limited sense but it probes the extreme low \HI\ mass tail of the ALFALFA survey. 
Stellar masses are obtained through SED fitting following \citet{Salim+2007}, 
assuming a \citet{Chabrier2003} IMF. Only 57 out of the 176 galaxies have stellar mass determination. These galaxies have \HI\ detections and high gas fractions, they are dwarf irregulars.

\subsection{Bronze category}
\label{HI-bronze}

\textbf{UNAM-KIAS catalog of isolated galaxies} \citep{Hernandez-Toledo+2010}: It is a magnitude-limited sample ($m_r>15.2$ mag) of galaxies 
from the SDSS DR5 that obey strict isolation criteria; it is composed of 1520 galaxies spanning all morphological types. 
The morphological classification was carried out by the authors. We have searched \HI\ information for these galaxies 
in HyperLeda (the 21-cm line magnitudes corrected for self-absorption, $m^{c}_{21}$). The 
\HI\ masses are calculated as $M_{H_{I}}[M_{\odot}]=2.356\times 10^{5}\cdot d_{L}^{2}\cdot F_{21}$, where 
$F_{21}[\mbox{Jy Kms$^{-1}$}] = 10^{0.4(17.40-m^{c}_{21})}$ and $d_{L}$ is the luminosity distance to the galaxy in Mpc. 
For the \HI\ non-detections, we have searched rms noise limits in the Digital archive of \HI\ 21 centimeter
line spectra of optically selected galaxies \citep{Springob+2005}, finding data only for 7 galaxies. Non-detected \HI\ upper mass 
limits are estimated as $M^{\rm lim}_{H_{I}}[M_{\odot}] = 1.5\cdot {\rm rms}\cdot\delta W$, where $\delta W$ is the full width of the 
\HI\ line obtained from the Tully-Fisher relation of \citet{Avila-Reese+2008} ($\delta W = 2V_{m}$ is assumed). For LTGs (ETGs), we 
find 272 (24) detections and 7 (0) non-detections. Stellar masses are taken from the group catalog of \citet{Yang+2007}, where 
the \citet{Bell+2003} mass-to-light ratios for a \citet{Kroupa2001} IMF were used. 

\textbf{Analysis of the interstellar Medium of Isolated GAlaxies} \citep[AMIGA;][]{Lisenfeld+2011}: It is a redshift-limited sample 
($1500\leq v_{rec}\mbox{ [km s$^{-1}$]}\leq 5000$) consisting of 273 isolated galaxies with reported multi-band imaging and CO data. 
We perform the same procedure described above for the UNAM-KIAS sample to estimate detected and non-detected \HI\ masses.
For LTGs (ETGs) galaxies, we find 203 (11) detections. Only 4 non-detections were found, all for ETGs.  
The stellar masses were calculated as described above for the UNGC sample. Morphologies were obtained using higher resolution images from SDSS or their own images.

\textbf{Low-mass Isolated galaxies} \citep[][]{Bradford+2015}: It is a sample of 148 isolated low-mass galaxies ($7\leq\log(\ms/\msun)\leq 9.5$) drawn from the SDSS NSA catalog \citep[see][]{Geha+2012}. Isolated galaxies are defined as those without massive hosts (at least 0.5 dex more massive than the
given galaxy) at projected distances less than 1.5 Mpc.  \HI\ measurements were obtained using the 305 m Arecibo and the 100 m Greenbank telescopes. 
Stellar masses are calculated in the NSA catalog using the kcorrect software of \citet{Blanton+2007} using the SDSS and GALEX photometric bands
and assuming a Chabrier 2003 IMF.  For the morphology, we use the \citet{Huertas-Company+2011} automatic classification, following the same procedure described above for the GASS survey, finding classification for 128 out of the 148 galaxies; all of them are of late type. Indeed, according 
to \citet{Geha+2012} all the isolated low-mass galaxies in the local Universe are star forming (late-type) objects.

\textbf{Herschel Reference Survey -- Virgo galaxies}: This is the same HRS sample described above but taking into account only galaxies
from the Virgo Cluster central regions A and B (59). Therefore, this sample is biased to contain galaxies in a very high density environment.

\textbf{ATLAS$^{3D}$ \HI\ sample -- Virgo core ETGs}: This is the same  ATLAS$^{3D}$ sample described above
but taking into account account only the Virgo core ETGs  (15). Therefore, this sample is biased to contain ETGs in a 
very high density environment.

\section{The compiled galaxy samples with CO (\H2) information}
\label{App:H2}

\subsection{Golden category}

\textbf{Herschel Reference Survey (HRS)-- field galaxies:} It is the same sample described in \S\S \ref{HI-golden} (excluding Virgo Cluster
core), with 155 galaxies  with available CO information (101 detections and 54 non detections). The  authors either used compiled CO observations 
from the literature or they carried out their own observations with the National Radio Astronomy Observatory (NRAO) 
Kitt Peak 12 m telescope \citep{Boselli+2014}. A MW constant or $H$-band luminosity-dependent \citep{Boselli+2002}  
\cotoh2~conversion factor is applied to calculate \mhm.

\textbf{CO Legacy Legacy Database for GASS} \citep[COLD GASS;][]{Saintonge+2011}: This is a
program aimed at observing CO(1-0) line fluxes at the IRAM 30 m telescope for galaxies from the
GASS survey described in \S\S \ref{HI-golden}.  
From the CO fluxes, the total CO luminosities, and hence the \H2\ masses, were calculated
for 349 galaxies. The authors apply the MW constant \cotoh2~conversion factor.

\textbf{ATLAS$^{3D}$ \H2\ sample -- field ETGs} \citep{Young+2011}: 
This is the same sample described in \S\S \ref{HI-golden} (excluding Virgo Cluster core)
but with observations in CO using the IRAM 30 m Radio Telescope. The sample amounts for 243 ETGs with CO observations.
The  authors use the constant MW \cotoh2~conversion factor.   

\subsection{Silver category}

\textbf{\citet{Stark+2013} compilation}: It corresponds to the same compiled galaxy sample described in \S\S \ref{HI-silver}. 
The authors observed 35 galaxies of the NFGS with the IRAM 30 m and the ARO 12 m telescopes to measure the CO $(J\rightarrow 2-1)$ (IRAM) 
and $(J\rightarrow 1-0)$ (IRAM \& ARO) lines. For the other galaxies, the \H2\ information from previous works was used.  
\citet{Stark+2013} use the MW constant \cotoh2\ factor for estimating \mhm.

\textbf{\citet{Leroy+2008} HERACLES sample}: It is the same sample described in \S\S \ref{HI-silver}.
The \H2\ information for the 23 galaxies (LTGs) comes from the CO $J\rightarrow 2-1$ maps from the HERA CO-Line 
Extragalactic Survey \citep[HERACLES][CO $J\rightarrow 2-1$ is related to CO $J\rightarrow 1-0$ by assuming the ratio 
$I_{\mbox{\tiny CO}}(2\rightarrow 1)/I_{\mbox{\tiny CO}}(1\rightarrow 0)=0.8$]{Leroy+2008}, and CO $J\rightarrow 1-0$ maps from
the Berkeley-Illinois-Maryland Association (BIMA) Survey of Nearby Galaxies \citep[BIMA SONG][]{Helfer+2003}.
The MW constant \cotoh2~conversion factor was used. 

\textbf{APEX Low-redshift Legacy Survey for MOlecular Gas:} \citep[ALLSMOG;][]{Bothwell+2014}
Using the APEX telescope, the $CO(2\rightarrow 1)$ emission line was measured to trace \H2\ in 42 late-type galaxies of masses
$8.5<$log(\ms/\msun)$<10$, in the redshift range $0.01<z<0.03$ and with metallicities $12+\log(O/H)>8.5$. Morphological classification was taken from NED. The stellar masses are derived based on SED fitting \citep{Kauffmann+2003} using the SDSS DR7 optical data. 
To obtain the $CO(1\rightarrow 0)$ line luminosities, the $CO(2\rightarrow 1)$ emission line is assumed to be fully thermalized.  
A MW constant or metallicity-dependent \citep{Wolfire+2010} \cotoh2\ conversion factor were applied to infer the \H2\ masses.

\textbf{Bauermeister et al. (2013) compilation:} We take from this literature compilation 8 galaxies in the low-redshift  
range $0.05\leq z \leq 0.1$.  All of them are star forming and we associate them to LTGs. Their stellar masses are in the range 
$4\times 10^{10}M_{\odot} \leq M_{\ast}\leq 1.6\times 10^{11}M_{\odot}$
and they were calculated by fitting SDSS $ugriz$ photometry to a grid of models spanning  a wide range of star formation histories.
The \H2\ masses are obtained by the authors from CO $J\rightarrow 1-0$ intensity maps with CARMA, using a MW constant \cotoh2~conversion factor. 

\subsection{Bronze category}

\textbf{Analysis of the interstellar Medium of Isolated GAlaxies} \citep[AMIGA;][]{Lisenfeld+2011}: This is the same sample described in 
\S\S \ref{HI-bronze}. The authors carried out their own observations of CO$(J\rightarrow 1-0)$ with the IRAM 30 m or the 14 m FCRAO telescopes for 
189 galaxies and 87 more were compiled from the literature. An aperture correction is applied to the CO data. A MW constant 
\cotoh2\ conversion factor is used to compute \mhm.

\textbf{Herschel Reference Survey -- Virgo core}: This is the same HRS sample described above but taking into account only 
the Virgo Cluster core regions A and B galaxies (62). Therefore, this sample is biased to contain galaxies in a very high density environment.

\textbf{ATLAS$^{3D}$ \H2\ sample -- Virgo core ETGs}: This is the same  ATLAS$^{3D}$ sample described above
but taking into account account only the Virgo core ETGs  (21). Therefore, this sample is biased to contain ETGs in a 
very high density environment.

 \section{The \cotoh2~conversion factor}
 \label{conv-factor}

\begin{figure*}[ht!]
\begin{center}
\includegraphics[trim = 0mm 100mm 0mm 40mm, clip, width=17.3cm, height=6.0cm]{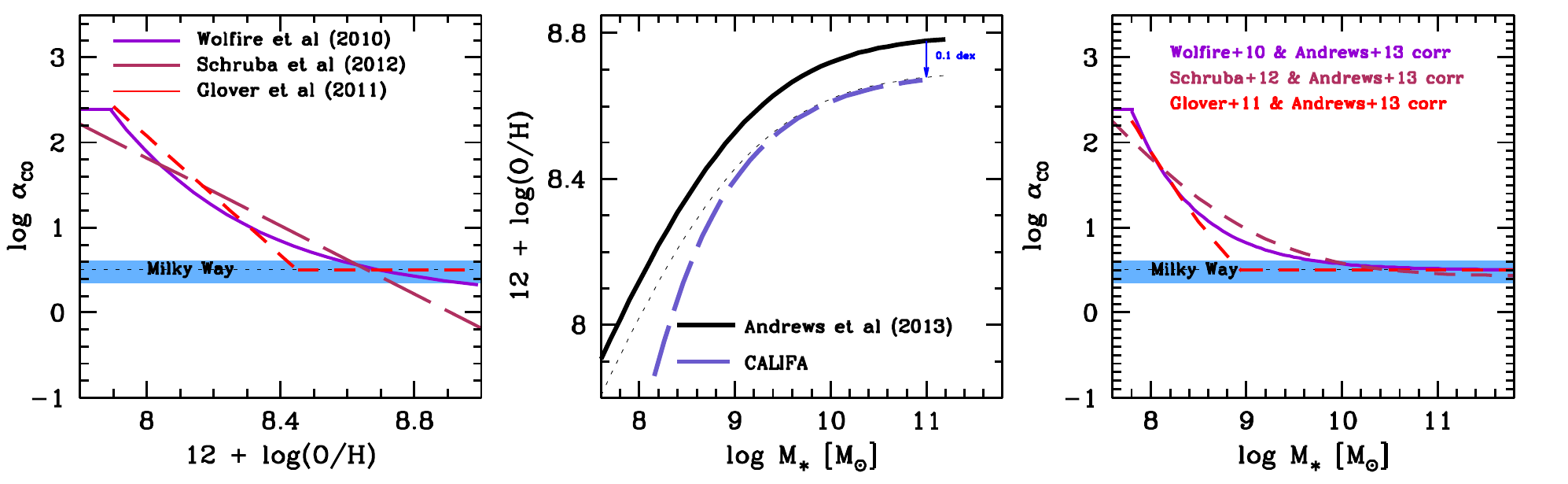}

\caption{ {\it Left panel: } Dependence of the \cotoh2\ factor on gas-phase metallicity as given by physical models \citep{Wolfire+2010,Glover+2011}
calibrated by observations and by a pure empirical approach \citep{Schruba+2012}. Observations do not allow to constrain these relations for 
metallicities lower than $12+\log_{10}(\mbox{O}/\mbox{H})\sim 7.9$   
 {\it Middle panel:} Dependence of metallicity on mass according to the CALIFA \citep{Sanchez+2013} and SDSS \citep{Andrews+2013} surveys. 
 We use an updated relation for CALIFA that includes more galaxies, specially at low masses (S. Sanchez, priv. communication); the
 masses were corrected from Salpeter to Chabrier IMF.  The dotted line is the SDSS relation lowered by 0.1 dex to correct for the aperture effect;
 notice how well it agrees with the CALIFA relation but it extends to lower masses, so this is the relation we use.  
 {\it Right panel:} Dependence of the \cotoh2\ factor on mass inferred from the \a_co--\ms\ and $Z-\ms$ dependences plotted in the other panels. 
 }
\label{conversion-factor} 
\end{center}
\end{figure*}

Several authors have shown that the \cotoh2~conversion factor depends on the gas phase metallicity, $Z$
\citep[see e.g.,][]{Boselli+2002,Schruba+2012,Narayanan+2012,Bolatto+2013}.  
In a recent review on the topic, among the several approaches for determining the dependence 
of \a_co\ on $Z$ in galaxies, \citet{Bolatto+2013} recommend to adopt a prescription based on a local physical model 
for the \H2\ and CO production and calibrate it with extragalactic observations.
In particular, they find that the prescription given in \citet{Wolfire+2010}, based on photodissociation models with shielding,
is the most consistent with the scarce observational data that provides \a_co\ vs. $Z$ in galaxies. According to  \citet{Wolfire+2010}:
\begin{equation}\label{eq:alphaCO_Wolfire2010_Bolatto2013}
\alpha_{\rm CO} = \alpha_{\rm CO,MW}\exp\left[\frac{+4.0\Delta A_{V}}{Z'\bar{A}_{V,MW}}\right]\exp\left[\frac{-4.0\Delta A_{V}}{\bar{A}_{V,MW}}\right]
\end{equation} 
where $\a_co_{\rm ,MW}=3.2$ (in units \msun\ pc$^{-2}$/K km s$^{-1}$) is the adopted conversion factor for the Milky Way, $Z'=Z/Z_{\odot}$
where $Z\equiv 12+\log_{10}(\mbox{O}/\mbox{H})$, $\Delta A_{V}\approx 1$, and $\bar{A}_{V,MW}$ is the mean extinction through a giant molecular cloud at 
Milky Way metallicity $Z_{\odot}$, with $\bar{A}_{V,MW}\approx 5$ for $\Sigma_{\rm GMC}\approx100\mbox{ }\msun\mbox{pc}^{-2}$. 
According to Eq. (\ref{eq:alphaCO_Wolfire2010_Bolatto2013}), $\alpha_{\rm CO} \approx \alpha_{\rm CO,MW}$ for $Z\gtrsim Z_{\odot}$.
The left panel of Fig. \ref{conversion-factor} shows the  \citet{Wolfire+2010} relation along with those of \citet{Glover+2011} and \citet{Schruba+2012}.

To relate \a_co\ with stellar mass, we use the mass-metallicity relation for galaxies in the local Universe. \citet{Sanchez+2013} and 
\citet{Andrews+2013} determined the mass-metallicity relation for galaxies using the CALIFA and SDSS surveys in the stellar mass range $8.4\leq \log(\ms/\msun)\leq 11.2$ and $7.4\leq \log(\ms/\msun)\leq 11.2$, respectively. The work by \citet{Sanchez+2013} provides a more reliable estimate of the mass-metallicity relation; recall that the SDDS galaxies are mapped by only one central fiber of fixed aperture, while CALIFA maps the whole galaxies with many integral field units. However, the mass range in the CALIFA sample is limited, while \citet{Andrews+2013} extends to very low masses. We use an updated version of the 
CALIFA mass-metallicity relation (S. F. Sanchez, priv. communication) and correct \ms\ to pass from the Salpeter IMF to the 
Chabrier one used in \citet{Andrews+2013}. At the mass range where both studies coincide, they agree modulo a shift in the SDSS relation by $\sim +0.1$ dex in metallicity with respect to the CALIFA one (see the middle panel of Fig. \ref{conversion-factor}). This is expected given that CALIFA covers the galaxies up to 2-3 effective radii while SDSS, in most of the cases, covers only the central regions which are typically more metallic than the outer ones \citep[see for a discussion][]{Sanchez+2013}. Thus, we use the relation as reported in \citet{Andrews+2013} but lowering it by 0.1 dex. They find that the function proposed by \citet{Moustakas+2011} fits well their observational results:
$$12+\log_{10}(\mbox{O}/\mbox{H}) =\left(12+\log_{10}(\mbox{O}/\mbox{H})_{\rm asm}\right) $$
\begin{equation}
\label{eq:mass-metallicity}
\hspace*{1.5cm} -\log_{10}\left(1+\left(\frac{M_{TO}}{M_{\ast}}\right)^{\gamma}\right),
\end{equation}
with $12+\log_{10}(\mbox{O}/\mbox{H})_{\rm asm}=8.798$ (we use 8.698, after subtracting 0.1 dex), $M_{TO}=8.901$, and $\gamma=0.640$.

\begin{figure*}[ht!]
\begin{center}
\includegraphics[trim = 0mm 83mm 30mm 40mm, clip, width=14cm]{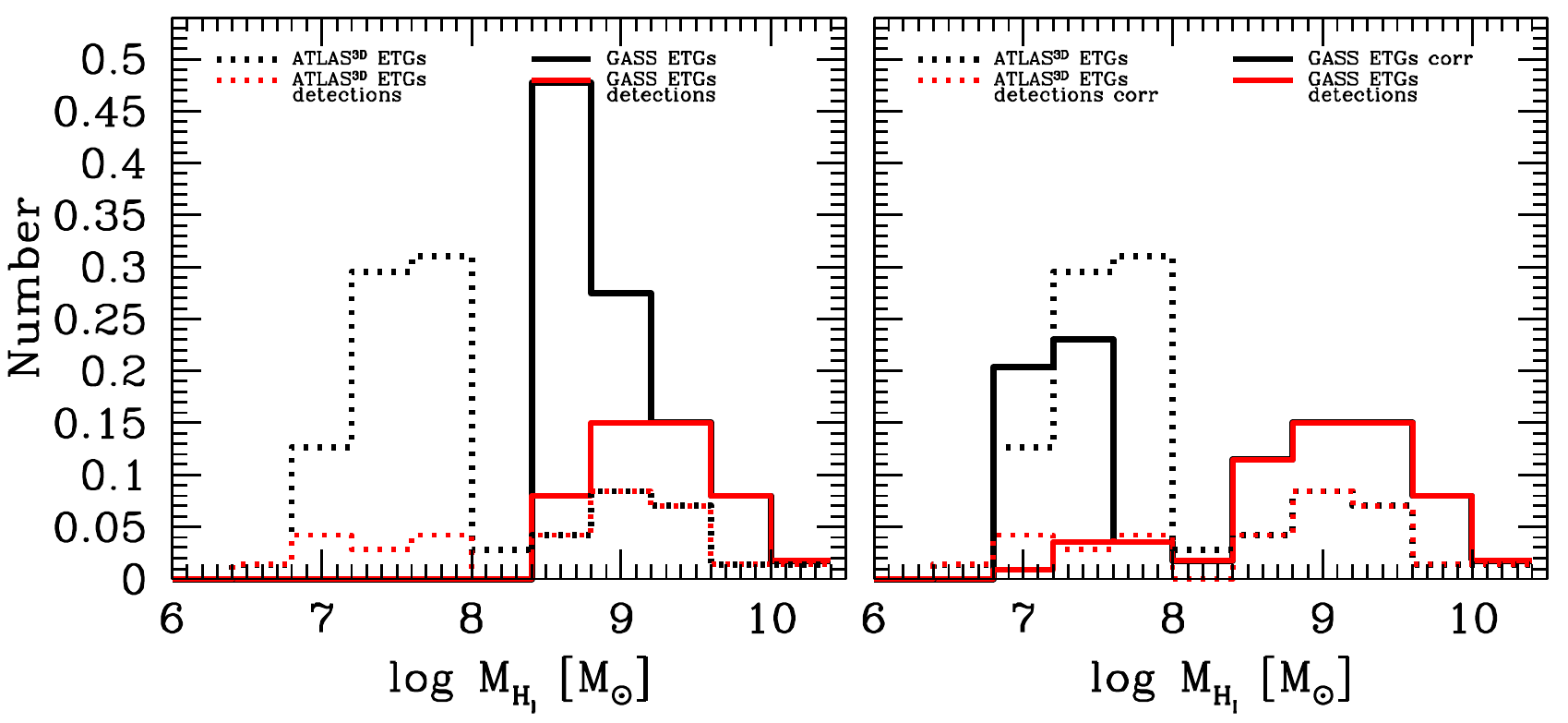}
\caption{ {\it Left panel: } Distributions of \HI\ masses for ETGs in the $10.10-10.65$ $\log\ms$ bin for
GASS (solid black line) and ATLAS$^{\rm 3D}$ (dashed black line). Non detections are also included, with values of \mha\ corresponding
to their upper limits (for ATLAS$^{\rm 3D}$, we use the upper limits increased already by a factor of two as explained in Section \ref{correlations}).
The red lines show the contribution of detected galaxies. The GASS distribution is clearly limited to much higher upper limits than
in ATLAS$^{\rm 3D}$, and this is mainly due to a distance selection effect.  
 {\it Right panel:} Same as in the left panel but after correcting the upper limits of GASS with respect to the observations of ATLAS$^{\rm 3D}$. 
 }
\label{MHI-histogram} 
\end{center}
\end{figure*}

Combining Eqs. (\ref{eq:alphaCO_Wolfire2010_Bolatto2013}) and (\ref{eq:mass-metallicity}), we are able now to obtain the mean \a_co--\ms~relLation. In fact,  metallicity in any calibration is one of the hardest astronomical quantities to measure with precision. However, for our purpose, given the large uncertainties and scatter, it is not relevant the exact calibration but the average dependence of the \a_co\ factor with mass. Following \citet{Bolatto+2013}, we actually normalize the \a_co--\ms\ dependence to $\a_co=\alpha_{\rm CO,MW}$ at $\ms=3\times 10^{10}$\msun, corresponding to a metallicity slightly lower than $Z_{\odot}$. For larger masses (metallicities), we assume this value to remain constant, and for lower masses, we use the mass dependence given by the combination of Eqs.  (\ref{eq:alphaCO_Wolfire2010_Bolatto2013}) and (\ref{eq:mass-metallicity}):  
\begin{equation}\label{eq:alphaCO-ms-Wolfire+10-Andrews+13}
\log(\alpha_{\rm CO}) = 0.15+0.35\left[1+0.1\left(\frac{3\times 10^{10}\msun}{\ms}\right)^{0.64}\right]
\end{equation}
This equation is valid roughly down to \ms $\sim10^{8}$\msun, which corresponds to metallicities $\sim 0.8$ dex below the solar one (or $12+\log({\rm O/H})\approx 7.9$); there are not observational determinations of \a_co\ at lower metallicities. Therefore, for \ms $< 10^{8}$ \msun, we use the same value 
of $\a_co$ at  $10^{8}$\msun, i.e., $\a_co\approx 250$. 
Besides, as highlighted in \citet{Bolatto+2013}, as one moves to increasingly low metallicities, the use of CO emission to quantify the \H2\ reservoir becomes more and more extrapolative, and eventually should appear a practical floor past which CO is not a useful tracer of total \H2\ mass; rather, CO will be a tracer of high column density peaks and well-shielded regions. 

The above mentioned \a_co--\ms\ dependence is applied to LTGs. The right panel of fig. \ref{conversion-factor} shows this dependence along with those calculated from the $\a_co-Z$ dependences from \citet{Glover+2011} and \citet{Schruba+2012}. For ETGs, which have typically higher metallicities than $Z_\odot$, we assume \a_co=$\alpha_{\rm CO,MW}$= const. at all masses.

\section{Corrections to the upper limits of ETGs} 
\label{upper-limit-corrections}

In Section \ref{correlations}, we have noted that the upper limits reported for the GASS (\HI) and COLD GASS (\H2) 
samples in the case of ETGs are significantly larger than those reported for the ATLAS$^{\rm 3D}$ or HRS samples. 
Following \citet{Serra+2012}, we have corrected the ATLAS$^{\rm 3D}$ upper limit values by a factor of two in order to 
take into account differences between the different telescopes and signal-to-noise thresholds used in this survey 
and in GASS (see Section \ref{correlations}). However, the main reason of the differences in the upper limits among 
these samples is a selection effect due to the different volumes covered by them. To illustrate this, in the left panel of 
Fig. \ref{MHI-histogram} we plot the histogram of \HI\ masses for ETGs in the $10.10-10.65$ $\log\ms$ bin for
GASS (solid black line) and ATLAS$^{\rm 3D}$ (dotted black line).  Non detections are also included, with values of \mha\ corresponding
to their upper limits. The red lines show the histograms of only detections. The number of GASS ETGs 
increases as \mha\ is lower and it has a peak at $\log$(\mha/\msun)$\approx 8.4-9.0$, contributed mainly by the upper 
limits and consistent with  the sensitivity limit of the ALFALFA survey at the distances of the GASS galaxies in the mentioned stellar mass range. 
For ATLAS$^{\rm 3D}$, with distances much closer than GASS, some ETGs are detected in \HI\ with masses lower than $\log$(\mha/\msun)=8.4,
but most of them are actually undetected, having upper limits 1--1.5 orders of magnitude lower than in the case of GASS, consistent
with the distance differences between both samples. 
The main difference between the \mha\ distributions of both samples is in their upper limits, and this is clearly due to a selection effect 
imposed by the different distance ranges of these samples.  Basically, if the undetected GASS ETGs would be at the distances of 
ATLAS$^{\rm 3D}$ ETGs, then probably most  of them would not be yet detected in \HI, having upper limits lower by 1--1.5 orders of magnitude.  
Thus, the high values of their upper limits imposed by the volume of GASS, is expected to introduce a bias in the determination of the 
gas-to-stellar mass correlations of ETGs. 

\begin{figure*}[ht!]
\begin{center}
\includegraphics[trim = 3mm 93mm 15mm 45mm, clip, width=17cm]{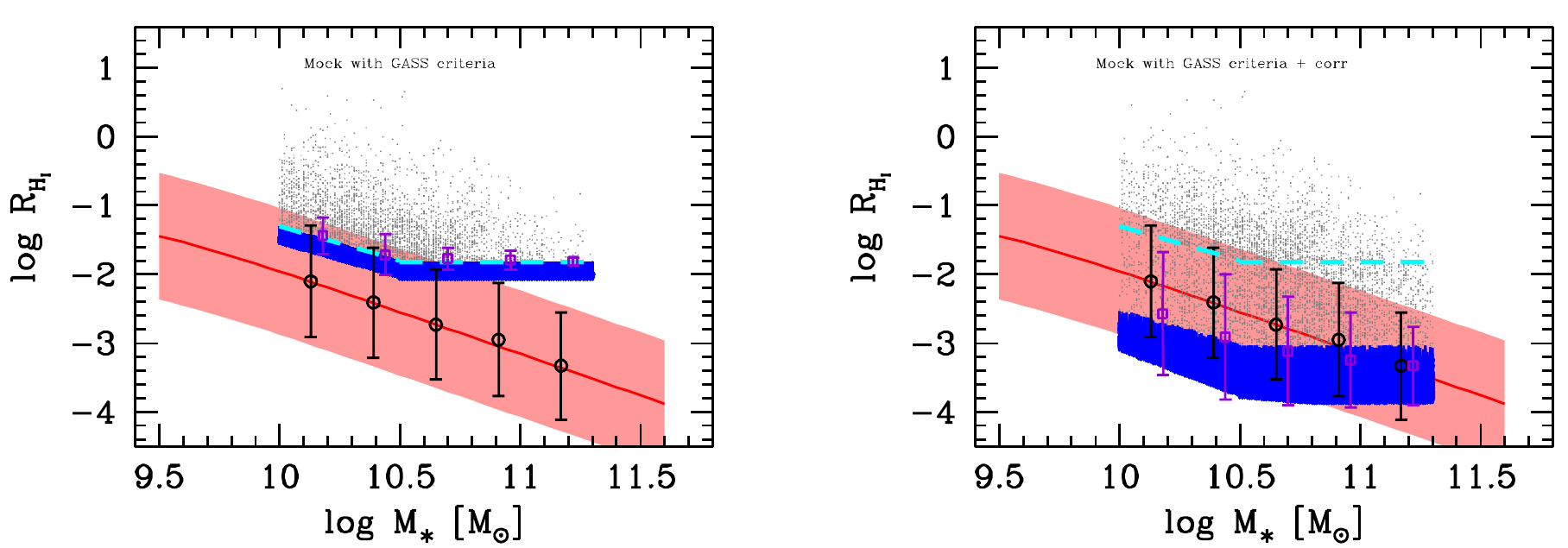}
\caption{ {\it Left panel:} ETGs from our $109<D<222$ Mpc volume mock catalog in the \RHI\ vs. \ms\ plane, following the selection 
and \RHI\ limits of GASS. All mock ETGs below the GASS \RHI\ limits (dashed line) are assumed as undetected and  
assigned an \RHI\ value equal to the \RHI\ limit (upper limit; blue arrows). The magenta squares with error bars are 
the mean and standard deviation calculated in different mass bins with the Kaplan-Meier estimator. The \RHI--\ms\ 
correlation for ETGs used in the generation of the mock catalog is plotted with the red solid line and 
shaded area. The circles with error bars are the mean and standard deviation calculated in different mass bins
for all the ETGs from the mock catalog. The mock catalog samples very well the input correlation but this is not anymore the 
case when the \RHI\ limit of GASS is imposed, even if using the Kaplan-Meier estimator to take into account the upper limits. 
 {\it Right panel:}  Same as in left panel but after applying our ATLAS$^{\rm 3D}$-based corrections to the upper limits of GASS (see text).
 The mean and standard deviation in the different mass bins, taking into account the (corrected) upper limits, follow now closely
 the input correlation.
 }
\label{mock-catalog} 
\end{center}
\end{figure*}

 In an attempt to correct for this selection effect in the upper limits, we will assume that the ETGs in the GASS 
 and ATLAS$^{\rm 3D}$ (and HRS, too) samples are representative of the same local ETG populations. Then, that the upper 
 limits for the ATLAS$^{\rm 3D}$ (or HRS) ETGs are
 significantly lower than those {\it of similar stellar mass} galaxies from GASS, is mainly due to the distance differences among 
 these samples. If the GASS ETGs would be as close as those of the ATLAS$^{\rm 3D}$ ones, then the upper limit 
 region in the plots of \HI-to-stellar mass ratio vs. \ms\ would be on average lower by a factor equal to the distance 
ratio to the square.  Thus, to homogenize the upper limits in \RHI\ given by the GASS and ATLAS$^{\rm 3D}$ 
samples, we lower the upper limits of the galaxies in the volume-limited sample with more distant galaxies (GASS) by
$(D_i/\bar{D}_{\rm ATLAS^{3D}})^2$, where $D_i$ is the distance of each GASS ETG and $\bar{D}_{\rm ATLAS^{3D}}=25$ Mpc is 
the average distance of the ATLAS$^{\rm 3D}$ ETGs.  In fact, according to the ATLAS$^{\rm 3D}$ observations, $25\%$ of ETGs below
the upper limit region of GASS were detected (see for an example Fig. \ref{MHI-histogram}). Therefore, we lower the GASS 
upper limits as mentioned above for 75\% of the galaxies, and for the remaining ones we assign randomly an \RHI\ value between 
its upper limit and the average upper limit of ATLAS$^{\rm 3D}$ galaxies at the corresponding stellar mass. 
The same procedure is applied to the COLD GASS ETGs for the \RH2\ upper limits, where the corresponding 
$\bar{D}_{\rm ATLAS^{3D}}$ for COLD GASS is 26 Mpc. 

The right panel of Fig. \ref{MHI-histogram}, shows the same histograms as in the left panel but now the upper limits of the GASS
sample were corrected as explained above. Observe how close result now the upper limit distributions of GASS and ATLAS$^{\rm 3D}$ 
galaxies after correcting by the distance selection effect.  
Further, we use a large mock galaxy catalog to test the procedure applied here to the GASS (or COLD GASS) upper limits 
for homogenizing them with those of nearby samples as ATLAS$^{\rm 3D}$. The mock catalog is a volume-limited sample (up to 313 Mpc) 
of $5\times 10^6$ galaxies that sample well the observational GSMF and LTG/ETG fractions as a function of \ms\ (see Section \ref{mass-functions}). 
We assign \HI\ masses to each LTG/ETG galaxy by using an {\it input} \RHI\ distribution for a given \ms\ (a \RHI--\ms\ relation and its scatter) 
for LTGs and ETGs. Distances are assigned assuming an isotropic distribution within a sphere of radius of the volume sampled. 
Note that we ignore any clustering properties of the galaxies. This is a safe assumption as we are only interested on the selection 
effects introduced by the detection limits of the GASS and ATLAS$^{\rm 3D}$ samples. 
Then, we select the ETGs more massive than $10^{10}$ \msun\ that are in the $109<D<222$ Mpc range (the GASS volume), and impose 
upper limits to the \RHI\ ratio as a function of mass as the one of GASS \citep[see][]{Catinella+2012}. Then,  we calculate the mean \RHI\ and its standard 
deviation taking into account the upper limits in mass bins as we did for the observational sample (using the Kaplan-Meier estimator). 
The question now is whether we recover or not the input \RHI--\ms\ correlation for ETGs. 

In the left panel of Fig. \ref{mock-catalog}, we plot our input \RHI--\ms\ correlation for ETGs (for this exercise, is described
by a double power-law function with the parameters given in Table \ref{parameters} and assuming a lognormal scatter) along 
with the values from the mock catalog in the $109<D<222$ Mpc volume and 
imposing the sensitivity limit of the GASS sample (dots). All the dots below this limit are plotted as upper limits (blue arrows); 
they populate the imposed sensitivity limit in the \RHI\ vs \ms\ diagram. The open circles with 
error bars are the mean and standard deviation calculated directly from the catalog in $\log\ms$ bins for ETGs in the
$109<D<222$ Mpc volume, while the magenta squares and error bars are the same means and standard deviations calculated 
with the Kaplan-Meier estimator for the case of imposing the GASS sensitivity limit. Thus, after imposing this limit, 
{\it the recovered correlation is far from the input one.}

Then, we apply the same corrections we have used for the real GASS data, based on the information from the ATLAS$^{\rm 3D}$ sample, i.e., the
GASS-like imposed upper limits to the mock catalog galaxies were lowered by $D_i^2$[Mpc$^2$]/25$^2$Mpc$^2$ in 75\% of the cases, and for the
remaining, a random detection value for \RHI\ was assigned as explained above. The right panel of Fig. \ref{mock-catalog}
shows the result of these corrections along with the mean and standard deviations calculated with the corrected data in the
same three mass bins as in the left panel (magenta squares with error bars).  Observe that after our corrections,
the calculated mean and standard deviation in each mass bin are in better agreement with those corresponding to the mock 
catalog without any selection, that is, the input \RHI--\ms\ correlation is reasonable well recovered, showing this the
necessity of applying the mentioned corrections. 

The effect of introducing or not the mentioned above correction to the GASS and COLD GASS upper limits on 
the determination of the \HI- and \H2-to-stellar mass correlations of ETGs are, of course, not so significant as in the experiment
shown in Fig. \ref{mock-catalog} because these samples are not the only ones used for that (subsections \ref{HI} and \ref{H2}). 
In Tables \ref{linear-parameters} and \ref{parameters} (cases ETG$^{\rm ndc}$), 
we present the fitted \HI-to-stellar mass correlation for ETGs in the case the upper limits of the GASS sample were not corrected by distance. 
The double power-law correlation, without the correction, changes slightly at the high-mass end: it would
be shallower but with a much larger scatter than when we took into account the correction; the latter is expected 
due to the strong segregation of the upper limits from COLD GASS and from the less distant ATLAS$^3D$ and HRS 
samples.  The single power-law would be shallower.
Similarly, in these Tables is also present the fitted \H2-to-stellar correlation for ETGs in the case the upper limits of the 
COLD GASS sample were not corrected by distance. The relations are actually almost the same when taking or not taking into account the correction 
but the scatter is larger at the high-mass end for the latter case, as expected due to the segregation of the upper limits from COLD GASS
and from the less distant ATLAS$^{3D}$ and HRS samples.

\section{Observational errors}
\label{observational-error}

To provide a rough estimate of the intrinsic scatter around the \RHI--\ms\ and \RH2--\ms\ correlations in subsections
\ref{RHI-Ms} and \ref{RH2-Ms}, estimates of the (statistical) observational errors, $\sigma_{\rm err}$, in the determination 
of \RHI\ and \RH2\ are necessary. For this, we need to know the respective observational 
uncertainties in the determination of the stellar, \HI, and \H2\ masses.

Most of the observational sources included in our compilation do not report the individual errors in the determination
of these masses, but they report conservative average estimates for them. 

For the stellar mass, the observational errors are typically estimated to be 0.1 dex \citep[see e.g.,][]{Conroy2013}.  
After homogenizing all the samples to a fixed IMF (Chabrier 2003) we have made the conservative assumption that 
other sources of systematic errors in the determination of \ms\ are negligible, see subsection \ref{errorMs}.
For the \HI\ mass, a combination of the statistical errors, distance uncertainties, and errors associated with the  absolute 
21cm flux scale calibration accounts for a total observational error of $\approx 0.1$ dex. Therefore, the average error 
in $\log\RHI$ is $\approx 0.14$ dex.  For the \H2\ mass, most of the works used in our compilation report average observational errors 
of $0.2-0.25$ dex. The uncertainty in the \a_co\ parameter has been taken into account, however,
it was probably significantly underestimated. In a recent review on the subject, \citet{Boselli+2014}
suggest that this uncertainty is actually of the order of 0.3 dex. Thus, considering that the observational errors in
the CO flux account for $30\%$ (0.11 dex;  \citealp[e.g.,][]{Boselli+2014}), and the uncertainty in the \a_co\ parameter is
0.3 dex, an estimate of the typical error in $\log\mhm$ is 0.32 dex.  The estimated error in $\log\RH2$
is then $\approx 0.34$ dex, using an error of 0.1 dex in $\log\ms$. 

\section{Calculation of the GSMF} 
\label{ourGSMF}

Here we outline how we construct our GSMF in a large mass range following \citet[][]{Kravtsov+2014}.
 For high masses, the SDSS-based GMSF presented in \citet{Bernardi+2013} is used. These authors have reanalyzed 
the photometry of the SDSS DR7, taking special care in the background estimate of extended luminous galaxies 
\citep[see also][]{Simard+2011,He+2013,Mendel+2014,DSouza+2015,Meert+2016}; 
after this reanalysis, the high-end of the luminosity (mass) function becomes shallower. 
Their GSMF is well fitted by a Schechter + sub exponential Schechter function. 
For small masses, the GSMFs determined by \citet[][from the GAMA survey]{Baldry+2012} are used.  These authors analyze low redshift samples that contain low luminosity galaxies, though a correction for surface brightness incompleteness was not applied. So,  their determinations 
at $\ms\lesssim10^8$ \msun\ are actually lower limits. This GSMF is well fitted by double Schechter function.
Both, high and low masses GSMFs assume \citet{Chabrier2003} IMF to estimate \ms. However, the masses in \citet{Bernardi+2013} were calculated by using the \citet{Bell+2003} mass-to-luminosity ratios, who 
employed the PEGASE stellar population synthesis models \citep{Fioc+1997}.  In \citet{Baldry+2012} the masses are calculated using the \citep[][BC03]{BC03} models. \citet{Conroy2013} has shown that 
the former are systematically larger than the latter by $\approx 0.10-0.14$ dex. Therefore, for the \citet{Bernardi+2013} GSMF, we 
dismiss uniformly \ms\ by 0.12 dex to homogenize the masses to the BC03 population synthesis model.

\begin{table}[h]
	\centering
	\caption{GSMF parameters.}
	\resizebox{8.0cm}{!} {
		\begin{tabular}{ccccccc}
			\hline
			\hline
			$\alpha_{1}$   & $\log(M_{1}^{\ast})$ & $\log(\phi_{1}^{\ast})$  &  $\alpha_{2}$  & $\log(M_{2}^{\ast})$  & $\log(\phi_{2}^{\ast})$ & $\beta$ \\
			& (\msun)  & (Mpc$^{-3}$ dex$^{-1}$) &  & (\msun) & (Mpc$^{-3}$ dex$^{-1}$)  & \\ \hline
			-1.47 & { 9.74} & -2.66 & { 0.07} & 8.84 & -2.66 & 0.37 \\ \hline
		\end{tabular}
	}
	\label{MFparameters}
\end{table}
 
Thus, we find that a fit to the \citet{Baldry+2012} and the 
\citet{Bernardi+2013} GSMF fit corrected by 0.12 dex in mass are combined to obtain a GSMF that spans from $\ms\approx 10^7$ to
$10^{12}$ \msun. The match of both fits (at the mass where the latter becomes higher than the former) takes places 
at $\ms\approx 10^{9.3}$ \msun. The obtained GSMF is well fitted by the combination of a Schechter function and a
sub exponential Schechter function. The respective parameters are given in Table \ref{MFparameters}. See Fig. \ref{GSMF} for the
plotted GSMF and its comparison to other GSMFs from the literature.

\bibliography{references}

\begin{thebibliography}
\expandafter\ifx\csname natexlab\endcsname\relax\def\natexlab#1{#1}\fi
\expandafter\ifx\csname href\endcsname\relax
  \def\href#1#2{}\fi
\expandafter\ifx\csname urllinklabel\endcsname\relax
  \def\urllinklabel{[LINK]}\fi
\expandafter\ifx\csname adsurllinklabel\endcsname\relax
  \def\adsurllinklabel{[ADS]}\fi

\bibitem[{{Accurso} {et~al.}(2017){Accurso}, {Saintonge}, {Catinella},
  {Cortese}, {Dav{\'e}}, {Dunsheath}, {Genzel}, {Gracia-Carpio}, {Heckman},
  {Jimmy}, {Kramer}, {Li}, {Lutz}, {Schiminovich}, {Schuster}, {Sternberg},
  {Sturm}, {Tacconi}, {Tran}, \& {Wang}}]{Accurso+2017}
{Accurso}, G., {Saintonge}, A., {Catinella}, B., {Cortese}, L., {Dav{\'e}}, R.,
  {Dunsheath}, S.~H., {Genzel}, R., {Gracia-Carpio}, J., {Heckman}, T.~M.,
  {Jimmy}, {Kramer}, C., {Li}, C., {Lutz}, K., {Schiminovich}, D., {Schuster},
  K., {Sternberg}, A., {Sturm}, E., {Tacconi}, L.~J., {Tran}, K.~V., \& {Wang},
  J. 2017, \mnras, 470, 4750


\bibitem[{{Andrews} \& {Martini}(2013)}]{Andrews+2013}
{Andrews}, B.~H. \& {Martini}, P. 2013, \apj, 765, 140


\bibitem[{{Avila-Reese} {et~al.}(2008){Avila-Reese}, {Zavala}, {Firmani}, \&
  {Hern{\'a}ndez-Toledo}}]{Avila-Reese+2008}
{Avila-Reese}, V., {Zavala}, J., {Firmani}, C., \& {Hern{\'a}ndez-Toledo},
  H.~M. 2008, \aj, 136, 1340


\bibitem[{{Baldry} {et~al.}(2012){Baldry}, {Driver}, {Loveday}, {Taylor},
  {Kelvin}, {Liske}, {Norberg}, {Robotham}, {Brough}, {Hopkins}, {Bamford},
  {Peacock}, {Bland-Hawthorn}, {Conselice}, {Croom}, {Jones}, {Parkinson},
  {Popescu}, {Prescott}, {Sharp}, \& {Tuffs}}]{Baldry+2012}
{Baldry}, I.~K., {Driver}, S.~P., {Loveday}, J., {Taylor}, E.~N., {Kelvin},
  L.~S., {Liske}, J., {Norberg}, P., {Robotham}, A.~S.~G., {Brough}, S.,
  {Hopkins}, A.~M., {Bamford}, S.~P., {Peacock}, J.~A., {Bland-Hawthorn}, J.,
  {Conselice}, C.~J., {Croom}, S.~M., {Jones}, D.~H., {Parkinson}, H.~R.,
  {Popescu}, C.~C., {Prescott}, M., {Sharp}, R.~G., \& {Tuffs}, R.~J. 2012,
  \mnras, 421, 621


\bibitem[{{Baldry} {et~al.}(2008){Baldry}, {Glazebrook}, \&
  {Driver}}]{Baldry+2008}
{Baldry}, I.~K., {Glazebrook}, K., \& {Driver}, S.~P. 2008, \mnras, 388, 945


\bibitem[{{Barnes} {et~al.}(2001){Barnes}, {Staveley-Smith}, {de Blok},
  {Oosterloo}, {Stewart}, {Wright}, {Banks}, {Bhathal}, {Boyce}, {Calabretta},
  {Disney}, {Drinkwater}, {Ekers}, {Freeman}, {Gibson}, {Green}, {Haynes}, {te
  Lintel Hekkert}, {Henning}, {Jerjen}, {Juraszek}, {Kesteven}, {Kilborn},
  {Knezek}, {Koribalski}, {Kraan-Korteweg}, {Malin}, {Marquarding}, {Minchin},
  {Mould}, {Price}, {Putman}, {Ryder}, {Sadler}, {Schr{\"o}der}, {Stootman},
  {Webster}, {Wilson}, \& {Ye}}]{Barnes+2001}
{Barnes}, D.~G., {Staveley-Smith}, L., {de Blok}, W.~J.~G., {Oosterloo}, T.,
  {Stewart}, I.~M., {Wright}, A.~E., {Banks}, G.~D., {Bhathal}, R., {Boyce},
  P.~J., {Calabretta}, M.~R., {Disney}, M.~J., {Drinkwater}, M.~J., {Ekers},
  R.~D., {Freeman}, K.~C., {Gibson}, B.~K., {Green}, A.~J., {Haynes}, R.~F.,
  {te Lintel Hekkert}, P., {Henning}, P.~A., {Jerjen}, H., {Juraszek}, S.,
  {Kesteven}, M.~J., {Kilborn}, V.~A., {Knezek}, P.~M., {Koribalski}, B.,
  {Kraan-Korteweg}, R.~C., {Malin}, D.~F., {Marquarding}, M., {Minchin}, R.~F.,
  {Mould}, J.~R., {Price}, R.~M., {Putman}, M.~E., {Ryder}, S.~D., {Sadler},
  E.~M., {Schr{\"o}der}, A., {Stootman}, F., {Webster}, R.~L., {Wilson}, W.~E.,
  \& {Ye}, T. 2001, \mnras, 322, 486


\bibitem[{{Bauermeister} {et~al.}(2013){Bauermeister}, {Blitz}, {Bolatto},
  {Bureau}, {Leroy}, {Ostriker}, {Teuben}, {Wong}, \&
  {Wright}}]{Bauermeister+2013}
{Bauermeister}, A., {Blitz}, L., {Bolatto}, A., {Bureau}, M., {Leroy}, A.,
  {Ostriker}, E., {Teuben}, P., {Wong}, T., \& {Wright}, M. 2013, \apj, 768,
  132


\bibitem[{{Behroozi} {et~al.}(2010){Behroozi}, {Conroy}, \&
  {Wechsler}}]{Behroozi+2010}
{Behroozi}, P.~S., {Conroy}, C., \& {Wechsler}, R.~H. 2010, \apj, 717, 379


\bibitem[{{Behroozi} {et~al.}(2013){Behroozi}, {Wechsler}, \&
  {Conroy}}]{Behroozi+2013}
{Behroozi}, P.~S., {Wechsler}, R.~H., \& {Conroy}, C. 2013, \apj, 770, 57


\bibitem[{{Bell} {et~al.}(2003){Bell}, {McIntosh}, {Katz}, \&
  {Weinberg}}]{Bell+2003}
{Bell}, E.~F., {McIntosh}, D.~H., {Katz}, N., \& {Weinberg}, M.~D. 2003, \apjs,
  149, 289


\bibitem[{{Bernardi} {et~al.}(2013){Bernardi}, {Meert}, {Sheth}, {Vikram},
  {Huertas-Company}, {Mei}, \& {Shankar}}]{Bernardi+2013}
{Bernardi}, M., {Meert}, A., {Sheth}, R.~K., {Vikram}, V., {Huertas-Company},
  M., {Mei}, S., \& {Shankar}, F. 2013, \mnras, 436, 697


\bibitem[{{Blanton} {et~al.}(2005{\natexlab{a}}){Blanton}, {Eisenstein},
  {Hogg}, {Schlegel}, \& {Brinkmann}}]{Blanton+2005c}
{Blanton}, M.~R., {Eisenstein}, D., {Hogg}, D.~W., {Schlegel}, D.~J., \&
  {Brinkmann}, J. 2005{\natexlab{a}}, \apj, 629, 143


\bibitem[{{Blanton} {et~al.}(2005{\natexlab{b}}){Blanton}, {Lupton},
  {Schlegel}, {Strauss}, {Brinkmann}, {Fukugita}, \& {Loveday}}]{Blanton+2005a}
{Blanton}, M.~R., {Lupton}, R.~H., {Schlegel}, D.~J., {Strauss}, M.~A.,
  {Brinkmann}, J., {Fukugita}, M., \& {Loveday}, J. 2005{\natexlab{b}}, \apj,
  631, 208


\bibitem[{{Blanton} \& {Moustakas}(2009)}]{Blanton+2009}
{Blanton}, M.~R. \& {Moustakas}, J. 2009, \araa, 47, 159


\bibitem[{{Blanton} \& {Roweis}(2007)}]{Blanton+2007}
{Blanton}, M.~R. \& {Roweis}, S. 2007, \aj, 133, 734


\bibitem[{{Blitz} \& {Rosolowsky}(2006)}]{Blitz+2006}
{Blitz}, L. \& {Rosolowsky}, E. 2006, \apj, 650, 933


\bibitem[{{Blyth} {et~al.}(2015){Blyth}, {van der Hulst}, {Verheijen},
  {Catinella}, {Fraternali}, {Haynes}, {Hess}, {Koribalski}, {Lagos}, {Meyer},
  {Obreschkow}, {Popping}, {Power}, {Verdes-Montenegro}, \&
  {Zwaan}}]{Blyth+2015}
{Blyth}, S., {van der Hulst}, T.~M., {Verheijen}, M.~A.~W., {Catinella}, B.,
  {Fraternali}, F., {Haynes}, M.~P., {Hess}, K.~M., {Koribalski}, B., {Lagos},
  C., {Meyer}, M., {Obreschkow}, D., {Popping}, A., {Power}, C.,
  {Verdes-Montenegro}, L.~L., \& {Zwaan}, M. 2015, Advancing Astrophysics with
  the Square Kilometre Array (AASKA14), 128


\bibitem[{{Bolatto} {et~al.}(2013){Bolatto}, {Wolfire}, \&
  {Leroy}}]{Bolatto+2013}
{Bolatto}, A.~D., {Wolfire}, M., \& {Leroy}, A.~K. 2013, \araa, 51, 207


\bibitem[{{Boselli} {et~al.}(2014{\natexlab{a}}){Boselli}, {Cortese}, \&
  {Boquien}}]{Boselli+2014}
{Boselli}, A., {Cortese}, L., \& {Boquien}, M. 2014{\natexlab{a}}, \aap, 564,
  A65


\bibitem[{{Boselli} {et~al.}(2014{\natexlab{b}}){Boselli}, {Cortese},
  {Boquien}, {Boissier}, {Catinella}, {Gavazzi}, {Lagos}, \&
  {Saintonge}}]{Boselli+2014a}
{Boselli}, A., {Cortese}, L., {Boquien}, M., {Boissier}, S., {Catinella}, B.,
  {Gavazzi}, G., {Lagos}, C., \& {Saintonge}, A. 2014{\natexlab{b}}, \aap, 564,
  A67


\bibitem[{{Boselli} {et~al.}(2014{\natexlab{c}}){Boselli}, {Cortese},
  {Boquien}, {Boissier}, {Catinella}, {Lagos}, \& {Saintonge}}]{Boselli+2014b}
{Boselli}, A., {Cortese}, L., {Boquien}, M., {Boissier}, S., {Catinella}, B.,
  {Lagos}, C., \& {Saintonge}, A. 2014{\natexlab{c}}, \aap, 564, A66


\bibitem[{{Boselli} {et~al.}(2010){Boselli}, {Eales}, {Cortese}, {Bendo},
  {Chanial}, {Buat}, {Davies}, {Auld}, {Rigby}, {Baes}, {Barlow}, {Bock},
  {Bradford}, {Castro-Rodriguez}, {Charlot}, {Clements}, {Cormier}, {Dwek},
  {Elbaz}, {Galametz}, {Galliano}, {Gear}, {Glenn}, {Gomez}, {Griffin}, {Hony},
  {Isaak}, {Levenson}, {Lu}, {Madden}, {O'Halloran}, {Okamura}, {Oliver},
  {Page}, {Panuzzo}, {Papageorgiou}, {Parkin}, {Perez-Fournon}, {Pohlen},
  {Rangwala}, {Roussel}, {Rykala}, {Sacchi}, {Sauvage}, {Schulz}, {Schirm},
  {Smith}, {Spinoglio}, {Stevens}, {Symeonidis}, {Vaccari}, {Vigroux},
  {Wilson}, {Wozniak}, {Wright}, \& {Zeilinger}}]{Boselli+2010}
{Boselli}, A., {Eales}, S., {Cortese}, L., {Bendo}, G., {Chanial}, P., {Buat},
  V., {Davies}, J., {Auld}, R., {Rigby}, E., {Baes}, M., {Barlow}, M., {Bock},
  J., {Bradford}, M., {Castro-Rodriguez}, N., {Charlot}, S., {Clements}, D.,
  {Cormier}, D., {Dwek}, E., {Elbaz}, D., {Galametz}, M., {Galliano}, F.,
  {Gear}, W., {Glenn}, J., {Gomez}, H., {Griffin}, M., {Hony}, S., {Isaak}, K.,
  {Levenson}, L., {Lu}, N., {Madden}, S., {O'Halloran}, B., {Okamura}, K.,
  {Oliver}, S., {Page}, M., {Panuzzo}, P., {Papageorgiou}, A., {Parkin}, T.,
  {Perez-Fournon}, I., {Pohlen}, M., {Rangwala}, N., {Roussel}, H., {Rykala},
  A., {Sacchi}, N., {Sauvage}, M., {Schulz}, B., {Schirm}, M., {Smith},
  M.~W.~L., {Spinoglio}, L., {Stevens}, J., {Symeonidis}, M., {Vaccari}, M.,
  {Vigroux}, L., {Wilson}, C., {Wozniak}, H., {Wright}, G., \& {Zeilinger}, W.
  2010, \pasp, 122, 261


\bibitem[{{Boselli} \& {Gavazzi}(2006)}]{Boselli+2006}
{Boselli}, A. \& {Gavazzi}, G. 2006, \pasp, 118, 517


\bibitem[{{Boselli} {et~al.}(2002){Boselli}, {Lequeux}, \&
  {Gavazzi}}]{Boselli+2002}
{Boselli}, A., {Lequeux}, J., \& {Gavazzi}, G. 2002, \aap, 384, 33


\bibitem[{{Bothwell} {et~al.}(2014){Bothwell}, {Wagg}, {Cicone}, {Maiolino},
  {M{\o}ller}, {Aravena}, {De Breuck}, {Peng}, {Espada}, {Hodge},
  {Impellizzeri}, {Mart{\'{\i}}n}, {Riechers}, \& {Walter}}]{Bothwell+2014}
{Bothwell}, M.~S., {Wagg}, J., {Cicone}, C., {Maiolino}, R., {M{\o}ller}, P.,
  {Aravena}, M., {De Breuck}, C., {Peng}, Y., {Espada}, D., {Hodge}, J.~A.,
  {Impellizzeri}, C.~M.~V., {Mart{\'{\i}}n}, S., {Riechers}, D., \& {Walter},
  F. 2014, \mnras, 445, 2599


\bibitem[{{Bradford} {et~al.}(2015){Bradford}, {Geha}, \&
  {Blanton}}]{Bradford+2015}
{Bradford}, J.~D., {Geha}, M.~C., \& {Blanton}, M.~R. 2015, \apj, 809, 146


\bibitem[{{Brown} {et~al.}(2015){Brown}, {Catinella}, {Cortese}, {Kilborn},
  {Haynes}, \& {Giovanelli}}]{Brown+2015}
{Brown}, T., {Catinella}, B., {Cortese}, L., {Kilborn}, V., {Haynes}, M.~P., \&
  {Giovanelli}, R. 2015, \mnras, 452, 2479


\bibitem[{{Brown} {et~al.}(2017){Brown}, {Catinella}, {Cortese}, {Lagos},
  {Dav{\'e}}, {Kilborn}, {Haynes}, {Giovanelli}, \&
  {Rafieferantsoa}}]{Brown+2017}
{Brown}, T., {Catinella}, B., {Cortese}, L., {Lagos}, C.~d.~P., {Dav{\'e}}, R.,
  {Kilborn}, V., {Haynes}, M.~P., {Giovanelli}, R., \& {Rafieferantsoa}, M.
  2017, \mnras, 466, 1275


\bibitem[{{Bruzual} \& {Charlot}(2003)}]{BC03}
{Bruzual}, G. \& {Charlot}, S. 2003, \mnras, 344, 1000


\bibitem[{Buckley \& James(1979)}]{BJ79}
Buckley, J. \& James, I. 1979, Biometrika, 66, 429
 \href{http://www.jstor.org/stable/2335161}{\urllinklabel}

\bibitem[{{Butcher} {et~al.}(2016){Butcher}, {Schneider}, {van Driel},
  {Lehnert}, \& {Minchin}}]{Butcher+2016}
{Butcher}, Z., {Schneider}, S., {van Driel}, W., {Lehnert}, M.~D., \&
  {Minchin}, R. 2016, \aap, 596, A60


\bibitem[{{Cappellari} {et~al.}(2011){Cappellari}, {Emsellem}, {Krajnovi{\'c}},
  {McDermid}, {Scott}, {Verdoes Kleijn}, {Young}, {Alatalo}, {Bacon}, {Blitz},
  {Bois}, {Bournaud}, {Bureau}, {Davies}, {Davis}, {de Zeeuw}, {Duc},
  {Khochfar}, {Kuntschner}, {Lablanche}, {Morganti}, {Naab}, {Oosterloo},
  {Sarzi}, {Serra}, \& {Weijmans}}]{Cappellari+2011}
{Cappellari}, M., {Emsellem}, E., {Krajnovi{\'c}}, D., {McDermid}, R.~M.,
  {Scott}, N., {Verdoes Kleijn}, G.~A., {Young}, L.~M., {Alatalo}, K., {Bacon},
  R., {Blitz}, L., {Bois}, M., {Bournaud}, F., {Bureau}, M., {Davies}, R.~L.,
  {Davis}, T.~A., {de Zeeuw}, P.~T., {Duc}, P.-A., {Khochfar}, S.,
  {Kuntschner}, H., {Lablanche}, P.-Y., {Morganti}, R., {Naab}, T.,
  {Oosterloo}, T., {Sarzi}, M., {Serra}, P., \& {Weijmans}, A.-M. 2011, \mnras,
  413, 813


\bibitem[{{Carilli} \& {Rawlings}(2004)}]{Carilli+2004}
{Carilli}, C.~L. \& {Rawlings}, S. 2004, \nar, 48, 979


\bibitem[{{Catinella} {et~al.}(2013){Catinella}, {Schiminovich}, {Cortese},
  {Fabello}, {Hummels}, {Moran}, {Lemonias}, {Cooper}, {Wu}, {Heckman}, \&
  {Wang}}]{Catinella+2013}
{Catinella}, B., {Schiminovich}, D., {Cortese}, L., {Fabello}, S., {Hummels},
  C.~B., {Moran}, S.~M., {Lemonias}, J.~J., {Cooper}, A.~P., {Wu}, R.,
  {Heckman}, T.~M., \& {Wang}, J. 2013, \mnras, 436, 34


\bibitem[{{Catinella} {et~al.}(2012){Catinella}, {Schiminovich}, {Kauffmann},
  {Fabello}, {Hummels}, {Lemonias}, {Moran}, {Wu}, {Cooper}, \&
  {Wang}}]{Catinella+2012}
{Catinella}, B., {Schiminovich}, D., {Kauffmann}, G., {Fabello}, S., {Hummels},
  C., {Lemonias}, J., {Moran}, S.~M., {Wu}, R., {Cooper}, A., \& {Wang}, J.
  2012, \aap, 544, A65


\bibitem[{{Chabrier}(2003)}]{Chabrier2003}
{Chabrier}, G. 2003, \pasp, 115, 763


\bibitem[{{Conroy}(2013)}]{Conroy2013}
{Conroy}, C. 2013, \araa, 51, 393


\bibitem[{{Conroy} \& {Wechsler}(2009)}]{Conroy+2009}
{Conroy}, C. \& {Wechsler}, R.~H. 2009, \apj, 696, 620


\bibitem[{{Cortese} {et~al.}(2012){Cortese}, {Boissier}, {Boselli}, {Bendo},
  {Buat}, {Davies}, {Eales}, {Heinis}, {Isaak}, \& {Madden}}]{Cortese+2012}
{Cortese}, L., {Boissier}, S., {Boselli}, A., {Bendo}, G.~J., {Buat}, V.,
  {Davies}, J.~I., {Eales}, S., {Heinis}, S., {Isaak}, K.~G., \& {Madden},
  S.~C. 2012, \aap, 544, A101


\bibitem[{{Cortese} {et~al.}(2011){Cortese}, {Catinella}, {Boissier},
  {Boselli}, \& {Heinis}}]{Cortese+2011}
{Cortese}, L., {Catinella}, B., {Boissier}, S., {Boselli}, A., \& {Heinis}, S.
  2011, \mnras, 415, 1797


\bibitem[{{Deng}(2013)}]{Deng2013}
{Deng}, X.-F. 2013, Research in Astronomy and Astrophysics, 13, 651


\bibitem[{{Dressler}(1980)}]{Dressler1980}
{Dressler}, A. 1980, \apj, 236, 351


\bibitem[{{D'Souza} {et~al.}(2015){D'Souza}, {Vegetti}, \&
  {Kauffmann}}]{DSouza+2015}
{D'Souza}, R., {Vegetti}, S., \& {Kauffmann}, G. 2015, \mnras, 454, 4027


\bibitem[{{Duffy} {et~al.}(2012){Duffy}, {Kay}, {Battye}, {Booth}, {Dalla
  Vecchia}, \& {Schaye}}]{Duffy+2012}
{Duffy}, A.~R., {Kay}, S.~T., {Battye}, R.~A., {Booth}, C.~M., {Dalla Vecchia},
  C., \& {Schaye}, J. 2012, \mnras, 420, 2799


\bibitem[{{Elmegreen}(1989)}]{Elmegreen1989}
{Elmegreen}, B.~G. 1989, \apj, 338, 178


\bibitem[{Feigelson \& Babu(2012)}]{Feigelson-Babu2012}
Feigelson, E.~D. \& Babu, G.~J. 2012, Modern Statistical Methods for Astronomy:
  With R Applications (Cambridge)


\bibitem[{{Feigelson} \& {Nelson}(1985)}]{Feigelson+1985}
{Feigelson}, E.~D. \& {Nelson}, P.~I. 1985, \apj, 293, 192


\bibitem[{{Fioc} \& {Rocca-Volmerange}(1997)}]{Fioc+1997}
{Fioc}, M. \& {Rocca-Volmerange}, B. 1997, \aap, 326, 950


\bibitem[{{Fu} {et~al.}(2010){Fu}, {Guo}, {Kauffmann}, \& {Krumholz}}]{Fu+2010}
{Fu}, J., {Guo}, Q., {Kauffmann}, G., \& {Krumholz}, M.~R. 2010, \mnras, 409,
  515


\bibitem[{{Fukugita} {et~al.}(2007){Fukugita}, {Nakamura}, {Okamura}, {Yasuda},
  {Barentine}, {Brinkmann}, {Gunn}, {Harvanek}, {Ichikawa}, {Lupton},
  {Schneider}, {Strauss}, \& {York}}]{Fukugita+2007}
{Fukugita}, M., {Nakamura}, O., {Okamura}, S., {Yasuda}, N., {Barentine},
  J.~C., {Brinkmann}, J., {Gunn}, J.~E., {Harvanek}, M., {Ichikawa}, T.,
  {Lupton}, R.~H., {Schneider}, D.~P., {Strauss}, M.~A., \& {York}, D.~G. 2007,
  \aj, 134, 579


\bibitem[{{Garnett}(2002)}]{Garnett2002}
{Garnett}, D.~R. 2002, \apj, 581, 1019


\bibitem[{{Gavazzi} {et~al.}(2005){Gavazzi}, {Boselli}, {van Driel}, \&
  {O'Neil}}]{Gavazzi+2005}
{Gavazzi}, G., {Boselli}, A., {van Driel}, W., \& {O'Neil}, K. 2005, \aap, 429,
  439


\bibitem[{{Geha} {et~al.}(2006){Geha}, {Blanton}, {Masjedi}, \&
  {West}}]{Geha+2006}
{Geha}, M., {Blanton}, M.~R., {Masjedi}, M., \& {West}, A.~A. 2006, \apj, 653,
  240


\bibitem[{{Geha} {et~al.}(2012){Geha}, {Blanton}, {Yan}, \&
  {Tinker}}]{Geha+2012}
{Geha}, M., {Blanton}, M.~R., {Yan}, R., \& {Tinker}, J.~L. 2012, \apj, 757, 85


\bibitem[{{Giovanelli} {et~al.}(2005){Giovanelli}, {Haynes}, {Kent},
  {Perillat}, {Saintonge}, {Brosch}, {Catinella}, {Hoffman}, {Stierwalt},
  {Spekkens}, {Lerner}, {Masters}, {Momjian}, {Rosenberg}, {Springob},
  {Boselli}, {Charmandaris}, {Darling}, {Davies}, {Garcia Lambas}, {Gavazzi},
  {Giovanardi}, {Hardy}, {Hunt}, {Iovino}, {Karachentsev}, {Karachentseva},
  {Koopmann}, {Marinoni}, {Minchin}, {Muller}, {Putman}, {Pantoja}, {Salzer},
  {Scodeggio}, {Skillman}, {Solanes}, {Valotto}, {van Driel}, \& {van
  Zee}}]{Giovanelli+2005}
{Giovanelli}, R., {Haynes}, M.~P., {Kent}, B.~R., {Perillat}, P., {Saintonge},
  A., {Brosch}, N., {Catinella}, B., {Hoffman}, G.~L., {Stierwalt}, S.,
  {Spekkens}, K., {Lerner}, M.~S., {Masters}, K.~L., {Momjian}, E.,
  {Rosenberg}, J.~L., {Springob}, C.~M., {Boselli}, A., {Charmandaris}, V.,
  {Darling}, J.~K., {Davies}, J., {Garcia Lambas}, D., {Gavazzi}, G.,
  {Giovanardi}, C., {Hardy}, E., {Hunt}, L.~K., {Iovino}, A., {Karachentsev},
  I.~D., {Karachentseva}, V.~E., {Koopmann}, R.~A., {Marinoni}, C., {Minchin},
  R., {Muller}, E., {Putman}, M., {Pantoja}, C., {Salzer}, J.~J., {Scodeggio},
  M., {Skillman}, E., {Solanes}, J.~M., {Valotto}, C., {van Driel}, W., \& {van
  Zee}, L. 2005, \aj, 130, 2598


\bibitem[{{Glover} \& {Mac Low}(2011)}]{Glover+2011}
{Glover}, S.~C.~O. \& {Mac Low}, M.-M. 2011, \mnras, 412, 337


\bibitem[{{Haynes} \& {Giovanelli}(1984)}]{Haynes+1984}
{Haynes}, M.~P. \& {Giovanelli}, R. 1984, \aj, 89, 758


\bibitem[{{Haynes} {et~al.}(2011){Haynes}, {Giovanelli}, {Martin}, {Hess},
  {Saintonge}, {Adams}, {Hallenbeck}, {Hoffman}, {Huang}, {Kent}, {Koopmann},
  {Papastergis}, {Stierwalt}, {Balonek}, {Craig}, {Higdon}, {Kornreich},
  {Miller}, {O'Donoghue}, {Olowin}, {Rosenberg}, {Spekkens}, {Troischt}, \&
  {Wilcots}}]{Haynes+2011}
{Haynes}, M.~P., {Giovanelli}, R., {Martin}, A.~M., {Hess}, K.~M., {Saintonge},
  A., {Adams}, E.~A.~K., {Hallenbeck}, G., {Hoffman}, G.~L., {Huang}, S.,
  {Kent}, B.~R., {Koopmann}, R.~A., {Papastergis}, E., {Stierwalt}, S.,
  {Balonek}, T.~J., {Craig}, D.~W., {Higdon}, S.~J.~U., {Kornreich}, D.~A.,
  {Miller}, J.~R., {O'Donoghue}, A.~A., {Olowin}, R.~P., {Rosenberg}, J.~L.,
  {Spekkens}, K., {Troischt}, P., \& {Wilcots}, E.~M. 2011, \aj, 142, 170


\bibitem[{{He} {et~al.}(2013){He}, {Xia}, {Hao}, {Jing}, {Mao}, \&
  {Li}}]{He+2013}
{He}, Y.~Q., {Xia}, X.~Y., {Hao}, C.~N., {Jing}, Y.~P., {Mao}, S., \& {Li}, C.
  2013, \apj, 773, 37


\bibitem[{{Helfer} {et~al.}(2003){Helfer}, {Thornley}, {Regan}, {Wong},
  {Sheth}, {Vogel}, {Blitz}, \& {Bock}}]{Helfer+2003}
{Helfer}, T.~T., {Thornley}, M.~D., {Regan}, M.~W., {Wong}, T., {Sheth}, K.,
  {Vogel}, S.~N., {Blitz}, L., \& {Bock}, D.~C.-J. 2003, \apjs, 145, 259


\bibitem[{{Hern{\'a}ndez-Toledo} {et~al.}(2010){Hern{\'a}ndez-Toledo},
  {V{\'a}zquez-Mata}, {Mart{\'{\i}}nez-V{\'a}zquez}, {Choi}, \&
  {Park}}]{Hernandez-Toledo+2010}
{Hern{\'a}ndez-Toledo}, H.~M., {V{\'a}zquez-Mata}, J.~A.,
  {Mart{\'{\i}}nez-V{\'a}zquez}, L.~A., {Choi}, Y.-Y., \& {Park}, C. 2010, \aj,
  139, 2525


\bibitem[{{Huang} {et~al.}(2012{\natexlab{a}}){Huang}, {Haynes}, {Giovanelli},
  \& {Brinchmann}}]{Huang+2012a}
{Huang}, S., {Haynes}, M.~P., {Giovanelli}, R., \& {Brinchmann}, J.
  2012{\natexlab{a}}, \apj, 756, 113


\bibitem[{{Huang} {et~al.}(2012{\natexlab{b}}){Huang}, {Haynes}, {Giovanelli},
  {Brinchmann}, {Stierwalt}, \& {Neff}}]{Huang+2012b}
{Huang}, S., {Haynes}, M.~P., {Giovanelli}, R., {Brinchmann}, J., {Stierwalt},
  S., \& {Neff}, S.~G. 2012{\natexlab{b}}, \aj, 143, 133


\bibitem[{{Huang} {et~al.}(2014){Huang}, {Haynes}, {Giovanelli}, {Hallenbeck},
  {Jones}, {Adams}, {Brinchmann}, {Chengalur}, {Hunt}, {Masters}, {Matsushita},
  {Saintonge}, \& {Spekkens}}]{Huang+2014}
{Huang}, S., {Haynes}, M.~P., {Giovanelli}, R., {Hallenbeck}, G., {Jones},
  M.~G., {Adams}, E.~A.~K., {Brinchmann}, J., {Chengalur}, J.~N., {Hunt},
  L.~K., {Masters}, K.~L., {Matsushita}, S., {Saintonge}, A., \& {Spekkens}, K.
  2014, \apj, 793, 40


\bibitem[{{Huertas-Company} {et~al.}(2011){Huertas-Company}, {Aguerri},
  {Bernardi}, {Mei}, \& {S{\'a}nchez Almeida}}]{Huertas-Company+2011}
{Huertas-Company}, M., {Aguerri}, J.~A.~L., {Bernardi}, M., {Mei}, S., \&
  {S{\'a}nchez Almeida}, J. 2011, \aap, 525, A157


\bibitem[{{Jansen} {et~al.}(2000{\natexlab{a}}){Jansen}, {Fabricant}, {Franx},
  \& {Caldwell}}]{Jansen+2000a}
{Jansen}, R.~A., {Fabricant}, D., {Franx}, M., \& {Caldwell}, N.
  2000{\natexlab{a}}, \apjs, 126, 331


\bibitem[{{Jansen} {et~al.}(2000{\natexlab{b}}){Jansen}, {Franx}, {Fabricant},
  \& {Caldwell}}]{Jansen+2000b}
{Jansen}, R.~A., {Franx}, M., {Fabricant}, D., \& {Caldwell}, N.
  2000{\natexlab{b}}, \apjs, 126, 271


\bibitem[{Johnston {et~al.}(2008)Johnston, Taylor, Bailes, Bartel, Baugh,
  Bietenholz, Blake, Braun, Brown, Chatterjee, Darling, Deller, Dodson,
  Edwards, Ekers, Ellingsen, Feain, Gaensler, Haverkorn, Hobbs, Hopkins,
  Jackson, James, Joncas, Kaspi, Kilborn, Koribalski, Kothes, Landecker, Lenc,
  Lovell, Macquart, Manchester, Matthews, McClure-Griffiths, Norris, Pen,
  Phillips, Power, Protheroe, Sadler, Schmidt, Stairs, Staveley-Smith, Stil,
  Tingay, Tzioumis, Walker, Wall, \& Wolleben}]{Johnston+2008}
Johnston, S., Taylor, R., Bailes, M., Bartel, N., Baugh, C., Bietenholz, M.,
  Blake, C., Braun, R., Brown, J., Chatterjee, S., Darling, J., Deller, A.,
  Dodson, R., Edwards, P., Ekers, R., Ellingsen, S., Feain, I., Gaensler, B.,
  Haverkorn, M., Hobbs, G., Hopkins, A., Jackson, C., James, C., Joncas, G.,
  Kaspi, V., Kilborn, V., Koribalski, B., Kothes, R., Landecker, T., Lenc, A.,
  Lovell, J., Macquart, J.-P., Manchester, R., Matthews, D., McClure-Griffiths,
  N., Norris, R., Pen, U.-L., Phillips, C., Power, C., Protheroe, R., Sadler,
  E., Schmidt, B., Stairs, I., Staveley-Smith, L., Stil, J., Tingay, S.,
  Tzioumis, A., Walker, M., Wall, J., \& Wolleben, M. 2008, Experimental
  Astronomy, 22, 151
 \href{http://dx.doi.org/10.1007/s10686-008-9124-7}{\urllinklabel}

\bibitem[{{Jones} {et~al.}(2018){Jones}, {Haynes}, {Giovanelli}, \&
  {Moorman}}]{Jones+2018}
{Jones}, M.~G., {Haynes}, M.~P., {Giovanelli}, R., \& {Moorman}, C. 2018,
  \mnras


\bibitem[{{Jones} {et~al.}(2016){Jones}, {Papastergis}, {Haynes}, \&
  {Giovanelli}}]{Jones+2016}
{Jones}, M.~G., {Papastergis}, E., {Haynes}, M.~P., \& {Giovanelli}, R. 2016,
  \mnras, 457, 4393


\bibitem[{{Kannappan} \& {Gawiser}(2007)}]{Kannappan&Gawiser2007}
{Kannappan}, S.~J. \& {Gawiser}, E. 2007, \apjl, 657, L5


\bibitem[{{Kannappan} {et~al.}(2009){Kannappan}, {Guie}, \&
  {Baker}}]{Kannappan+2009}
{Kannappan}, S.~J., {Guie}, J.~M., \& {Baker}, A.~J. 2009, \aj, 138, 579


\bibitem[{{Kannappan} {et~al.}(2013){Kannappan}, {Stark}, {Eckert}, {Moffett},
  {Wei}, {Pisano}, {Baker}, {Vogel}, {Fabricant}, {Laine}, {Norris}, {Jogee},
  {Lepore}, {Hough}, \& {Weinberg-Wolf}}]{Kannappan+2013}
{Kannappan}, S.~J., {Stark}, D.~V., {Eckert}, K.~D., {Moffett}, A.~J., {Wei},
  L.~H., {Pisano}, D.~J., {Baker}, A.~J., {Vogel}, S.~N., {Fabricant}, D.~G.,
  {Laine}, S., {Norris}, M.~A., {Jogee}, S., {Lepore}, N., {Hough}, L.~E., \&
  {Weinberg-Wolf}, J. 2013, \apj, 777, 42


\bibitem[{Kaplan \& Meier(1958)}]{Kaplan+1958}
Kaplan, E.~L. \& Meier, P. 1958, Journal of the American Statistical
  Association, 53, pp. 457
 \href{http://www.jstor.org/stable/2281868}{\urllinklabel}

\bibitem[{{Karachentsev} {et~al.}(2014){Karachentsev}, {Kaisina}, \&
  {Makarov}}]{Karachentsev+2014}
{Karachentsev}, I.~D., {Kaisina}, E.~I., \& {Makarov}, D.~I. 2014, \aj, 147, 13


\bibitem[{{Karachentsev} {et~al.}(2013){Karachentsev}, {Makarov}, \&
  {Kaisina}}]{Karachentsev+2013}
{Karachentsev}, I.~D., {Makarov}, D.~I., \& {Kaisina}, E.~I. 2013, \aj, 145,
  101


\bibitem[{{Kauffmann} {et~al.}(2003){Kauffmann}, {Heckman}, {White}, {Charlot},
  {Tremonti}, {Peng}, {Seibert}, {Brinkmann}, {Nichol}, {SubbaRao}, \&
  {York}}]{Kauffmann+2003}
{Kauffmann}, G., {Heckman}, T.~M., {White}, S.~D.~M., {Charlot}, S.,
  {Tremonti}, C., {Peng}, E.~W., {Seibert}, M., {Brinkmann}, J., {Nichol},
  R.~C., {SubbaRao}, M., \& {York}, D. 2003, \mnras, 341, 54


\bibitem[{{Kauffmann} {et~al.}(2004){Kauffmann}, {White}, {Heckman},
  {M{\'e}nard}, {Brinchmann}, {Charlot}, {Tremonti}, \&
  {Brinkmann}}]{Kauffmann+2004}
{Kauffmann}, G., {White}, S.~D.~M., {Heckman}, T.~M., {M{\'e}nard}, B.,
  {Brinchmann}, J., {Charlot}, S., {Tremonti}, C., \& {Brinkmann}, J. 2004,
  \mnras, 353, 713


\bibitem[{{Kennicutt} {et~al.}(2003){Kennicutt}, {Armus}, {Bendo}, {Calzetti},
  {Dale}, {Draine}, {Engelbracht}, {Gordon}, {Grauer}, {Helou}, {Hollenbach},
  {Jarrett}, {Kewley}, {Leitherer}, {Li}, {Malhotra}, {Regan}, {Rieke},
  {Rieke}, {Roussel}, {Smith}, {Thornley}, \& {Walter}}]{Kennicutt+2003}
{Kennicutt}, Jr., R.~C., {Armus}, L., {Bendo}, G., {Calzetti}, D., {Dale},
  D.~A., {Draine}, B.~T., {Engelbracht}, C.~W., {Gordon}, K.~D., {Grauer},
  A.~D., {Helou}, G., {Hollenbach}, D.~J., {Jarrett}, T.~H., {Kewley}, L.~J.,
  {Leitherer}, C., {Li}, A., {Malhotra}, S., {Regan}, M.~W., {Rieke}, G.~H.,
  {Rieke}, M.~J., {Roussel}, H., {Smith}, J.-D.~T., {Thornley}, M.~D., \&
  {Walter}, F. 2003, \pasp, 115, 928


\bibitem[{{Keres} {et~al.}(2003){Keres}, {Yun}, \& {Young}}]{Keres+2003}
{Keres}, D., {Yun}, M.~S., \& {Young}, J.~S. 2003, \apj, 582, 659


\bibitem[{{Klypin} {et~al.}(2015){Klypin}, {Karachentsev}, {Makarov}, \&
  {Nasonova}}]{Klypin+2015}
{Klypin}, A., {Karachentsev}, I., {Makarov}, D., \& {Nasonova}, O. 2015,
  \mnras, 454, 1798


\bibitem[{{Kravtsov} {et~al.}(2014){Kravtsov}, {Vikhlinin}, \&
  {Meshscheryakov}}]{Kravtsov+2014}
{Kravtsov}, A., {Vikhlinin}, A., \& {Meshscheryakov}, A. 2014, ArXiv e-prints


\bibitem[{{Kroupa}(2001)}]{Kroupa2001}
{Kroupa}, P. 2001, \mnras, 322, 231


\bibitem[{{Kroupa} {et~al.}(1993){Kroupa}, {Tout}, \& {Gilmore}}]{Kroupa+1993}
{Kroupa}, P., {Tout}, C.~A., \& {Gilmore}, G. 1993, \mnras, 262, 545


\bibitem[{{Krumholz} {et~al.}(2009){Krumholz}, {McKee}, \&
  {Tumlinson}}]{Krumholz+2009}
{Krumholz}, M.~R., {McKee}, C.~F., \& {Tumlinson}, J. 2009, \apj, 693, 216


\bibitem[{{Lagos} {et~al.}(2011){Lagos}, {Baugh}, {Lacey}, {Benson}, {Kim}, \&
  {Power}}]{Lagos+2011}
{Lagos}, C.~D.~P., {Baugh}, C.~M., {Lacey}, C.~G., {Benson}, A.~J., {Kim},
  H.-S., \& {Power}, C. 2011, \mnras, 418, 1649


\bibitem[{{Lagos} {et~al.}(2015){Lagos}, {Crain}, {Schaye}, {Furlong}, {Frenk},
  {Bower}, {Schaller}, {Theuns}, {Trayford}, {Bah{\'e}}, \& {Dalla
  Vecchia}}]{Lagos+2015}
{Lagos}, C.~d.~P., {Crain}, R.~A., {Schaye}, J., {Furlong}, M., {Frenk}, C.~S.,
  {Bower}, R.~G., {Schaller}, M., {Theuns}, T., {Trayford}, J.~W., {Bah{\'e}},
  Y.~M., \& {Dalla Vecchia}, C. 2015, \mnras, 452, 3815


\bibitem[{{Lagos} {et~al.}(2014){Lagos}, {Davis}, {Lacey}, {Zwaan}, {Baugh},
  {Gonzalez-Perez}, \& {Padilla}}]{Lagos+2014}
{Lagos}, C.~d.~P., {Davis}, T.~A., {Lacey}, C.~G., {Zwaan}, M.~A., {Baugh},
  C.~M., {Gonzalez-Perez}, V., \& {Padilla}, N.~D. 2014, \mnras, 443, 1002


\bibitem[{Lee \& Wang(2003)}]{Lee-Wang2003}
Lee, E.~T. \& Wang, J.~W. 2003, Statistical Methods for Survival Data Analysis
  (Wiley)


\bibitem[{{Lemonias} {et~al.}(2013){Lemonias}, {Schiminovich}, {Catinella},
  {Heckman}, \& {Moran}}]{Lemonias+2013}
{Lemonias}, J.~J., {Schiminovich}, D., {Catinella}, B., {Heckman}, T.~M., \&
  {Moran}, S.~M. 2013, \apj, 776, 74


\bibitem[{{Leroy} {et~al.}(2008){Leroy}, {Walter}, {Brinks}, {Bigiel}, {de
  Blok}, {Madore}, \& {Thornley}}]{Leroy+2008}
{Leroy}, A.~K., {Walter}, F., {Brinks}, E., {Bigiel}, F., {de Blok}, W.~J.~G.,
  {Madore}, B., \& {Thornley}, M.~D. 2008, \aj, 136, 2782


\bibitem[{{Lisenfeld} {et~al.}(2011){Lisenfeld}, {Espada}, {Verdes-Montenegro},
  {Kuno}, {Leon}, {Sabater}, {Sato}, {Sulentic}, {Verley}, \&
  {Yun}}]{Lisenfeld+2011}
{Lisenfeld}, U., {Espada}, D., {Verdes-Montenegro}, L., {Kuno}, N., {Leon}, S.,
  {Sabater}, J., {Sato}, N., {Sulentic}, J., {Verley}, S., \& {Yun}, M.~S.
  2011, \aap, 534, A102


\bibitem[{{Maddox} {et~al.}(2015){Maddox}, {Hess}, {Obreschkow}, {Jarvis}, \&
  {Blyth}}]{Maddox+2015}
{Maddox}, N., {Hess}, K.~M., {Obreschkow}, D., {Jarvis}, M.~J., \& {Blyth},
  S.-L. 2015, \mnras, 447, 1610


\bibitem[{{Mandelbaum} {et~al.}(2006){Mandelbaum}, {Seljak}, {Kauffmann},
  {Hirata}, \& {Brinkmann}}]{Mandelbaum+2006}
{Mandelbaum}, R., {Seljak}, U., {Kauffmann}, G., {Hirata}, C.~M., \&
  {Brinkmann}, J. 2006, \mnras, 368, 715


\bibitem[{{Mandelbaum} {et~al.}(2016){Mandelbaum}, {Wang}, {Zu}, {White},
  {Henriques}, \& {More}}]{Mandelbaum+2016}
{Mandelbaum}, R., {Wang}, W., {Zu}, Y., {White}, S., {Henriques}, B., \&
  {More}, S. 2016, \mnras, 457, 3200


\bibitem[{{Martin} {et~al.}(2010){Martin}, {Papastergis}, {Giovanelli},
  {Haynes}, {Springob}, \& {Stierwalt}}]{Martin+2010}
{Martin}, A.~M., {Papastergis}, E., {Giovanelli}, R., {Haynes}, M.~P.,
  {Springob}, C.~M., \& {Stierwalt}, S. 2010, \apj, 723, 1359


\bibitem[{{McGaugh}(2005)}]{McGaugh2005}
{McGaugh}, S.~S. 2005, \apj, 632, 859


\bibitem[{{Meert} {et~al.}(2016){Meert}, {Vikram}, \& {Bernardi}}]{Meert+2016}
{Meert}, A., {Vikram}, V., \& {Bernardi}, M. 2016, \mnras, 455, 2440


\bibitem[{{Mendel} {et~al.}(2014){Mendel}, {Simard}, {Palmer}, {Ellison}, \&
  {Patton}}]{Mendel+2014}
{Mendel}, J.~T., {Simard}, L., {Palmer}, M., {Ellison}, S.~L., \& {Patton},
  D.~R. 2014, \apjs, 210, 3


\bibitem[{{Meyer} {et~al.}(2004){Meyer}, {Zwaan}, {Webster}, {Staveley-Smith},
  {Ryan-Weber}, {Drinkwater}, {Barnes}, {Howlett}, {Kilborn}, {Stevens},
  {Waugh}, {Pierce}, {Bhathal}, {de Blok}, {Disney}, {Ekers}, {Freeman},
  {Garcia}, {Gibson}, {Harnett}, {Henning}, {Jerjen}, {Kesteven}, {Knezek},
  {Koribalski}, {Mader}, {Marquarding}, {Minchin}, {O'Brien}, {Oosterloo},
  {Price}, {Putman}, {Ryder}, {Sadler}, {Stewart}, {Stootman}, \&
  {Wright}}]{Meyer+2004}
{Meyer}, M.~J., {Zwaan}, M.~A., {Webster}, R.~L., {Staveley-Smith}, L.,
  {Ryan-Weber}, E., {Drinkwater}, M.~J., {Barnes}, D.~G., {Howlett}, M.,
  {Kilborn}, V.~A., {Stevens}, J., {Waugh}, M., {Pierce}, M.~J., {Bhathal}, R.,
  {de Blok}, W.~J.~G., {Disney}, M.~J., {Ekers}, R.~D., {Freeman}, K.~C.,
  {Garcia}, D.~A., {Gibson}, B.~K., {Harnett}, J., {Henning}, P.~A., {Jerjen},
  H., {Kesteven}, M.~J., {Knezek}, P.~M., {Koribalski}, B.~S., {Mader}, S.,
  {Marquarding}, M., {Minchin}, R.~F., {O'Brien}, J., {Oosterloo}, T., {Price},
  R.~M., {Putman}, M.~E., {Ryder}, S.~D., {Sadler}, E.~M., {Stewart}, I.~M.,
  {Stootman}, F., \& {Wright}, A.~E. 2004, \mnras, 350, 1195


\bibitem[{{Moffett} {et~al.}(2016){Moffett}, {Ingarfield}, {Driver},
  {Robotham}, {Kelvin}, {Lange}, {Me{\v s}tri{\'c}}, {Alpaslan}, {Baldry},
  {Bland-Hawthorn}, {Brough}, {Cluver}, {Davies}, {Holwerda}, {Hopkins},
  {Kafle}, {Kennedy}, {Norberg}, \& {Taylor}}]{Moffett+2016}
{Moffett}, A.~J., {Ingarfield}, S.~A., {Driver}, S.~P., {Robotham}, A.~S.~G.,
  {Kelvin}, L.~S., {Lange}, R., {Me{\v s}tri{\'c}}, U., {Alpaslan}, M.,
  {Baldry}, I.~K., {Bland-Hawthorn}, J., {Brough}, S., {Cluver}, M.~E.,
  {Davies}, L.~J.~M., {Holwerda}, B.~W., {Hopkins}, A.~M., {Kafle}, P.~R.,
  {Kennedy}, R., {Norberg}, P., \& {Taylor}, E.~N. 2016, \mnras, 457, 1308


\bibitem[{{More} {et~al.}(2011){More}, {van den Bosch}, {Cacciato}, {Skibba},
  {Mo}, \& {Yang}}]{More+2011}
{More}, S., {van den Bosch}, F.~C., {Cacciato}, M., {Skibba}, R., {Mo}, H.~J.,
  \& {Yang}, X. 2011, \mnras, 410, 210


\bibitem[{{Moster} {et~al.}(2013){Moster}, {Naab}, \& {White}}]{Moster+2013}
{Moster}, B.~P., {Naab}, T., \& {White}, S.~D.~M. 2013, \mnras, 428, 3121


\bibitem[{{Moster} {et~al.}(2010){Moster}, {Somerville}, {Maulbetsch}, {van den
  Bosch}, {Macci{\`o}}, {Naab}, \& {Oser}}]{Moster+2010}
{Moster}, B.~P., {Somerville}, R.~S., {Maulbetsch}, C., {van den Bosch}, F.~C.,
  {Macci{\`o}}, A.~V., {Naab}, T., \& {Oser}, L. 2010, \apj, 710, 903


\bibitem[{{Moustakas} {et~al.}(2013){Moustakas}, {Coil}, {Aird}, {Blanton},
  {Cool}, {Eisenstein}, {Mendez}, {Wong}, {Zhu}, \& {Arnouts}}]{Moustakas+2013}
{Moustakas}, J., {Coil}, A.~L., {Aird}, J., {Blanton}, M.~R., {Cool}, R.~J.,
  {Eisenstein}, D.~J., {Mendez}, A.~J., {Wong}, K.~C., {Zhu}, G., \& {Arnouts},
  S. 2013, \apj, 767, 50


\bibitem[{{Moustakas} {et~al.}(2011){Moustakas}, {Zaritsky}, {Brown}, {Cool},
  {Dey}, {Eisenstein}, {Gonzalez}, {Jannuzi}, {Jones}, {Kochanek}, {Murray}, \&
  {Wild}}]{Moustakas+2011}
{Moustakas}, J., {Zaritsky}, D., {Brown}, M., {Cool}, R., {Dey}, A.,
  {Eisenstein}, D.~J., {Gonzalez}, A.~H., {Jannuzi}, B., {Jones}, C.,
  {Kochanek}, C.~S., {Murray}, S.~S., \& {Wild}, V. 2011, ArXiv e-prints


\bibitem[{{Narayanan} {et~al.}(2012){Narayanan}, {Krumholz}, {Ostriker}, \&
  {Hernquist}}]{Narayanan+2012}
{Narayanan}, D., {Krumholz}, M.~R., {Ostriker}, E.~C., \& {Hernquist}, L. 2012,
  \mnras, 421, 3127


\bibitem[{{Noordermeer} {et~al.}(2005){Noordermeer}, {van der Hulst},
  {Sancisi}, {Swaters}, \& {van Albada}}]{Noordermeer+2005}
{Noordermeer}, E., {van der Hulst}, J.~M., {Sancisi}, R., {Swaters}, R.~A., \&
  {van Albada}, T.~S. 2005, \aap, 442, 137


\bibitem[{{Obreschkow} \& {Rawlings}(2009)}]{Obreschkow+2009}
{Obreschkow}, D. \& {Rawlings}, S. 2009, \mnras, 394, 1857


\bibitem[{{Papastergis} {et~al.}(2012){Papastergis}, {Cattaneo}, {Huang},
  {Giovanelli}, \& {Haynes}}]{Papastergis+2012}
{Papastergis}, E., {Cattaneo}, A., {Huang}, S., {Giovanelli}, R., \& {Haynes},
  M.~P. 2012, \apj, 759, 138


\bibitem[{{Paturel} {et~al.}(2003){Paturel}, {Petit}, {Prugniel}, {Theureau},
  {Rousseau}, {Brouty}, {Dubois}, \& {Cambresy}}]{Paturel+2003}
{Paturel}, G., {Petit}, C., {Prugniel}, P., {Theureau}, G., {Rousseau}, J.,
  {Brouty}, M., {Dubois}, P., \& {Cambresy}, L. 2003, VizieR Online Data
  Catalog, 7237, 0


\bibitem[{{Pforr} {et~al.}(2012){Pforr}, {Maraston}, \& {Tonini}}]{Pforr+2012}
{Pforr}, J., {Maraston}, C., \& {Tonini}, C. 2012, \mnras, 422, 3285


\bibitem[{Press {et~al.}(1996)Press, Teukolsky, Vetterling, \&
  Flannery}]{Press+1996}
Press, W.~H., Teukolsky, S.~A., Vetterling, W.~T., \& Flannery, B.~P. 1996,
  Numerical Recipes in Fortran 90 (2Nd Ed.): The Art of Parallel Scientific
  Computing (New York, NY, USA: Cambridge University Press)


\bibitem[{{Rodr{\'{\i}}guez-Puebla} {et~al.}(2013){Rodr{\'{\i}}guez-Puebla},
  {Avila-Reese}, \& {Drory}}]{Rodriguez-Puebla+2013}
{Rodr{\'{\i}}guez-Puebla}, A., {Avila-Reese}, V., \& {Drory}, N. 2013, \apj,
  767, 92


\bibitem[{{Rodr{\'{\i}}guez-Puebla} {et~al.}(2015){Rodr{\'{\i}}guez-Puebla},
  {Avila-Reese}, {Yang}, {Foucaud}, {Drory}, \& {Jing}}]{Rodriguez-Puebla+2015}
{Rodr{\'{\i}}guez-Puebla}, A., {Avila-Reese}, V., {Yang}, X., {Foucaud}, S.,
  {Drory}, N., \& {Jing}, Y.~P. 2015, \apj, 799, 130


\bibitem[{{Rodr{\'{\i}}guez-Puebla} {et~al.}(2017){Rodr{\'{\i}}guez-Puebla},
  {Primack}, {Avila-Reese}, \& {Faber}}]{Rodriguez-Puebla+2017}
{Rodr{\'{\i}}guez-Puebla}, A., {Primack}, J.~R., {Avila-Reese}, V., \& {Faber},
  S.~M. 2017, \mnras, 470, 651


\bibitem[{{Saintonge} {et~al.}(2011){Saintonge}, {Kauffmann}, {Kramer},
  {Tacconi}, {Buchbender}, {Catinella}, {Fabello}, {Graci{\'a}-Carpio}, {Wang},
  {Cortese}, {Fu}, {Genzel}, {Giovanelli}, {Guo}, {Haynes}, {Heckman},
  {Krumholz}, {Lemonias}, {Li}, {Moran}, {Rodriguez-Fernandez}, {Schiminovich},
  {Schuster}, \& {Sievers}}]{Saintonge+2011}
{Saintonge}, A., {Kauffmann}, G., {Kramer}, C., {Tacconi}, L.~J., {Buchbender},
  C., {Catinella}, B., {Fabello}, S., {Graci{\'a}-Carpio}, J., {Wang}, J.,
  {Cortese}, L., {Fu}, J., {Genzel}, R., {Giovanelli}, R., {Guo}, Q., {Haynes},
  M.~P., {Heckman}, T.~M., {Krumholz}, M.~R., {Lemonias}, J., {Li}, C.,
  {Moran}, S., {Rodriguez-Fernandez}, N., {Schiminovich}, D., {Schuster}, K.,
  \& {Sievers}, A. 2011, \mnras, 415, 32


\bibitem[{{Salim} {et~al.}(2007){Salim}, {Rich}, {Charlot}, {Brinchmann},
  {Johnson}, {Schiminovich}, {Seibert}, {Mallery}, {Heckman}, {Forster},
  {Friedman}, {Martin}, {Morrissey}, {Neff}, {Small}, {Wyder}, {Bianchi},
  {Donas}, {Lee}, {Madore}, {Milliard}, {Szalay}, {Welsh}, \&
  {Yi}}]{Salim+2007}
{Salim}, S., {Rich}, R.~M., {Charlot}, S., {Brinchmann}, J., {Johnson}, B.~D.,
  {Schiminovich}, D., {Seibert}, M., {Mallery}, R., {Heckman}, T.~M.,
  {Forster}, K., {Friedman}, P.~G., {Martin}, D.~C., {Morrissey}, P., {Neff},
  S.~G., {Small}, T., {Wyder}, T.~K., {Bianchi}, L., {Donas}, J., {Lee}, Y.-W.,
  {Madore}, B.~F., {Milliard}, B., {Szalay}, A.~S., {Welsh}, B.~Y., \& {Yi},
  S.~K. 2007, \apjs, 173, 267


\bibitem[{{S{\'a}nchez} {et~al.}(2013){S{\'a}nchez}, {Rosales-Ortega},
  {Jungwiert}, {Iglesias-P{\'a}ramo}, {V{\'{\i}}lchez}, {Marino}, {Walcher},
  {Husemann}, {Mast}, {Monreal-Ibero}, {Cid Fernandes}, {P{\'e}rez},
  {Gonz{\'a}lez Delgado}, {Garc{\'{\i}}a-Benito}, {Galbany}, {van de Ven},
  {Jahnke}, {Flores}, {Bland-Hawthorn}, {L{\'o}pez-S{\'a}nchez}, {Stanishev},
  {Miralles-Caballero}, {D{\'{\i}}az}, {S{\'a}nchez-Blazquez}, {Moll{\'a}},
  {Gallazzi}, {Papaderos}, {Gomes}, {Gruel}, {P{\'e}rez}, {Ruiz-Lara},
  {Florido}, {de Lorenzo-C{\'a}ceres}, {Mendez-Abreu}, {Kehrig}, {Roth},
  {Ziegler}, {Alves}, {Wisotzki}, {Kupko}, {Quirrenbach}, {Bomans}, \& {Califa
  Collaboration}}]{Sanchez+2013}
{S{\'a}nchez}, S.~F., {Rosales-Ortega}, F.~F., {Jungwiert}, B.,
  {Iglesias-P{\'a}ramo}, J., {V{\'{\i}}lchez}, J.~M., {Marino}, R.~A.,
  {Walcher}, C.~J., {Husemann}, B., {Mast}, D., {Monreal-Ibero}, A., {Cid
  Fernandes}, R., {P{\'e}rez}, E., {Gonz{\'a}lez Delgado}, R.,
  {Garc{\'{\i}}a-Benito}, R., {Galbany}, L., {van de Ven}, G., {Jahnke}, K.,
  {Flores}, H., {Bland-Hawthorn}, J., {L{\'o}pez-S{\'a}nchez}, A.~R.,
  {Stanishev}, V., {Miralles-Caballero}, D., {D{\'{\i}}az}, A.~I.,
  {S{\'a}nchez-Blazquez}, P., {Moll{\'a}}, M., {Gallazzi}, A., {Papaderos}, P.,
  {Gomes}, J.~M., {Gruel}, N., {P{\'e}rez}, I., {Ruiz-Lara}, T., {Florido}, E.,
  {de Lorenzo-C{\'a}ceres}, A., {Mendez-Abreu}, J., {Kehrig}, C., {Roth},
  M.~M., {Ziegler}, B., {Alves}, J., {Wisotzki}, L., {Kupko}, D.,
  {Quirrenbach}, A., {Bomans}, D., \& {Califa Collaboration}. 2013, \aap, 554,
  A58


\bibitem[{{Schruba} {et~al.}(2012){Schruba}, {Leroy}, {Walter}, {Bigiel},
  {Brinks}, {de Blok}, {Kramer}, {Rosolowsky}, {Sandstrom}, {Schuster},
  {Usero}, {Weiss}, \& {Wiesemeyer}}]{Schruba+2012}
{Schruba}, A., {Leroy}, A.~K., {Walter}, F., {Bigiel}, F., {Brinks}, E., {de
  Blok}, W.~J.~G., {Kramer}, C., {Rosolowsky}, E., {Sandstrom}, K., {Schuster},
  K., {Usero}, A., {Weiss}, A., \& {Wiesemeyer}, H. 2012, \aj, 143, 138


\bibitem[{{Serra} {et~al.}(2012){Serra}, {Oosterloo}, {Morganti}, {Alatalo},
  {Blitz}, {Bois}, {Bournaud}, {Bureau}, {Cappellari}, {Crocker}, {Davies},
  {Davis}, {de Zeeuw}, {Duc}, {Emsellem}, {Khochfar}, {Krajnovi{\'c}},
  {Kuntschner}, {Lablanche}, {McDermid}, {Naab}, {Sarzi}, {Scott}, {Trager},
  {Weijmans}, \& {Young}}]{Serra+2012}
{Serra}, P., {Oosterloo}, T., {Morganti}, R., {Alatalo}, K., {Blitz}, L.,
  {Bois}, M., {Bournaud}, F., {Bureau}, M., {Cappellari}, M., {Crocker}, A.~F.,
  {Davies}, R.~L., {Davis}, T.~A., {de Zeeuw}, P.~T., {Duc}, P.-A., {Emsellem},
  E., {Khochfar}, S., {Krajnovi{\'c}}, D., {Kuntschner}, H., {Lablanche},
  P.-Y., {McDermid}, R.~M., {Naab}, T., {Sarzi}, M., {Scott}, N., {Trager},
  S.~C., {Weijmans}, A.-M., \& {Young}, L.~M. 2012, \mnras, 422, 1835


\bibitem[{{Simard} {et~al.}(2011){Simard}, {Mendel}, {Patton}, {Ellison}, \&
  {McConnachie}}]{Simard+2011}
{Simard}, L., {Mendel}, J.~T., {Patton}, D.~R., {Ellison}, S.~L., \&
  {McConnachie}, A.~W. 2011, \apjs, 196, 11


\bibitem[{{Somerville} \& {Dav{\'e}}(2015)}]{Somerville+2015}
{Somerville}, R.~S. \& {Dav{\'e}}, R. 2015, \araa, 53, 51


\bibitem[{{Springob} {et~al.}(2005){Springob}, {Haynes}, {Giovanelli}, \&
  {Kent}}]{Springob+2005}
{Springob}, C.~M., {Haynes}, M.~P., {Giovanelli}, R., \& {Kent}, B.~R. 2005,
  \apjs, 160, 149


\bibitem[{{Stark} {et~al.}(2013){Stark}, {Kannappan}, {Wei}, {Baker}, {Leroy},
  {Eckert}, \& {Vogel}}]{Stark+2013}
{Stark}, D.~V., {Kannappan}, S.~J., {Wei}, L.~H., {Baker}, A.~J., {Leroy},
  A.~K., {Eckert}, K.~D., \& {Vogel}, S.~N. 2013, \apj, 769, 82


\bibitem[{{Stewart} {et~al.}(2009){Stewart}, {Bullock}, {Wechsler}, \&
  {Maller}}]{Stewart+2009}
{Stewart}, K.~R., {Bullock}, J.~S., {Wechsler}, R.~H., \& {Maller}, A.~H. 2009,
  \apj, 702, 307


\bibitem[{{Swaters} \& {Balcells}(2002)}]{Swaters&Balcells2002}
{Swaters}, R.~A. \& {Balcells}, M. 2002, \aap, 390, 863


\bibitem[{{Toomre}(1964)}]{Toomre}
{Toomre}, A. 1964, \apj, 139, 1217


\bibitem[{{van Driel} {et~al.}(2016){van Driel}, {Butcher}, {Schneider},
  {Lehnert}, {Minchin}, {Blyth}, {Chemin}, {Hallet}, {Joseph}, {Kotze},
  {Kraan-Korteweg}, {Olofsson}, \& {Ramatsoku}}]{vanDriel+2016}
{van Driel}, W., {Butcher}, Z., {Schneider}, S., {Lehnert}, M.~D., {Minchin},
  R., {Blyth}, S.-L., {Chemin}, L., {Hallet}, N., {Joseph}, T., {Kotze}, P.,
  {Kraan-Korteweg}, R.~C., {Olofsson}, A.~O.~H., \& {Ramatsoku}, M. 2016, \aap,
  595, A118


\bibitem[{{Verheijen}(1997)}]{Verheijen+1997}
{Verheijen}, M.~A.~W. 1997, PhD thesis, PhD thesis, Univ.~Groningen, The
  Netherlands , (1997)


\bibitem[{{Walter} {et~al.}(2008){Walter}, {Brinks}, {de Blok}, {Bigiel},
  {Kennicutt}, {Thornley}, \& {Leroy}}]{Walter+2008}
{Walter}, F., {Brinks}, E., {de Blok}, W.~J.~G., {Bigiel}, F., {Kennicutt},
  Jr., R.~C., {Thornley}, M.~D., \& {Leroy}, A. 2008, \aj, 136, 2563


\bibitem[{{Wei} {et~al.}(2010){Wei}, {Kannappan}, {Vogel}, \&
  {Baker}}]{Wei+2010a}
{Wei}, L.~H., {Kannappan}, S.~J., {Vogel}, S.~N., \& {Baker}, A.~J. 2010, \apj,
  708, 841


\bibitem[{{Welch} {et~al.}(2010){Welch}, {Sage}, \& {Young}}]{Welch+2010}
{Welch}, G.~A., {Sage}, L.~J., \& {Young}, L.~M. 2010, \apj, 725, 100


\bibitem[{{Willick} {et~al.}(1997){Willick}, {Strauss}, {Dekel}, \&
  {Kolatt}}]{Willick+1997}
{Willick}, J.~A., {Strauss}, M.~A., {Dekel}, A., \& {Kolatt}, T. 1997, \apj,
  486, 629


\bibitem[{{Wolfire} {et~al.}(2010){Wolfire}, {Hollenbach}, \&
  {McKee}}]{Wolfire+2010}
{Wolfire}, M.~G., {Hollenbach}, D., \& {McKee}, C.~F. 2010, \apj, 716, 1191


\bibitem[{{Wright} {et~al.}(2017){Wright}, {Robotham}, {Driver}, {Alpaslan},
  {Andrews}, {Baldry}, {Bland-Hawthorn}, {Brough}, {Brown}, {Colless}, {da
  Cunha}, {Davies}, {Graham}, {Holwerda}, {Hopkins}, {Kafle}, {Kelvin},
  {Loveday}, {Maddox}, {Meyer}, {Moffett}, {Norberg}, {Phillipps}, {Rowlands},
  {Taylor}, {Wang}, \& {Wilkins}}]{Wright+2017}
{Wright}, A.~H., {Robotham}, A.~S.~G., {Driver}, S.~P., {Alpaslan}, M.,
  {Andrews}, S.~K., {Baldry}, I.~K., {Bland-Hawthorn}, J., {Brough}, S.,
  {Brown}, M.~J.~I., {Colless}, M., {da Cunha}, E., {Davies}, L.~J.~M.,
  {Graham}, A.~W., {Holwerda}, B.~W., {Hopkins}, A.~M., {Kafle}, P.~R.,
  {Kelvin}, L.~S., {Loveday}, J., {Maddox}, S.~J., {Meyer}, M.~J., {Moffett},
  A.~J., {Norberg}, P., {Phillipps}, S., {Rowlands}, K., {Taylor}, E.~N.,
  {Wang}, L., \& {Wilkins}, S.~M. 2017, \mnras, 470, 283


\bibitem[{{Yang} {et~al.}(2007){Yang}, {Mo}, {van den Bosch}, {Pasquali}, {Li},
  \& {Barden}}]{Yang+2007}
{Yang}, X., {Mo}, H.~J., {van den Bosch}, F.~C., {Pasquali}, A., {Li}, C., \&
  {Barden}, M. 2007, \apj, 671, 153


\bibitem[{{Young} {et~al.}(1995){Young}, {Xie}, {Tacconi}, {Knezek}, {Viscuso},
  {Tacconi-Garman}, {Scoville}, {Schneider}, {Schloerb}, {Lord}, {Lesser},
  {Kenney}, {Huang}, {Devereux}, {Claussen}, {Case}, {Carpenter}, {Berry}, \&
  {Allen}}]{Young+1995}
{Young}, J.~S., {Xie}, S., {Tacconi}, L., {Knezek}, P., {Viscuso}, P.,
  {Tacconi-Garman}, L., {Scoville}, N., {Schneider}, S., {Schloerb}, F.~P.,
  {Lord}, S., {Lesser}, A., {Kenney}, J., {Huang}, Y.-L., {Devereux}, N.,
  {Claussen}, M., {Case}, J., {Carpenter}, J., {Berry}, M., \& {Allen}, L.
  1995, \apjs, 98, 219


\bibitem[{{Young} {et~al.}(2011){Young}, {Bureau}, {Davis}, {Combes},
  {McDermid}, {Alatalo}, {Blitz}, {Bois}, {Bournaud}, {Cappellari}, {Davies},
  {de Zeeuw}, {Emsellem}, {Khochfar}, {Krajnovi{\'c}}, {Kuntschner},
  {Lablanche}, {Morganti}, {Naab}, {Oosterloo}, {Sarzi}, {Scott}, {Serra}, \&
  {Weijmans}}]{Young+2011}
{Young}, L.~M., {Bureau}, M., {Davis}, T.~A., {Combes}, F., {McDermid}, R.~M.,
  {Alatalo}, K., {Blitz}, L., {Bois}, M., {Bournaud}, F., {Cappellari}, M.,
  {Davies}, R.~L., {de Zeeuw}, P.~T., {Emsellem}, E., {Khochfar}, S.,
  {Krajnovi{\'c}}, D., {Kuntschner}, H., {Lablanche}, P.-Y., {Morganti}, R.,
  {Naab}, T., {Oosterloo}, T., {Sarzi}, M., {Scott}, N., {Serra}, P., \&
  {Weijmans}, A.-M. 2011, \mnras, 414, 940


\bibitem[{{Zhang} {et~al.}(2009){Zhang}, {Li}, {Kauffmann}, {Zou}, {Catinella},
  {Shen}, {Guo}, \& {Chang}}]{Zhang+2009}
{Zhang}, W., {Li}, C., {Kauffmann}, G., {Zou}, H., {Catinella}, B., {Shen}, S.,
  {Guo}, Q., \& {Chang}, R. 2009, \mnras, 397, 1243


\bibitem[{{Zibetti} {et~al.}(2009){Zibetti}, {Charlot}, \&
  {Rix}}]{Zibetti+2009}
{Zibetti}, S., {Charlot}, S., \& {Rix}, H.-W. 2009, \mnras, 400, 1181


\bibitem[{{Zu} \& {Mandelbaum}(2015)}]{Zu+2015}
{Zu}, Y. \& {Mandelbaum}, R. 2015, \mnras, 454, 1161


\bibitem[{{Zwaan} {et~al.}(2005){Zwaan}, {Meyer}, {Staveley-Smith}, \&
  {Webster}}]{Zwaan+2005}
{Zwaan}, M.~A., {Meyer}, M.~J., {Staveley-Smith}, L., \& {Webster}, R.~L. 2005,
  \mnras, 359, L30


\end{thebibliography}

\end{document}